# Subnanometer thermodynamic overlayers on bimetallic nanoparticles

Caitlin A. McCandler[1,3,#], Xianzhuo Lao[2,#], Jie Liu[2], Huipu Liu[2], Kristin A. Persson[1,*], Peng-Cheng Chen[2,4,*]

[1]Department of Materials Science and Engineering, University of California Berkeley, Berkeley, CA 94720, USA.

[2]College of Smart Materials and Future Energy, State Key Laboratory of Coatings for Advanced Equipment, Fudan University, Shanghai 200438, China.

[3]Initiative for Computational Catalysis, Flatiron Institute, New York, NY 10010, USA.

[4]International Institute for Intelligent Nanorobots and Nanosystems, Fudan University, Shanghai 200438, China.

#These authors contributed equally to this work.

Correspondence to: pcchen@fudan.edu.cn (P.-C.C.); kristinpersson@berkeley.edu (K.A.P.)

**Abstract:**

Nanoparticle properties are governed by their surfaces, yet their atomic-scale surface structures remain challenging to predict. Here, we report that bimetallic nanoparticles can form a thermodynamically controlled "shell-dimer" architecture in which one metal forms an overlayer only a few atomic layers thick on another. Using Au-Rh as a model system, atomic-resolution imaging and molecular dynamics show that an ultrathin Au overlayer forms on Rh, stabilized by competition among surface, interfacial, and strain energies. Anisotropic strain limits its growth to the subnanometer scale, and changes in surface chemistry can destabilize the overlayer altogether. Across a range of bimetallic nanoparticles, we found that overlayer formation is associated with elemental immiscibility and lattice mismatch. These findings provide a basis for understanding and controlling surface structures in multimetallic nanoparticles.

**Main Text:**

The physical and chemical properties of nanoparticles are often governed disproportionately by their outermost surface layers. This is especially evident in catalysis, where small changes in surface composition, coordination, and structure can strongly alter activity, selectivity, and stability (*1-4*), but the same principle extends to plasmonics, sensing, magnetic heating, and biomedical therapy (*5-7*). The nanoparticle surface is therefore not simply the termination of a bulk crystal, but a distinct functional region whose structure and properties can differ substantially from the nanoparticle interior. For example, the electronic structure of an exposed atom depends on its local environment, so nanoparticles terminated by the same element can exhibit different reactivity depending on its subsurface composition, overlayer thickness, and local coordination (*8, 9*). Strain arising from surface relaxation and lattice mismatch between adjoining phases can modify adsorption energetics and catalytic activity, with effects that can propagate across several atomic layers (*1, 2, 4, 10-13*). Nanoparticle surfaces are also enriched in defects and can restructure in response to adsorbates or supports, adopting atomic arrangements distinct from the ideal bulk crystal structure (*14, 15*).

The strong dependence of nanoparticle function on surface structure has motivated extensive efforts to synthesize nanoparticles with ultrathin shell overlayers and other engineered surface architectures (*1-8, 14, 16-21*). However, these surface structures may not remain stable under operating conditions. *In situ* studies have shown that nanoparticles can undergo surface segregation, alloying, dealloying, encapsulation, and other structural transformations, driven by temperature, surface-bound molecules or atoms (adsorbates), electrochemical potential, and support interactions (*19, 22-26*). For example, changes in temperature or gas composition can switch which element preferentially segregates to the nanoparticle surface and forms specific overlayers, by altering the relative stability of adsorbate-free and passivated surfaces (*15, 22-24, 27-30*). In addition, adsorbates can promote surface alloying in otherwise phase-separated nanoparticles (*19, 31, 32*). These observations demonstrate that the surface structure of nanoparticles reflects a balance between bulk mixing thermodynamics, surface and interfacial energetics, and strain (*29-31, 33-35*). However, a general framework for predicting the surface architecture of bimetallic nanoparticles as a function of composition and environmental adsorbates remains lacking.

Here, we experimentally investigated the phase behavior and surface structure of a series of bimetallic nanoparticles and identified a thermodynamic nanoparticle architecture that we term a shell-dimer, in which one metal forms an anisotropically strained overlayer several atomic layers thick on the other metal. We show that the stability, thickness, and anisotropy of the overlayers arise from competition among mixing enthalpy, surface passivation, interfacial tension, and epitaxial strain. This thermodynamic framework explains how competing nanoscale energetic contributions stabilize finite-thickness overlayers on nanoparticles and establishes design principles for controlling the surface structure of bimetallic nanoparticles.

**Au overlayer with anisotropic strains on AuRh nanoparticles**

Using Au–Rh as a model system, we found that Au and Rh phase-separate into two half-sphere domains within a single surface-clean nanoparticle, with the Rh domain encapsulated by a conformal Au overlayer only 1–4 atomic layers thick, resulting in an AuRh shell–dimer architecture (Figs. 1 and S1-S6). The surface-clean nanoparticles were synthesized by direct thermal annealing of a mixture of $AuCl_3$ and $RhCl_3$ to convert the precursor salts into nanoparticles with thermodynamically equilibrated structures (Fig. S1). Aberration-corrected high-angle annular dark-field scanning transmission electron microscopy (HAADF-STEM) characterization of the nanoparticles directly resolves the Au overlayer and the underlying Au–Rh domain structure, aided by the atomic-number-dependent contrast between Au and Rh. Energy-dispersive X-ray spectroscopy (EDS) elemental mapping further confirms the phase separation of Au and Rh and the encapsulation of the Rh domain by the Au overlayer (Figs. 1B, 1C, and S2-S6).

Notably, the Au overlayers exhibit anisotropic strains, i.e., compressive strains parallel to the nanoparticle surface and tensile strains perpendicular to the particle surface (Figs. 1D-1F). Due to the difference between Au and Rh in lattice parameters (lattice constants: Au 4.07 Å, Rh 3.80 Å), the Au atoms of the overlayers are compressed in the direction parallel to the particle surface so that the Au lattice adapts to the underlying Rh lattice, forming a coherent Au-Rh interface. The Au atomic layer right above the Rh domain is the most compressed, exhibiting a projected atomic distance of 0.235 nm in the (111) planes viewed along the [110] zone axis (Figs. 1E and S7, measured from HAADF-STEM images). As the Au layers get closer to the particle surface, the interatomic distances in each layer slightly expand, terminating at the surface Au layer with a projected atomic distance of 0.238 nm, which is still smaller than the 0.249 nm of bulk Au. The results suggest that the epitaxial constraint of the Rh domain leads to the in-plane compressive strain of the Au overlayer. In contrast, the outermost atomic layers of the Au domain retain the bulk Au interatomic spacing (Fig. 1E), except for a slight contraction in the surface layer due to compression at the nanoparticle surface.

Different from the compressed state of the Au overlayer in the direction parallel to the particle surface, the interplanar spacing between the Au atomic layers in the overlayer is substantially extended. Specifically, the spacing between the Au (111) planes increases with increasing distance from the Rh domain. As shown in Fig. 1F, the interplanar spacings between the (111) layers of the Rh domain, the Rh domain and the 3$^{rd}$ Au layer, the 3$^{rd}$ Au and 2$^{nd}$ Au layers, and the 2$^{nd}$ Au and surface Au layers are 0.220, 0.234, 0.247, and 0.254 nm, respectively. The spacing between the outermost Au (111) atomic layers is significantly larger than that of bulk Au (0.236 nm), confirming the presence of extreme tensile strains in the direction perpendicular to the particle surface. In comparison, the spacing between the outermost Au (111) layers on the Au domain is close to that of bulk Au (Fig. 1F). Experimentally, we found that the thickness of the Au overlayer on Rh is always limited to no more than four atomic layers, presumably due to the substantial anisotropic strains developed in the multilayer structure. In addition, to exclude the possibility that the overlayer structure was kinetically trapped, we synthesized AuRh nanoparticles by preparing monometallic nanoparticles first, followed by adding the second metal salts to generate the bimetallic nanoparticles. Regardless of the preparation sequence, HAADF-STEM characterization of these AuRh nanoparticles show the same shell-dimer architecture with a subnanometer Au overlayer on Rh (Fig. S8), confirming the thermodynamic stability of the overlayer structure.

**Thermodynamic origin of the overlayer structure**

To identify the energetic factors stabilizing the experimentally observed shell-dimer architecture, we compared three phase-separated Au–Rh nanoparticle morphologies using molecular dynamics (MD) simulations: a conventional dimer, a shell-dimer, and a core–shell nanoparticle (Fig. 2). Representative geometries of each morphology were constructed, simulated at elevated temperature, and then quenched to room temperature (see the supplementary materials for details). The final relaxed geometries are shown in Fig. 2A. Among them, the shell-dimer had the lowest potential energy, while the core–shell and dimer structures were higher in energy by 25 and 36 meV per atom, respectively. These results indicate that neither simple phase separation nor symmetric encapsulation is thermodynamically optimal.

The per-atom potential-energy distributions further reveal the origins of these energetic penalties (Fig. 2B). Red coloring marks atoms with elevated energies relative to their corresponding bulk phases, with the largest penalties occurring at surfaces and interfaces, particularly at the exposed Rh surface and the Au–Rh interface. The dimer is destabilized by the high energy of exposed Rh, whereas Au encapsulation lowers this surface-energy contribution in both the shell-dimer and core–shell structures. The energy difference between these two encapsulated morphologies appears to arise primarily from strain accumulation within the Au shell of the core–shell nanoparticle, which promotes defect formation and loss of crystallographic coherence (Fig. 2C). In contrast, the shell-dimer accommodates the Au–Rh lattice mismatch through a localized directional strain field near the interface while retaining a comparatively ordered Au domain (Figs. 2C, S9, and S10). Sequential cross-sections of the three nanoparticle models illustrate the three-dimensional strain distribution and confirm that the Au overlayer remains continuous around the Rh domain (Fig. 2D). Overall, the simulations show that the shell-dimer is stabilized by a balance among surface, interfacial, strain, and defect energies: Au coverage removes most of the high-energy Rh surface, while the finite overlayer thickness limits strain accumulation and preserves crystallographic coherence.

Given that the Au overlayer is thermodynamically stabilized due to the interplay of surface, interface, strain, and defect energies, we next systematically analyzed AuRh nanoparticles with different Au content to further understand the formation of Au overlayers. As shown in Figs. 3A and S11-S14, nanoparticles containing trace amounts of Au or Rh tend to form alloy nanoparticles with no observable surface enrichment of Au. When the Au content is increased to ~20% (atomic percent), the limited amount of Au atoms mainly contribute to maximizing the Au coverage on Rh, while individual Au domains are rarely observed in the nanoparticles. Further increasing the Au content from ~40% to ~90% leads to an excess of Au atoms for covering the Rh domain. Additional Au atoms preferentially segregate into a single domain within the nanoparticle, rather than extending the Au overlayer, owing to the strain accumulated in the latter. For AuRh nanoparticles of Au content between 20%-90%, the Au overlayer coverage is consistently ~90% (Figs. S11 and S12). The overlayers on these nanoparticles all consist of 1-4 atomic layers bearing anisotropic strains, with 2 and 3 atomic layers showing the highest frequency (Figs. S11, S15, and S16).

## Overlayer structure regulated by adsorbates

In practical synthesis and application scenarios of nanoparticles, the particles are often bound by ligands or other adsorbates that originate from the environment. To examine whether adsorbates affect the overlayers, we introduced six commonly used colloidal nanoparticle ligands into the preparation process: polyvinylpyrrolidone (PVP), sodium dodecyl sulfate (SDS), poly(ethylene glycol) dithiol (HS-PEG-HS), polyethylene glycol (PEG), hexadecyltrimethylammonium bromide (CTAB), and trisodium citrate (Citrate) (Figs. 3B-3F and S17-S24). The ligand molecules are partially decomposed during the thermal annealing process, leaving residue adsorbates that could alter the surface energy of Au and Rh. X-ray photoelectron spectroscopy (XPS) characterization of the surface of modified nanoparticles presents clear signals of elements from corresponding adsorbates used (Figs. 3C, 3D, and S18-S22). Unlike surface-clean AuRh nanoparticles, in which Au uniformly encapsulates Rh to form a subnanometer overlayer, nanoparticles modified by adsorbates exhibit dramatic changes with respect to the overlayer structure. In particular, the Au skin breaks up into discrete patches and the overlayer coverage decreases, exposing the underlying Rh region (Figs. 3B and S18-S24). Quantitative analysis of the Au overlayers further corroborates that the adsorbates lead to a decrease in Au overlayer coverage (Fig. 3E). Importantly, the order of Au overlayer coverage is PVP > SDS > HS-PEG-HS > PEG > CTAB > Citrate, with negligible Au overlayer coverage in the case of Citrate, suggesting that the chemical binding of adsorbates is an important factor affecting the formation of metal overlayers. Fig. 3F provides a detailed breakdown of the percentage of overlayers with different thicknesses. On surface-clean nanoparticles, Au overlayers are 1-4 atomic layers thick. The overlayer thickness decreases upon introduction of different adsorbates, and it shows the same decreasing order as the decrease of coverage caused by the adsorbates. Specifically, PVP has a minimal effect on the thickness and structure of Au overlayers (Fig. S18). When using the HS-PEG-HS and PEG, the proportion of bilayer Au increases, trilayer Au decreases, and tetralayer Au almost disappears (Figs. S20, S23, and S24). When using the CTAB and Citrate, the Rh domains in the nanoparticles are mainly capped by monolayer Au or exposed without any Au coverage (Figs. S21 and S22). Collectively, with the increase of adsorbate binding affinity, Au overlayers are gradually eliminated from the nanoparticles.

From an energy perspective, the addition of adsorbates alters the relative surface energy of the two metals, weakening or even reversing the original driving force of Au segregation on the Rh surface, thereby triggering a transformation in the thermodynamically stable nanoparticle configuration. To rationalize these adsorbate-dependent structural changes, we developed a thermodynamic model that considers the principal energetic contributions governing phase stability in bimetallic nanoparticles, i.e., bulk mixing, interfacial energy, surface energy, and strain. The tendency of two metals to mix or phase-separate is described by the regular-solution interaction parameter, $w$, derived from the enthalpy of mixing, together with the interfacial tension, $\gamma$, between compositionally distinct domains (Figs. 4A and S25). These quantities are strongly correlated: elemental pairs with unfavorable bulk mixing generally also exhibit larger positive interfacial energies, whereas more miscible pairs have smaller or even negative interfacial energies. Bulk and interfacial thermodynamics alone, however, do not determine nanoparticle structure because surface contributions become increasingly important as particle size decreases. The surface energy, $\alpha$, therefore contributes substantially to the total energy of nanoscale systems. Adsorption of environmental species, including gases and ligands, can substantially change surface energies and consequently the relative stability of competing nanoparticle phases (Fig. S26). Density functional theory (DFT) calculations illustrate this passivation effect, showing that hydrogen (H) and methyl ($CH_3$) adsorption can strongly modify the surface energies of several metals (Figs. 4B and S27-S30). The chemical environment therefore provides a direct handle for tuning the preference of each element to occupy the nanoparticle surface. We first used Au-Rh as a representative case study to show the influence of adsorbates on the competition among nanoparticle phases (Fig. 4C). In the absence of passivation, the substantially lower surface energy of Au relative to Rh provides a strong driving force for Au to segregate to the exterior, despite the energetic cost of forming an Au-Rh interface. As the surfaces become progressively passivated by $CH_3$, the relative surface energies of Au and Rh become more similar, weakening the driving force for Au encapsulation and ultimately stabilizing the dimer configuration. H adsorption produces a related effect: because H binds substantially more strongly to Rh than to Au, increasing the hydrogen chemical potential selectively lowers the Rh surface energy and, at sufficiently high $H_2$ pressures, is predicted to drive a transition in the nanoparticle structure to the dimer phase. These results provide a thermodynamic explanation for the experimentally observed breakup and thinning of the Au overlayer upon addition of strongly binding adsorbates. Rather than simply perturbing an existing shell, adsorbates reorder the relative free energies of competing nanoparticle structures.

Across different bimetallic systems, the predicted phase behavior depends strongly on the intrinsic miscibility of the elemental pair (Figs. 4D and S31-S33). Relatively miscible systems remain alloyed over a broad range of surface energies, whereas increasing immiscibility favors phase-separated structures and expands the range of conditions under which dimer and overlayer morphologies become thermodynamically accessible. Thus, bulk miscibility and environmentally dependent surface energies together provide a general framework for determining whether a bimetallic nanoparticle favors alloying, surface segregation, or formation of an overlayer.

**Generality of overlayer structure in bimetallic nanoparticles**

To test the predicted overlayer formation in bimetallic systems, we further investigated a range of surface-clean nanoparticles spanning immiscible, partially miscible, and miscible elemental pairs (Figs. 5 and S34-S53). Importantly, thermodynamic overlayers are only observed in immiscible or partially miscible elements, consistent with the predicted phase diagrams (Figs. 4C, 4D, 5A, 5B, and S31-S33). For nanoparticles composed of immiscible elements such as AuRu, AuNi, AuCo, AgRh, and AgCu (Fig. S35-S48), HAADF-STEM images and EDS mapping clearly show the phase separation of the elements as well as the formation of subnanometer overlayers on the nanoparticles. Au overlayers can be generated on Rh, Ni, Ru, and Co domains, and Ag overlayers can be generated on Rh and Cu domains (Fig. 5A). Au and Ag, with their low surface energies, act as capping layers on the other metal domains to minimize the total surface and interfacial energy of the nanoparticles. Meanwhile, the immiscibility between the elements stabilizes the overlayer structure by preventing interfacial alloying. The actual thickness of the overlayers depends on each specific system, due to the varying degrees of lattice mismatch and the strains accumulated accordingly. For partially miscible systems such as CuRh (Figs. 5B and S49-S51), surface segregation of the Cu element with lower surface energy can be observed in the nanoparticles, while the interface between the overlayer and core domain is not distinct. In contrast to the former two scenarios, miscible elements such as AuPd (Figs. 5C and S51-S53) exhibit complete alloying on the surface with no obvious surface segregation, as predicted by our theoretical analysis (Fig. 4).

Further strain analysis of the immiscible binary systems shows consistent anisotropic strain behavior in the overlayers to that observed in AuRh (Figs. 5D, S37, S43, S46, and S48). For example, in AuCo, AuNi, and AuRu nanoparticles, the Au overlayers present compressive strains parallel to the particle surface as the lattice parameters of Co, Ni, and Ru are smaller than that of Au (Fig. S48). The interatomic distances within each Au layer slightly increase with increasing distance from the Rh domain but remain smaller than the interatomic distances of bulk Au. In the direction perpendicular to the nanoparticle surface, the Au overlayers present significant tensile strain (Fig. 5D). Atomic column displacement analysis of the AuCo and AuRh nanoparticles, referenced to the Co and Rh lattice, respectively, visually show the accumulation of tensile strains in the direction perpendicular to the nanoparticle surface (Fig. 5E). Notably, in all the Au-based binary systems, the interplanar spacing in the Au overlayers progressively increases from the layer immediately above the other metal domains toward the surface layer, ultimately reaching ~0.25 nm (Figs. 5D and 5E). This value is presumably the maximum strain that can be developed in the Au overlayers, which stops it from further growing and resulting in the subnanometer skin structure. Taken together, elemental immiscibility is an essential factor to form thermodynamically stable overlayers on nanoparticles, and the overlayer thickness is dependent in part on the lattice mismatch between the component metals. As the miscibility of constituent elements increases from complete immiscibility to partial miscibility to complete miscibility, the surface structure of resulting nanoparticles changes from a subnanometer-thick overlayer to surface segregation to complete surface alloying, providing a universal principle for the precise design of bimetallic nanoparticles with specific surface structures.

## Conclusions

We have identified a thermodynamically stable shell-dimer architecture with subnanometer-thick overlayers that expands the range of surface structures accessible in bimetallic nanoparticles beyond conventional alloying, segregation, and core–shell structures. Our thermodynamic framework explains how surface, interfacial, and strain energies together drive overlayer formation and make continued growth increasingly unfavorable, providing a unified mechanistic basis for identifying elemental combinations that favor this surface structure. These ultrathin, anisotropically strained overlayers create surface environments that differ from those of the individual constituent metals, providing new opportunities to tune surface properties through bimetallic composition. The generality of this thermodynamic surface structure extends beyond the specific binary systems demonstrated here and provides a vast design space for nanoparticles with tunable properties relevant to applications such as catalysis and plasmonics.

**Acknowledgements: Funding:** This work was supported by the National Natural Science Foundation of China (92477117 and 22375049). This research used resources of the National Energy Research Scientific Computing Center (NERSC), a U.S. Department of Energy Office of Science User Facility operated under contract no. DE-AC02-05CH11231. P.-C.C. acknowledges support from Xiaomi Young Scholar Award. C.A.M. acknowledges the Kavli ENSI Graduate Student Fellowship for financial support. The Flatiron Institute is a division of the Simons Foundation. **Author Contributions:** Conceptualization: C.A.M., X.L., K.A.P., P.-C.C.; Nanoparticle synthesis and characterization: X.L., J.L., H.L., P.-C.C.; Simulation: C.A.M., K.A.P.; Data analysis: C.A.M., X.L., J.L., K.A.P., P.-C.C.; Supervision: K.A.P., P.-C.C.; Manuscript writing and editing: C.A.M., X.L., K.A.P., P.-C.C. **Competing interests:** The authors declare no competing financial interest. **Data and materials availability**: All data are available in the main text or the supplementary materials.

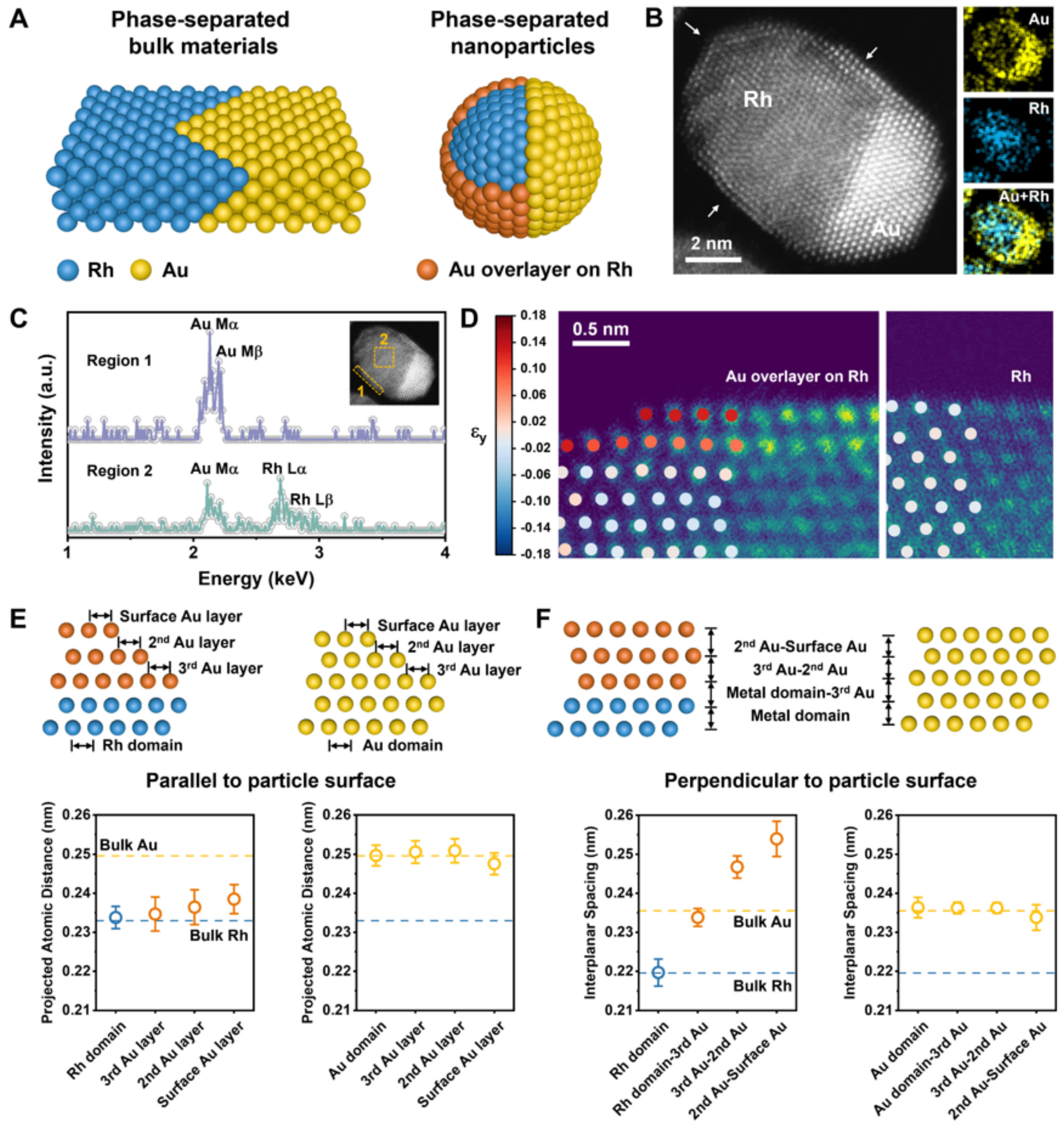

**Fig. 1. Au overlayer with anisotropic strains formed on surface-clean AuRh nanoparticles.** (**A**) Schematic illustration of phase-separated bulk materials and nanoparticles. (**B**) HAADF-STEM image and EDS elemental mapping of a representative AuRh nanoparticle. (**C**) EDS spectra of different regions of the nanoparticle. (**D**) Comparison between the HAADF-STEM images of a Rh domain covered by Au overlayer and an exposed Rh domain without Au overlayer. Estimated positions of the Au and Rh atomic columns are indicated by solid dots. Strain in the y direction is calculated by the displacements of atomic columns with respect to an ideal Rh lattice. (**E**) Strain analysis of the Au and Rh domains in the direction parallel to the nanoparticle surface. Schemes depict the projected atomic distance measured within the (111) planes viewed along the [110] zone axis. The (111) planes are parallel to the nanoparticle surface. (**F**) Strain analysis of the Au and Rh domains in the direction perpendicular to the nanoparticle surface. Schemes depict the interplanar spacing measured between the (111) planes viewed along the [110] zone axis.

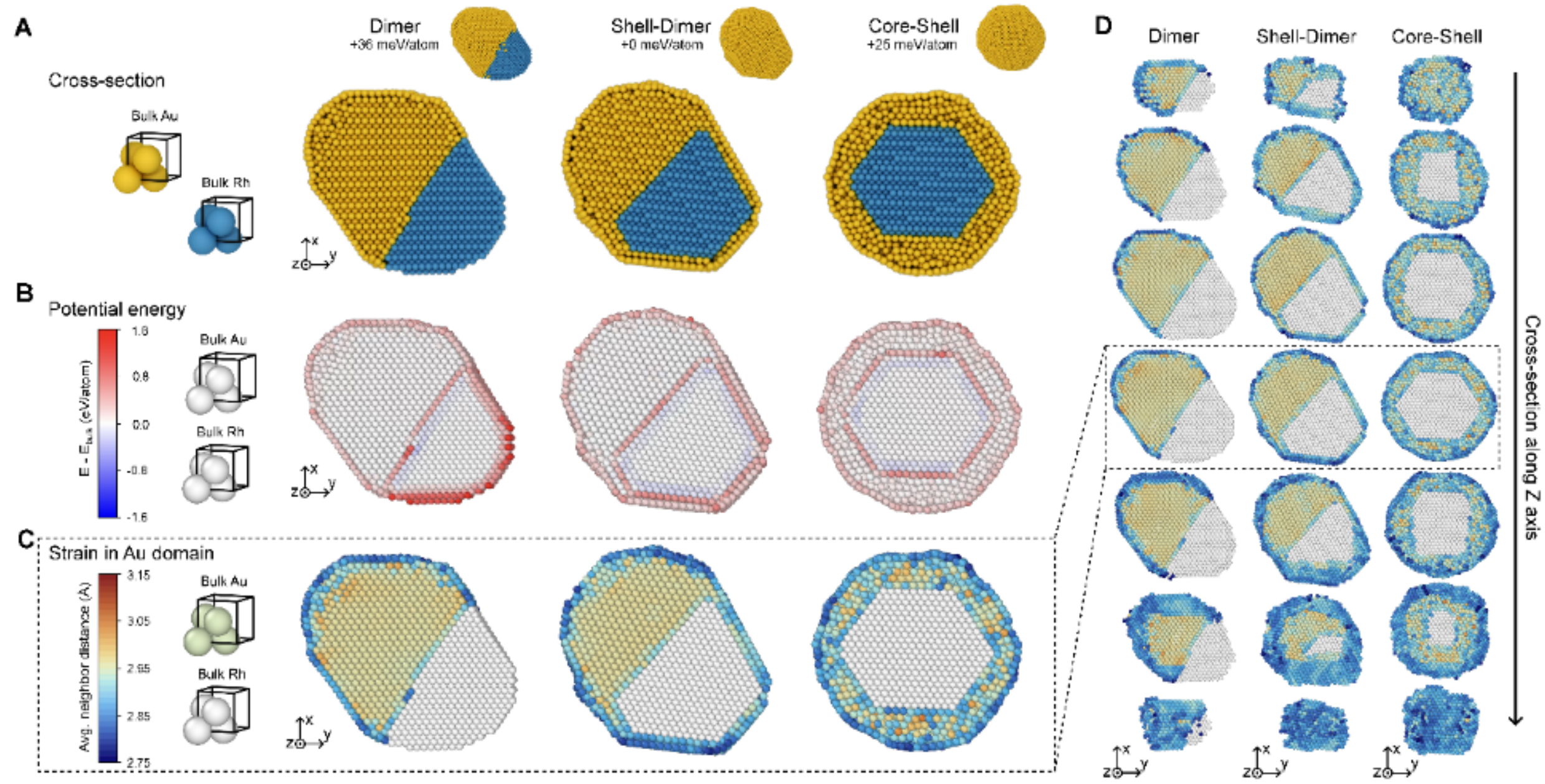

**Fig. 2. Driving forces governing the stability of the shell-dimer morphology.** (**A**) Cross-sectional views of representative Au–Rh dimer, shell-dimer, and core–shell nanoparticles, along with their relative energies calculated from MD simulations. Au and Rh atoms are shown in yellow and blue, respectively. (**B**) Per-atom potential energies relative to the corresponding bulk phases highlight energetic penalties at surfaces, interfaces, and structural defects. The dimer is destabilized by the high surface energy of exposed Rh, whereas the core–shell structure is destabilized by defects that prevent the Au shell from forming a coherent single crystal. (**C**) Distribution of strain within the Au domain, quantified by the average Au–Au nearest-neighbor distance. Rh atoms are shown in white, and Au atoms are colored according to their average Au–Au neighbor distance. The heterogeneous strain distribution further illustrates the structural penalty associated with the core–shell morphology. (**D**) Sequential cross-sections along the z axis show the three-dimensional organization of each morphology and the continuity of the Au overlayer in the shell-dimer.

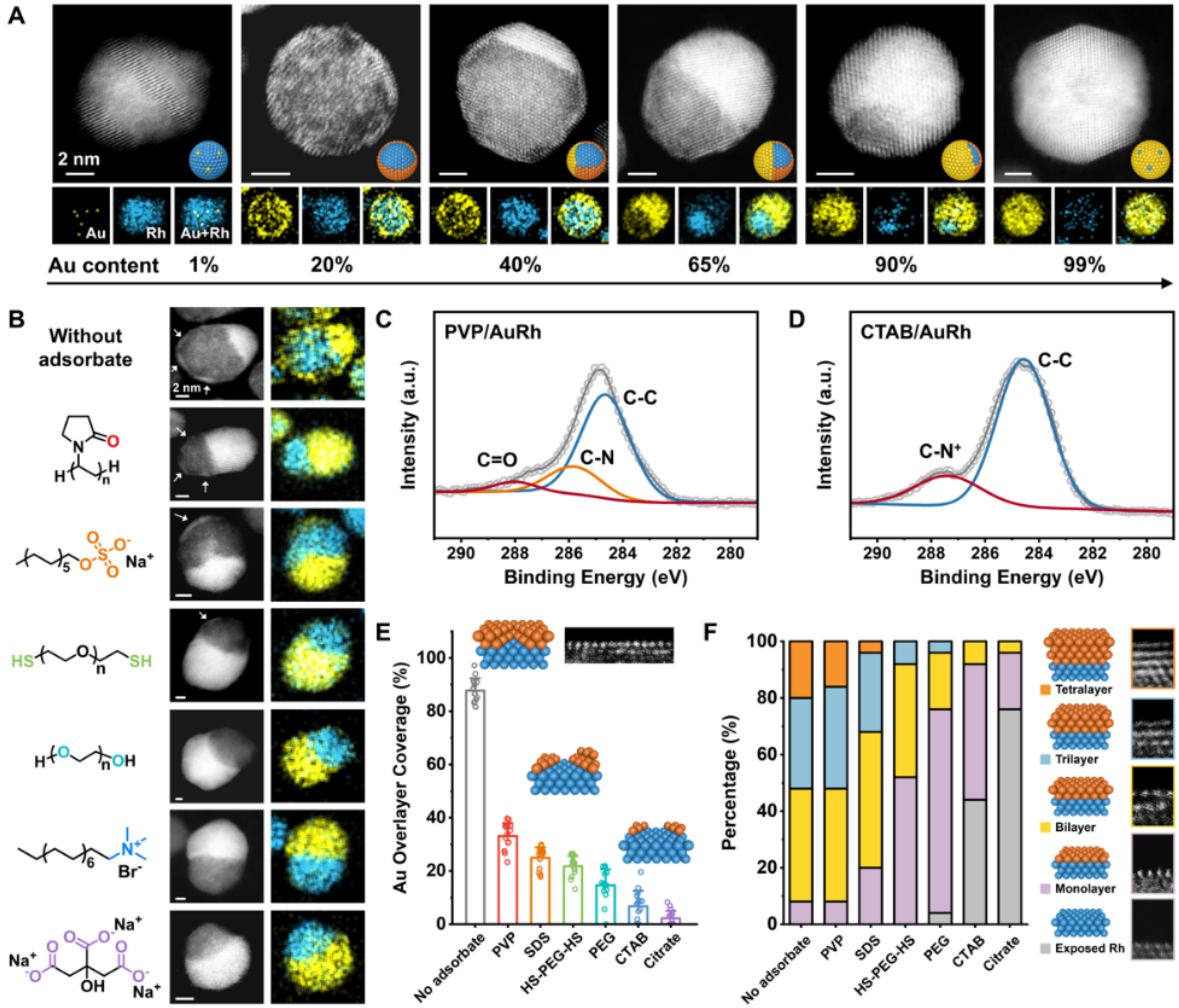

**Fig. 3. Effect of composition and adsorbates on the structure of Au overlayers.** (**A**) HAADF-STEM images and EDS mapping of AuRh nanoparticles with different compositions. (**B**) HAADF-images of surface-clean AuRh nanoparticles and AuRh nanoparticles decorated by adsorbates originating from PVP, SDS, HS-PEG-HS, PEG, CTAB, and citrate ligands. (**C**,**D**) C 1s XPS spectra of AuRh nanoparticles decorated by adsorbates originating from (C) PVP and (D) CTAB. (**E**) Coverage of Au overlayer on the Rh domains of AuRh nanoparticles decorated by adsorbates originating from different ligands. (**F**) Distribution of overlayers with different number of Au atomic layers. Exposed Rh denotes nanoparticles without Au overlayers.

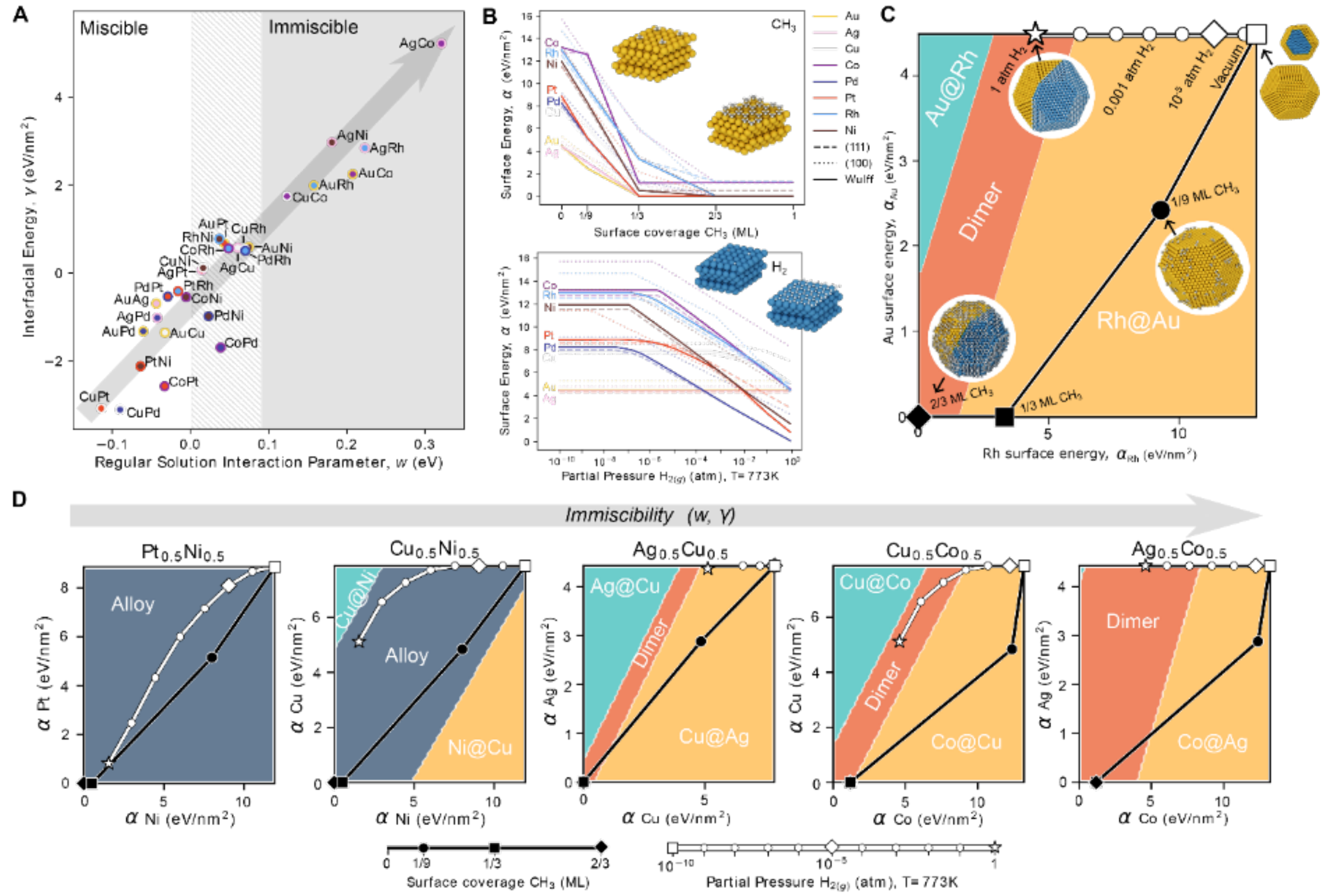

**Fig. 4. Thermodynamic phase diagrams of binary metal nanoparticles.** (**A**) Miscibility of binary metal pairs, characterized by the interfacial tension, γ, and the regular-solution interaction parameter, *w*, which is determined from the enthalpy of mixing of the two elements. (**B**) Surface energies can be modified through adsorption of passivating species. Calculated surface energies of representative metals are shown as a function of H and $CH_3$ coverage for several surface terminations. (**C**) Phase diagram for Au–Rh nanoparticles showing the predicted stability regions of alloy, dimer, and shell morphologies, as a function of the surface passivation of each metal. Core-shell and shell-dimer structures are grouped as shell morphologies and denoted A@B, where A is the interior metal and B is the shell metal. Under ultra-high vacuum conditions, where no passivating species are present, Au–Rh nanoparticles favor a shell morphology in which Au coats the Rh-rich domain. Passivation by $CH_3$ on the Au and Rh surfaces (black line) or by H on the Rh surface (white line) instead stabilizes the dimer morphology. (**D**) Phase diagrams for additional representative bimetallic systems show that increasing immiscibility promotes the emergence of shell structures and broadens the stability region of the dimer phase. These predictions demonstrate that the relative stability of dimer and overlayer morphologies can be tuned through surface chemistry under different environmental and synthesis conditions.

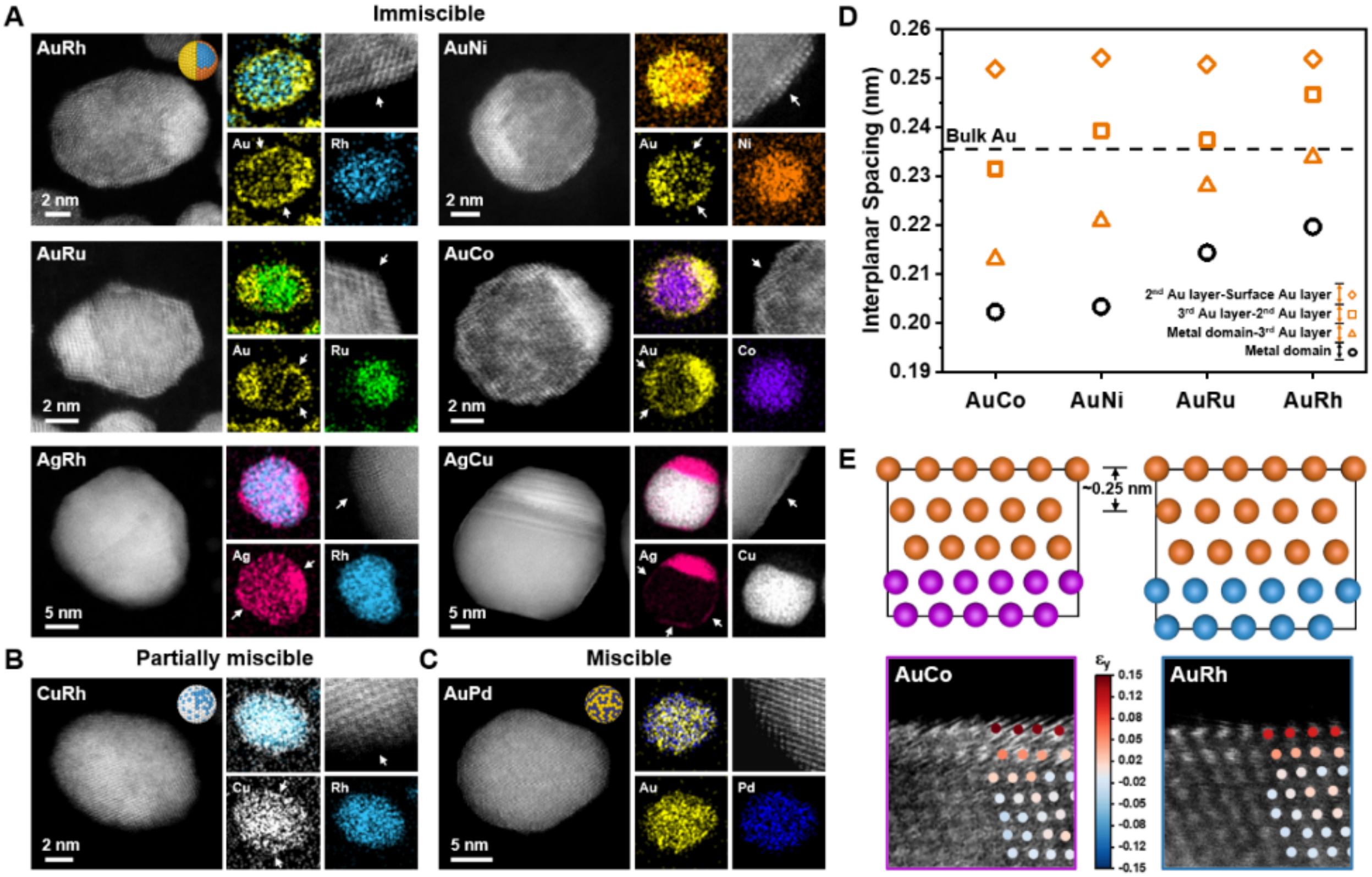

**Fig. 5. Generality of the overlayer structure on bimetallic nanoparticles.** (**A**) HAADF-STEM images and EDS elemental mapping of surface-clean AuRh, AuNi, AuRu, AuCo, AgRh, and AgCu nanoparticles, showing overlayer structure on nanoparticles composed of immiscible elements. White arrows indicate the overlayers on the nanoparticles. (**B**) HAADF-STEM images and EDS elemental mapping of surface-clean CuRh nanoparticles showing surface segregation of one element in nanoparticles composed of partially miscible elements. White arrows indicate the surface segregation of Cu in the nanoparticle. (**C**) HAADF-STEM images and EDS elemental mapping of surface-clean AuPd nanoparticles showing surface alloying in nanoparticles composed of miscible elements. (**D**) Strain analysis of surface-clean AuCo, AuNi, AuRu, and AuRh nanoparticles in the direction perpendicular to the metal (Co/Ni/Ru/Rh) domains. (**E**) Schematic illustration and HAADF-STEM images of the Co domain and Rh domain in AuCo and AuRh nanoparticles. Estimated positions of the atomic columns are indicated by solid dots. Strain in the y direction is calculated by the displacements of atomic columns with respect to an ideal Co or Rh lattice.

# Supplementary Materials

## Subnanometer thermodynamic overlayers on bimetallic nanoparticles

Caitlin A. McCandler[1,3,#], Xianzhuo Lao[2,#], Jie Liu[2], Huipu Liu[2], Kristin A. Persson[1,*], Peng-Cheng Chen[2,4,*]

[1]Department of Materials Science and Engineering, University of California Berkeley, Berkeley, CA 94720, USA.

[2]College of Smart Materials and Future Energy, State Key Laboratory of Coatings for Advanced Equipment, Fudan University, Shanghai 200438, China.

[3]Initiative for Computational Catalysis, Flatiron Institute, New York, NY 10010, USA.

[4]International Institute for Intelligent Nanorobots and Nanosystems, Fudan University, Shanghai 200438, China.

#These authors contributed equally to this work.

Correspondence to: pcchen@fudan.edu.cn (P.-C.C.); kristinpersson@berkeley.edu (K.A.P.)

## Materials and Methods

Materials

Metal compounds (>99.9% trace metal basis), gold (III) chloride ($AuCl_3$), silver nitrate ($AgNO_3$), ruthenium (III) chloride hydrate ($RuCl_3 \cdot xH_2O$), and nickel (II) chloride hexahydrate ($NiCl_2 \cdot 6H_2O$) were purchased from Sigma-Aldrich, Inc. Rhodium chloride hydrate ($RhCl_3 \cdot xH_2O$), cobalt (II) chloride hexahydrate ($CoCl_2 \cdot 6H_2O$), copper (II) nitrate trihydrate ($Cu(NO_3)_2 \cdot 3H_2O$), and palladium (II) chloride ($PdCl_2$) were purchased from Aladdin Scientific Corp. Ethanol ($C_2H_5OH$, ≥99.5%), polyvinylpyrrolidone (PVP, $M_w$ 40000), sodium dodecyl sulfate (SDS), hexadecyltrimethylammonium bromide (CTAB, ≥99.0%), and trisodium citrate dihydrate (≥99.0%) were purchased from Sinopharm Chemical Reagent Co, Ltd. Poly(ethylene glycol) dithiol (HS-PEG-HS, $M_w$ 3400) was purchased from Sigma-Aldrich, Inc. Polyethylene glycol (PEG, $M_w$ 40000) was purchased from Adamas-Life Science, Inc. All the chemicals were used without further purification.

Synthesis of AuRh, AuRu, AuNi, AuCo, AgRh, AgCu, CuRh, and AuPd nanoparticles

Nanoparticles were prepared by thermal decomposition of metal salts. In a typical experiment, $AuCl_3$ and $RhCl_3 \cdot xH_2O$ were separately dissolved in ethanol and mixed. The mixture was thermally annealed at 500°C for 12 hours, followed by cooling to 25°C in 6 hours. AuRu, AuNi, AuCo, AgRh, AgCu, CuRh, and AuPd nanoparticles were prepared using the same method, except that the metal salts were replaced by $RuCl_3 \cdot xH_2O$, $NiCl_2 \cdot 6H_2O$, $CoCl_2 \cdot 6H_2O$, $AgNO_3$, $Cu(NO_3)_2 \cdot 3H_2O$, or $PdCl_2$. Adsorbate-modified AuRh nanoparticles were prepared by adding ligand molecules into the salt mixture solution.

Characterization

Scanning transmission electron microscopy (STEM) characterization of the nanoparticles was performed on a dual-energy dispersive X-ray spectroscopy (EDS) detector equipped JEOL ARM200F transmission electron microscope. The dark-field images were taken with an annular dark-field detector at an electron acceleration voltage of 200 kV. Nanoparticle composition was determined using the equipped dual EDS detectors on the JEOL ARM200F with a 200 kV electron acceleration voltage. The $L_\alpha$ peaks of Ru, Rh, Pd, Ag, and Au, and the $K_\alpha$ peaks of Co, Ni, and Cu in the EDS spectra were used for elemental mapping and composition quantification. Atomic compositions of the nanoparticles were calculated based on the collected EDS spectra with the standardless Cliff-Lorimer correction method.

Strain analysis of the overlayer structure

To visually show the strain state of the nanoparticle surface, the center coordinates ($x$, $y$) of the atom columns in a high-angle annular dark-field scanning transmission electron microscopy (HAADF-STEM) image are first identified and then the relative offset of each atom column is assessed with reference to the standard position of atom columns in an unstrained lattice, thus yielding the offset percentages ($z_x$ and $z_y$) in the $x$ and $y$ directions. The displacement plot is obtained based on the atom column's coordinates and the offset percentages in the $x$ or $y$ directions. In the strain analysis map, positive values represent tensile strains, negative values represent compressive strains, and zero values represent an unstrained state.

Analysis of overlayer coverage on the nanoparticles

In HAADF-STEM images, atomic brightness is proportional to atomic number ($I \propto Z^{1.7}$). Gold and rhodium have significantly different atomic numbers. Thus, the coverage of Au on Rh along the nanoparticle contour line can be distinguished by the atomic brightness difference between Au and Rh (Fig. S12). Experimentally, the coverage was quantified by setting the length of the Rh domain contour as the denominator, and the total length of the Au overlayer on the Rh domain contour as the numerator. The brightness threshold used to differentiate Au and Rh is the median value between the Au atomic column brightness and the Rh atomic column brightness.

Density Functional Theory (DFT) calculations

Bulk and surface properties were calculated for Au, Ag, Cu, Co, Ni, Pd, Pt and Rh and every binary combination of these metals using density functional theory (DFT). The DFT calculations were performed with spin-polarization and a plane wave basis set, implemented in the Vienna Ab-initio Simulation Package (VASP) (*36*). The exchange and correlation energies were calculated using the Perdew-Burke-Ernzerhof (PBE) form of the generalized gradient approximation (GGA) (*37*). All calculations were geometrically relaxed, and non-periodic systems (i.e. surfaces and gasses) were provided with at least 10Å of vacuum spacing to reduce self-interaction between periodic images. Calculations of bulk materials used 10x10x10 k-point sampling, while calculations of surfaces and interfaces used 4x4x1 k-point sampling, reducing the sampling along the direction perpendicular to the interface to 1 point, and calculations of isolated gaseous species used one k-point, i.e., the Γ point. Gaussian smearing was applied with a width of 0.2 eV (0.03 eV for gaseous species). A cutoff energy of 520 eV (1000 eV for gaseous species) was applied for the plane wave basis set and the electron–ion interactions were described by the projector augmented wave (PAW) method (*38*). Dipole corrections were also applied to calculations of gaseous species. When calculation convergence was not achieved due to magnetic convergence, the structures were pre-relaxed with non-spin-polarized geometry optimizations and then recalculated with spin-polarized calculations.

Molecular dynamics simulations

Molecular dynamics (MD) simulations were performed for AuRh nanoparticles using a machine-learned interatomic potential (MLIP) fitted to Au-Rh-H interactions (*39*). The MLIP has an Atomic Cluster Expansion (ACE) (*40*) architecture which considers up to 4-body interactions, has 3222 fitted parameters, and was trained on 13311 examples of nanoparticles, surfaces, and bulk systems at the PBE/GGA level of theory. The reported test error with respect to systems within 1 eV/atom of the energy hull was 9.41 meV/atom in energies and 54.22 meV/Å in forces. MD simulations were performed in the LAMMPS simulation software and rendered with Ovito (*41-43*). Representative geometries were created for the dimer, shell-dimer and core-shell particles which each had 7191 Au atoms and 4197 Rh atoms (63.1 at.% Au). The shell-dimer geometry was constructed to match the geometry observed in the particle imaged in Figure 1, the dimer geometry was constructed to resemble the shell-dimer geometry, and the core-shell geometry was constructed using the *Wulffpack* tool (*44*). Simulations were performed with Langevin dynamics in the NVT ensemble (damping parameter = 1 ps, timestep = 2 fs) and with Monte Carlo (MC) atom swaps between Au and Rh atoms. First the particles were thermally annealed at 773 K for ~3 ns with 200 attempted MC swaps every 100 steps. Then the particles were cooled to 300 K for ~1 ns and finally geometrically relaxed. The dimer geometry was only heated for 300 ps because the Au atoms started to coat the Rh domain, effectively transforming into the shell-dimer geometry.

**Supplementary Text**

Interfacial tension calculation

The interfacial tension, $\gamma$, is calculated with DFT by comparing the energies of the unary bulk elements to the energy of a layered slab structure of the two elements with layers normal to either the (111) or (100) facet. Each interface calculation had 8 layers (stacked in the z-direction) of each element with 9 atoms per layer. DFT calculations of bulk systems are performed in an infinitely periodic system, meaning that epitaxial strain arising from the difference in lattice parameters and moduli of each element will be applied uniformly in the plane parallel to the interface (x-y plane). It is important to decouple the energetic effects of the bonding at the interface with the artificially introduced strain; the calculation of the interfacial strain should yield the same answer even with a greater number of layers. We determine the strain-free interfacial tension of each surface following the method described by Xu *et al.* for Mg/MgO interfaces (*45*).

In real systems, epitaxial strain arises at interfaces, which can be alleviated by the formation of defects and dislocations such that strain energy tends to dissipate away from interfaces. In nanoparticles, strain dissipates even more easily than in the bulk due to the elimination of strain at their surfaces. In a simulation of Au-Rh interfaces, we can see that the strain is very large in the layers that are directly in contact, with a large dissipation in the other layers (Figure 2). Following this observation, we approximate the total interfacial energy to be the combination of the strain-free interfacial energy and the strain energy of a single layer of each element. A more complex treatment of the strain effects may consider the size-dependence of strain dissipation at surfaces, the contraction of nanoparticle surfaces to reduce total surface area, dislocations, non-epitaxial interfaces, icosahedral and decahedral twinned interfaces, and atomically diffuse interfaces (*10*).

Binary nanoparticle structure stability

Building on previous thermodynamic models of binary nanoparticle phase stability (*31, 33*), we considered four idealized binary nanoparticle structures: alloy, dimer, core-shell, and shell-dimer (Fig. S26). These serve as reference states for comparing thermodynamic stability; intermediate structures such as partially mixed structures and ordered intermetallics are not explicitly included.

For a nanoparticle in equilibrium with an environment at specified temperature and pressure, the relative stability of the candidate structures was determined from their Gibbs free energies. The free-energy change associated with forming each nanoparticle from the corresponding bulk elemental reference states was decomposed into contributions from bulk mixing, formation of external surfaces, and formation of internal interfaces:

$$\Delta G = \Delta H_m + \Delta H_s + \Delta H_i - T\Delta S_m \qquad (1)$$

Here, the subscripts *m*, *s*, and *i* denote contributions from mixing, surfaces, and interfaces, respectively. For a binary system containing elements A and B, the model is parameterized by the surface energies of A and B, $\alpha_A$ and $\alpha_B$; the interfacial tension between A- and B-rich domains, $\gamma_{AB}$; and the regular-solution interaction parameter, *w*, describing the energetic preference for unlike relative to like atomic neighbors. Together, these terms define the free energy associated with each binary nanoparticle morphology relative to the bulk elemental phases.

*1. Bulk mixing contribution*

The alloy phase was approximated as a random substitutional mixture of A and B and described using the regular-solution model. For an FCC lattice, each atom has Z = 12 nearest neighbors. If $x_A$ and $x_B$ are the atomic fractions of A and B, respectively, with $x_A + x_B = 1$, the enthalpy of mixing is determined by the regular-solution interaction parameter *w*. Positive values of *w* correspond to energetically unfavorable A-B interactions and therefore favor phase separation at 0 K, whereas negative values favor mixing.

The interaction parameter was obtained from DFT calculations of ordered bulk structures at three compositions. The energies of $L1_2$ structures at $A_{0.25}B_{0.75}$ and $A_{0.75}B_{0.25}$, together with the $L1_0$ structure at $A_{0.50}B_{0.50}$, were compared with the energies of the corresponding unary bulk phases to determine the enthalpy of mixing and parameterize *w*. The enthalpic contribution for a randomly mixed particle containing n atoms was then evaluated using the regular-solution approximation:

$$\Delta H_m = \frac{nZ}{2} x_A x_B w \quad (2)$$

The configurational entropy was approximated assuming ideal random mixing:

$$\Delta S_m = -R\big(x_A \ln(x_A) + x_B \ln(x_B)\big) \quad (3)$$

The phase-separated dimer, core-shell, and shell-dimer structures were treated as containing compositionally distinct A- and B-rich domains and therefore do not receive this random-alloy mixing contribution.

*2. Surface energy contribution*

Each candidate architecture exposes different fractions of elements A and B at the nanoparticle surface. The surface energies $\alpha_A$ and $\alpha_B$ correspond to the Wulff-averaged surface energies calculated for the appropriate environmental conditions. Their calculation from clean and adsorbate-covered DFT slabs is described in the section *Surface energy calculations of passivated metals*.

To obtain closed-form expressions for the surface and interfacial areas, the nanoparticle geometries were approximated using spheres and spherical caps. The total nanoparticle was assigned a radius *r* corresponding to a sphere containing the total number of A and B atoms at the prescribed composition. The random alloy was treated as a homogeneous sphere. The core-shell morphology was represented by two concentric spheres, and both B@A and A@B configurations were evaluated. The dimer was represented as two spherical caps separated by a planar interface, with relative cap volumes determined by composition. The shell-dimer retained the phase-separated internal geometry of the dimer but included an overlayer of one element covering the externally exposed surface of the other domain.

For each geometry, the corresponding surface contribution was written as:

$$\Delta H_s[\text{alloy}] = 4\pi r^2(\alpha_A x_A + \alpha_B x_B) \quad (4)$$

$$\Delta H_s[\text{heterodimer}] = 2\pi r^2\big(\alpha_A(1-\chi) + \alpha_B(1+\chi)\big) \quad (5)$$

$$\chi = \left(-\frac{1}{2} - \frac{\sqrt{3}}{2}i\right)\left(2x_A - 1 + 2\sqrt{x_A\sqrt{x_A - 1}}\right)^{1/3} + \left(-\frac{1}{2} + \frac{\sqrt{3}}{2}i\right)\left(2x_A - 1 - 2\sqrt{x_A\sqrt{x_A - 1}}\right)^{1/3} \quad (6)$$

$$\Delta H_s[\text{core-shell B@A}] = 4\pi r^2 \alpha_A \quad (7)$$

$$\Delta H_s[\text{shell-dimer B@A}] = 4\pi r^2 \alpha_A \quad (8)$$

Because adsorption can substantially modify $\alpha_A$ and $\alpha_B$, the preferred morphology may change with adsorbate identity, coverage, temperature, or gas pressure even when the overall particle composition is unchanged.

*3. Interfacial energy contribution*

Phase-separated structures additionally incur an energetic contribution associated with the interface between A- and B-rich domains. Interfacial tensions were calculated with DFT for coherent (111) and (100) interfaces and include the strain treatment described in the section *Interfacial tension calculation*. The dimer contains a single approximately planar interface separating the two domains. Its interfacial contribution was therefore calculated using the lower-energy of the (111) and (100) coherent interfaces:

$$\Delta H_i[\text{heterodimer}] = \pi r^2 (1 - \chi^2) \min\left(\gamma_{(111)}, \gamma_{(100)}\right) \quad (9)$$

For the core-shell morphology, the internal interface surrounds the core. The equilibrium shape of this interface was approximated using a Wulff construction based on the calculated (111) and (100) interfacial tensions:

$$\Delta H_i[\text{core-shell B@A}] = 4\pi r^2 (x_B)^{2/3} \mathrm{Wulff}\left(\gamma_{(111)}, \gamma_{(100)}\right) \quad (10)$$

The shell-dimer contains two types of A-B contact: the interface separating the two principal domains and the additional interface formed where the surface overlayer covers the second element. The planar domain interface was treated in the same manner as the dimer, whereas the overlayer interface was treated using the Wulff-averaged interfacial tension. The total interfacial contribution is therefore:

$$\Delta H_i[\text{shell-dimer B@A}] = \pi r^2 (1 - \chi^2) \min\left(\gamma_{(111)}, \gamma_{(100)}\right) + 2\pi r^2 (\chi + 1) \mathrm{Wulff}\left(\gamma_{(111)}^{\text{shell}}, \gamma_{(100)}^{\text{shell}}\right) \quad (11)$$

When both calculated interfacial tensions are positive, the Wulff construction provides an effective interfacial tension for the interface. If either calculated interfacial tension is negative, a conventional Wulff construction is not defined; in these cases, the lower calculated interfacial tension was used.

For each binary composition, particle size, temperature, and set of surface energies, the total free energy of each candidate morphology was obtained by summing its mixing, surface, and interfacial contributions according to Eq. 1. The structure with the lowest calculated free energy was assigned as the thermodynamically preferred phase. Repeating this procedure while varying the surface energies of A and B generated the nanoparticle phase diagrams presented in Fig. 4 and Figs. S31-33.

Surface energy calculations of passivated metals

The surface energies of clean and adsorbate-covered facets were calculated from DFT slab energies using an *ab initio* thermodynamic framework. Slab geometries were generated with the SlabGenerator implemented in *pymatgen* for each surface of interest (*46*). The (100) and (111) facets, which are generally among the lowest-energy surfaces of FCC metals, were used as the primary facets for the surface-energy and adsorption calculations. Each slab contained at least six atomic layers and nine metal atoms per surface, and a vacuum region of at least 15 Å separated periodic slab images. For a clean, symmetric slab exposing two equivalent surfaces of facet (hkl), the surface energy, $\alpha_{(hkl)}^{\text{clean}}$, was calculated as

$$\alpha_{(hkl)}^{\text{clean}} = (E_{(hkl)} - n\mu_{\text{bulk}})/(2A_{(hkl)}) \quad (12)$$

where $E_{(hkl)}$ is the DFT total energy of the clean slab, n is the number of metal atoms in the slab, $\mu_{\text{bulk}}$ is the DFT energy per atom of the corresponding bulk metal, and $A_{(hkl)}$ is the area of one surface. The factor of two accounts for the two equivalent surfaces of the clean slab. Adsorbates were placed on only one side of each slab, while the opposite face remained clean. Initial configurations of adsorbates were the on-top, bridging, and hollow sites, and these geometries were allowed to relax to find the most stable binding motif. For a slab containing m adsorbates each having a chemical potential of $\mu_{ads}$, the total surface energy is

$$\alpha_{(hkl)} = \frac{E_{(hkl)}^{ads} - n\mu_{bulk} - m\mu_{ads}}{A_{(hkl)}} - \alpha_{(hkl)}^{clean} \tag{13}$$

For each facet and coverage, the lowest-energy relaxed configuration was used to construct the coverage-dependent surface-energy curves (Figs. S27 and S28). At a specified adsorbate chemical potential, the equilibrium surface energy was calculated by minimizing overall calculated coverages and configurations. The selected passivated surface energies of the principal facets were used as inputs to the Wulff construction to estimate the morphology and effective surface energy of each passivated unary metal:

$$\alpha = \mathrm{Wulff}\left(\alpha_{(111)}, \alpha_{(100)}\right) \tag{14}$$

This treatment permits adsorption to alter both the magnitude of the surface energy and the relative stability of the exposed facets. The calculated values should nevertheless be interpreted with the known limitations of DFT for adsorption on metal surfaces. In particular, GGA calculations can misidentify the preferred adsorption site for some adsorbate–surface combinations (*47*), and nanoscale surface strain may modify adsorption strengths relative to ideal, unstrained slabs (*10*). These uncertainties do not change the calculation procedure described above but should be considered when interpreting small energetic differences between configurations.

Hydrogen-passivated surfaces

Hydrogen adsorption was treated as being in equilibrium with an $H_2$ gas reservoir. The chemical potential of an adsorbed H atom was therefore set equal to one-half of the chemical potential of gaseous $H_2$ (*48*). Additionally, the chemical potential of hydrogen gas, $\mu_{H2}$, at any pressure and temperature is related to its standard state (1 atm) chemical potential, $\mu_{H2}^{0}$, assuming it follows the ideal gas law.

$$\mu_{H2} = \mu_{H2}^{0} + k_b \, T \ln(P_{H_2(g)}) \tag{15}$$

$$\mu_{H} = \frac{1}{2}\,\mu_{H2}^{0} + \frac{1}{2}\,k_b \, T \ln(P_{H_2(g)}) \tag{16}$$

Substitution of Equation 16 into Equation 13 gives the thermodynamically preferred hydrogen coverage and the corresponding facet surface energy at a specified temperature and $H_2$ partial pressure (Fig. S27 and Fig. 4B). Hydrogen adsorption was evaluated over the full range of relevant coverages by sequentially populating on-top, bridge, and hollow sites and relaxing each structure. For many surfaces, the surface energy varied approximately linearly below one monolayer, consistent with the H atoms not interacting with one another. At higher surface coverages, deviations from linearity become more pronounced due to increased electronic interactions, changes in local coordination, and structural relaxation.

Methyl-passivated surfaces

$CH_3$ was used as a model for surface-bound alkyl groups that may remain from solution-phase nanoparticle synthesis. The methyl group is the smallest alkyl adsorbate and therefore does not reproduce the full steric environment of longer ligands; however, it captures the local metal–carbon bonding of an alkyl headgroup while allowing a systematic comparison across metals, facets, and coverages. In contrast to hydrogen, the methyl coverage was not assumed to equilibrate directly with a gas-phase $CH_3$ reservoir under the synthesis conditions. The methyl adsorption energy was referenced to an isolated, gas-phase $CH_3$ species, corresponding to desorption of the adsorbed methyl group from the surface:

$$* CH_3 \rightarrow CH_3 + * \qquad (17)$$

Accordingly, the chemical potential used for methyl in the surface-energy expression was that of the isolated gas-phase $CH_3$ species:

$$\mu^0_{CH_3} = \mu^0_{CH_3(g)} \qquad (18)$$

This reference chemical potential was inserted into Equation 13 to calculate the surface energy associated with each $CH_3$ coverage (Fig. S28 and Fig. 4B). Because $CH_3$ is substantially bulkier than H, lateral steric interactions and adsorbate-induced strain became important at lower coverages. Consequently, the $CH_3$ surface-energy curves commonly show stronger nonlinearity, and high-coverage structures can become substantially less favorable as the adsorbates crowd the surface. Some $CH_3$-covered structures yielded formally negative surface energies. Within the grand-canonical expression, a negative value means that formation of additional adsorbate-covered surface would lower the free energy if the adsorbate reservoir were unlimited and the same coverage could be maintained as new surface area formed. However, this limit is not physically accessible in a finite-ligand synthesis. Therefore, for subsequent Wulff construction calculations that required non-negative facet energies, negative values were bound at zero to avoid unphysical constructions.

Synthesis of nanoparticles without organic adsorbates

Surface-clean AuRh nanoparticles were synthesized by thermal annealing of metal chloride salts to decompose the salts into metallic nanoparticles. HAADF-STEM characterization shows that the resulting nanoparticles include monometallic Au nanoparticles, monometallic Rh nanoparticles, and bimetallic AuRh nanoparticles with various compositions (Fig. S1). The bimetallic AuRh nanoparticles exhibit a characteristic structure with two half-sphere domains. The lower-brightness domain corresponds to Rh and the higher-brightness domain corresponds to Au. Zoom-in characterization of every AuRh nanoparticle confirms that the Rh domains are always encapsulated by Au overlayers.

Au overlayers on AuRh nanoparticles with different compositions

The presence of Au overlayers was observed in nanoparticles across a range of compositions, with a consistent surface coverage of approximately 90% and thicknesses ranging from 1 to 4 atomic layers (Fig. S11). In the direction parallel to the nanoparticle surface, Au atoms exhibit compressive strain to accommodate the Rh lattice. As the Au layers get closer to the nanoparticle surface, the projected interatomic distance increase slightly but remain smaller than the value of bulk Au. In the direction perpendicular to the nanoparticle surface, the spacing between Au (111) planes increase with increasing distance from the Rh domain, eventually exceeding the value of bulk Au.

Oxidation of nanoparticles exposed to air

Nanoparticles containing Cu, Ni, or Co could be oxidized by air. To rule out the possibility of oxide formation in the nanoparticles, we exposed AuCo nanoparticles in air for 30 min before STEM characterization. As shown in Fig. 47, Co are easily oxidized upon air exposure since the surface-clean nanoparticles are not protected by organic ligands. The resulting $CoO_x$ presents lowest brightness compared to Au and Co, making it easily distinguishable in the STEM images. AuCo nanoparticles containing oxide exhibit architectures distinct from the shell-dimer architecture we have observed in metallic nanoparticles. Specifically, metallic Au and Co form crystalline domains with smooth surface (Fig. S47A-S47C), while cobalt oxide is poorly crystalline and form protruded islands that cover the surface of Au and Co (Fig. S47D-S47F), suggesting that cobalt oxide has a surface energy lower than that of Au and Co and will enrich on the outmost surface.

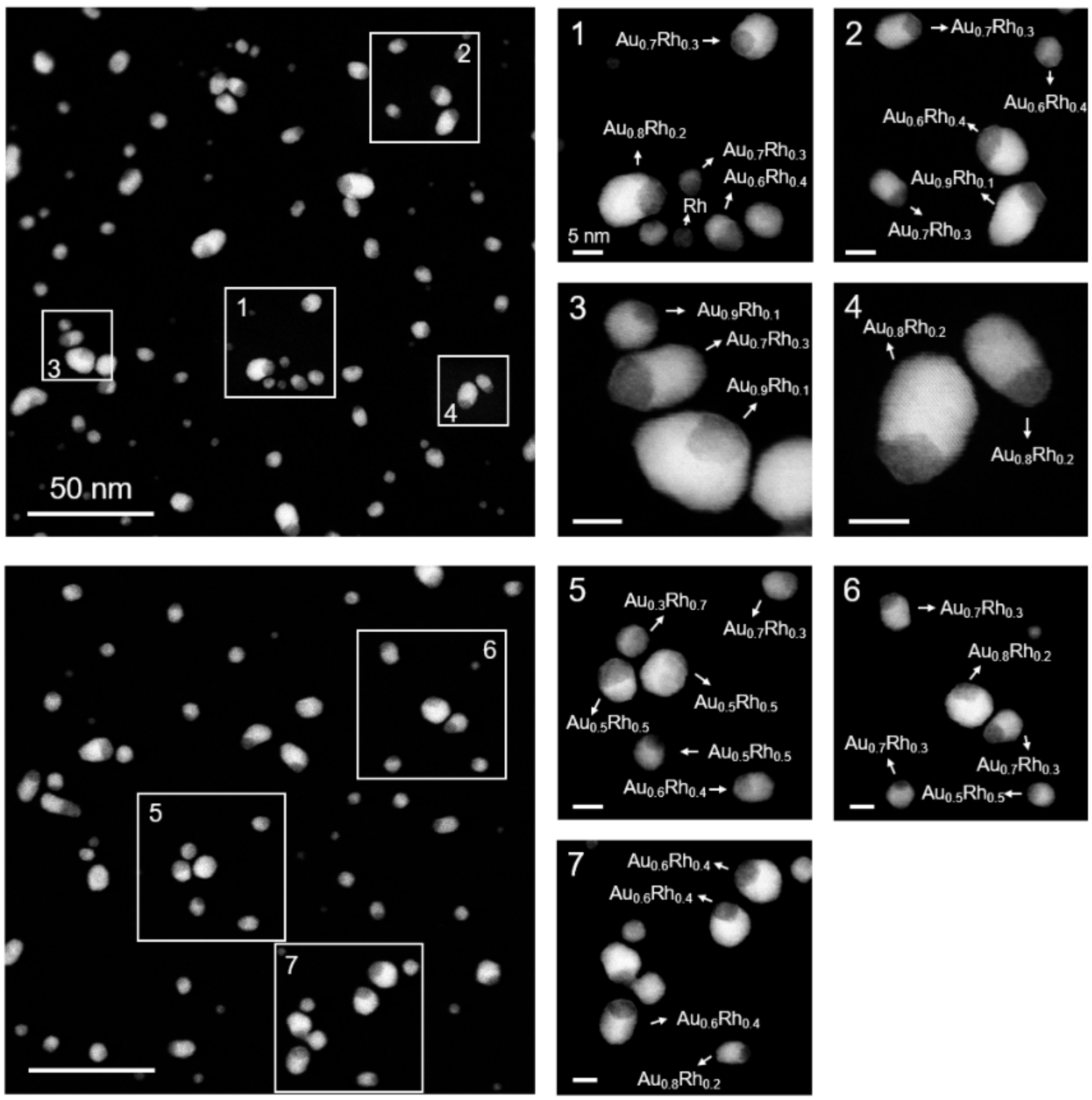

**Fig. S1.**
HAADF-STEM images of nanoparticles synthesized by thermal decomposition of the mixture of $AuCl_3$ and $RhCl_3$. The resulting nanoparticles include Au, Rh, and AuRh with varying compositions.

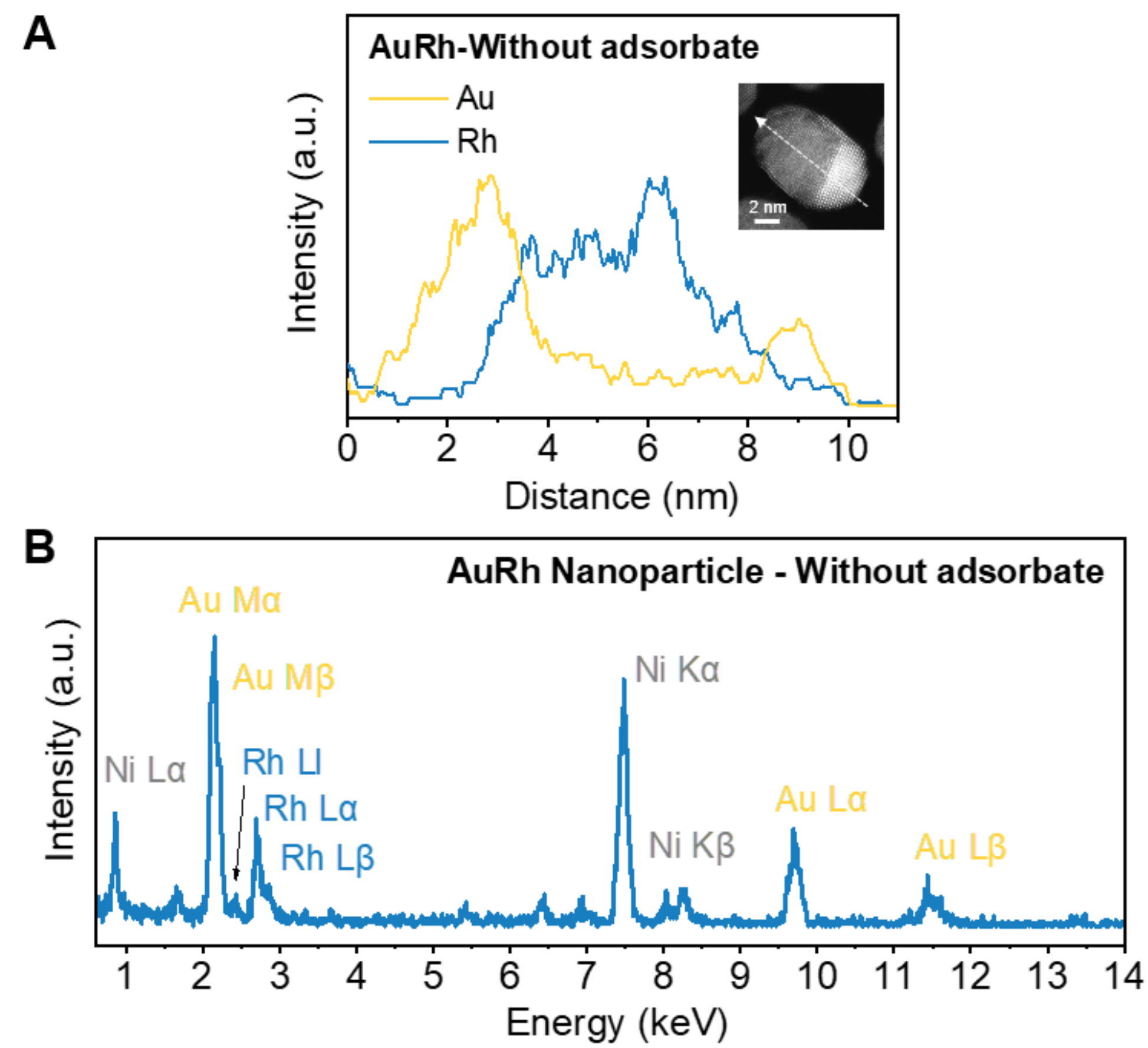

**Fig. S2.**
(A) EDS line scan and (B) EDS spectrum of AuRh nanoparticle. The white arrow in the inset HAADF-STEM image indicates the trace of the EDS line scan.

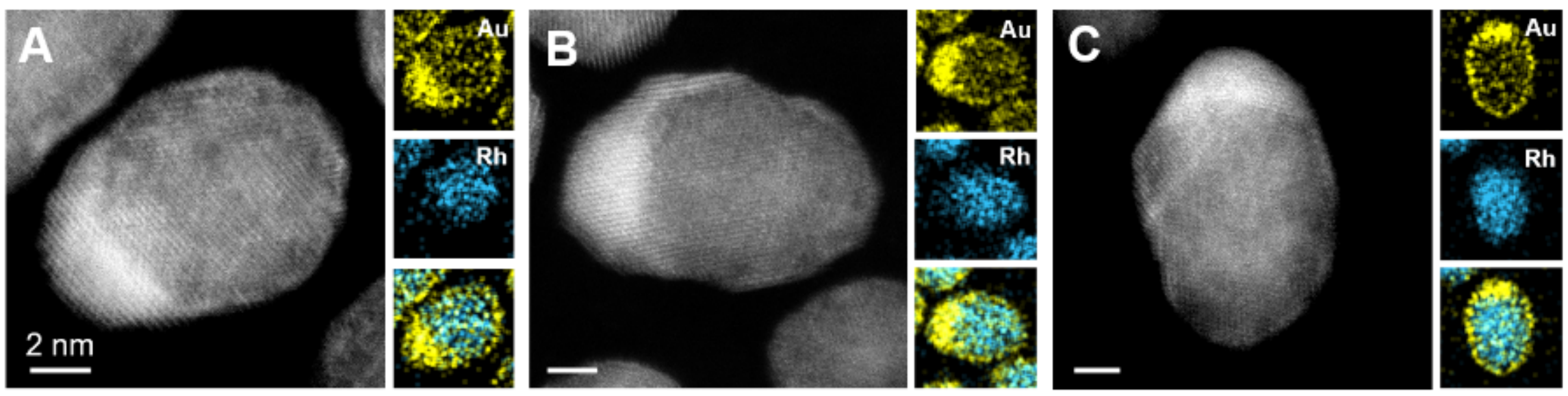

**Fig. S3.**
HAADF-STEM images and EDS elemental mapping of AuRh nanoparticles with Au overlayers on the Rh domain. The elemental mapping shows that Au and Rh are immiscible with each other, forming a phase-separated structure with an ultrathin Au overlayer on the surface of the Rh domain.

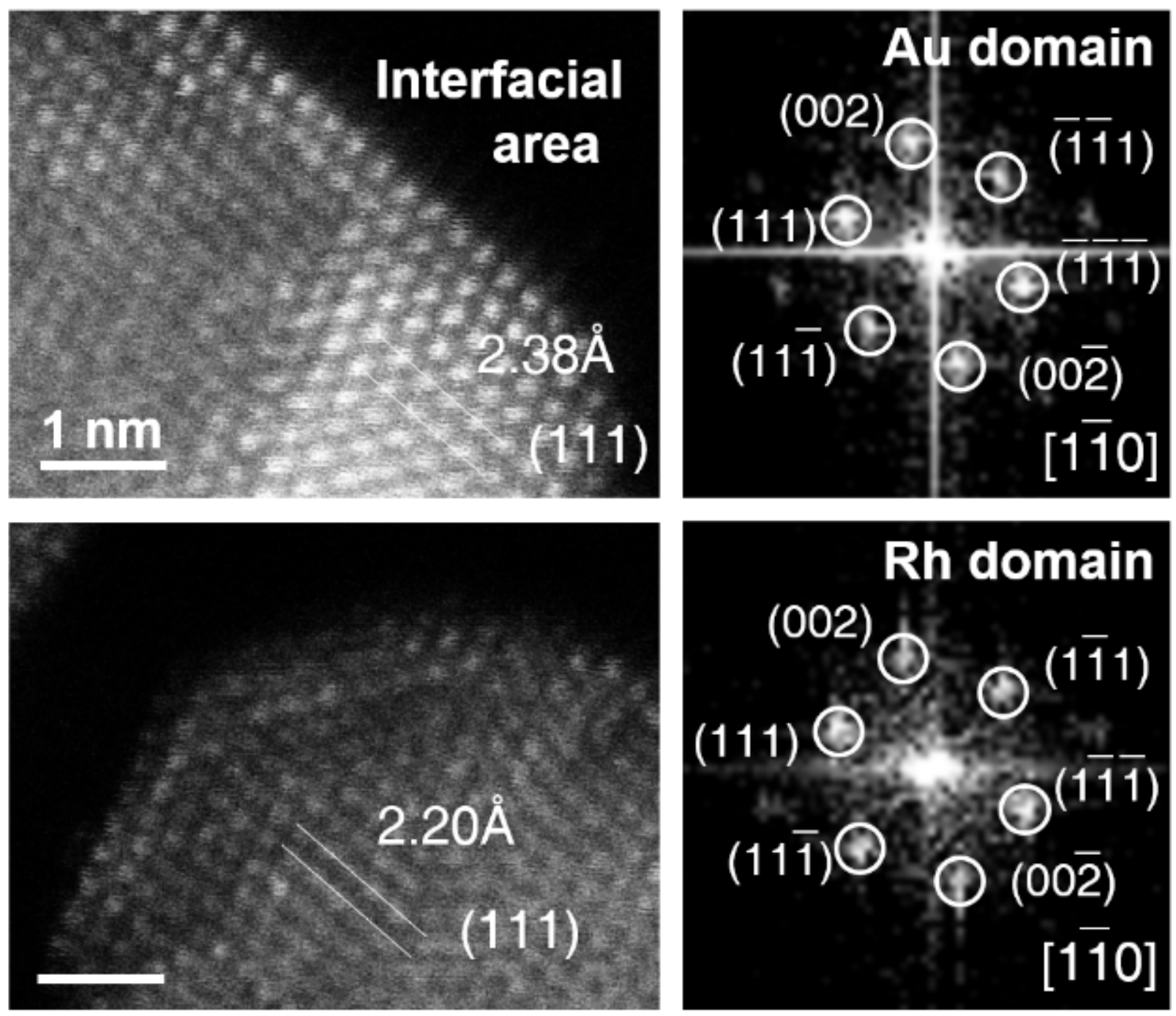

**Fig. S4.**
Atomic resolution HAADF-STEM images and fast Fourier transform (FFT) analysis of the AuRh nanoparticle in Fig. 1B. Both Au and Rh adopt a face-centered cubic (fcc) structure, with distinct unit cell parameters (Au: 4.07 Å; Rh: 3.80 Å). The lattice spacings of Au and Rh domains can be resolved in the STEM images. The measured (111) interplanar spacings are approximately 2.38 Å for Au and 2.20 Å for Rh. Fast Fourier Transform (FFT) analysis of the Au and Rh domains exhibit characteristic diffraction spots consistent with an fcc structure oriented along [110] zone axis, confirming that both Au and Rh domains are crystalline and adopt the fcc phase.

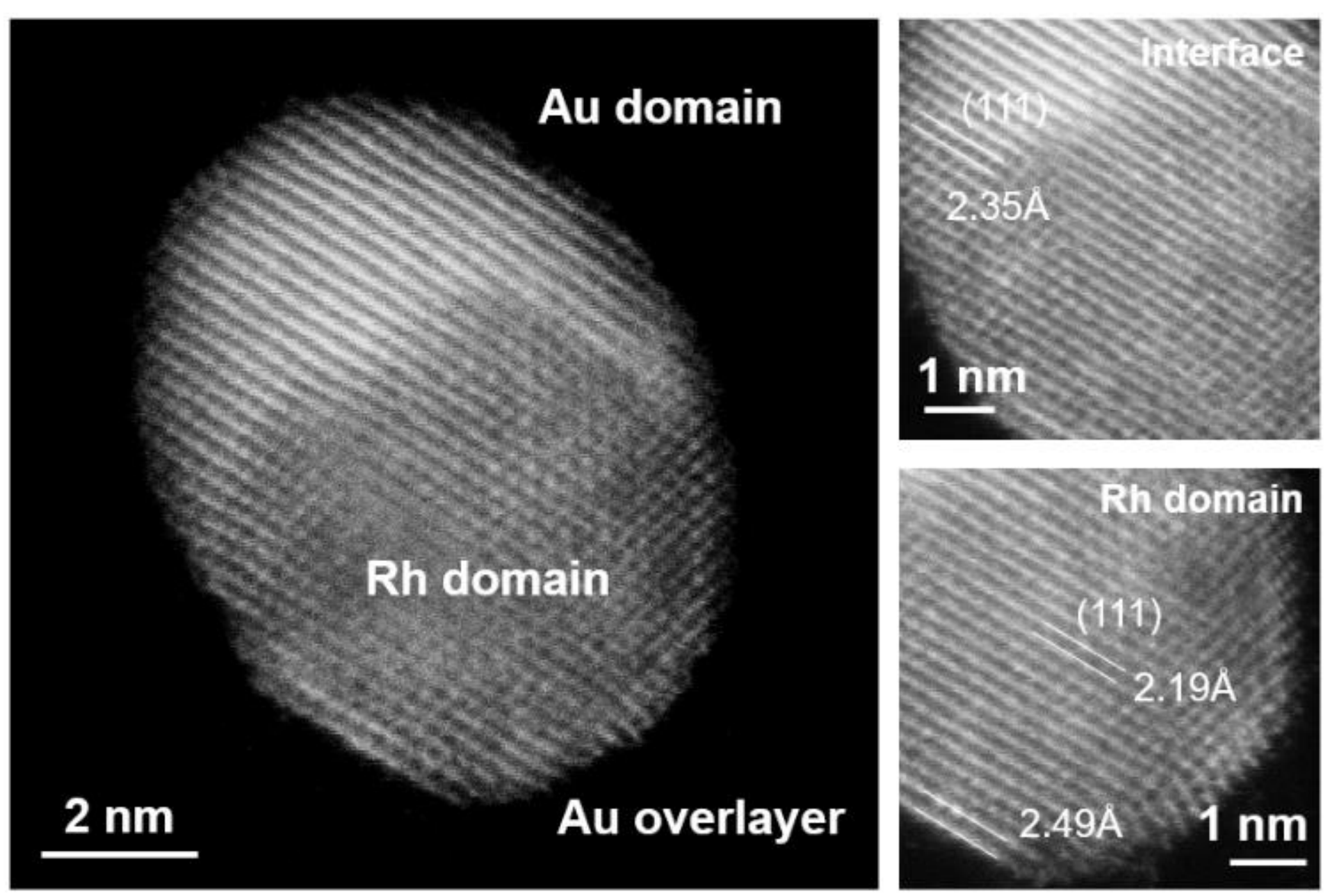

**Fig. S5.**
Atomic resolution HAADF-STEM image of a AuRh nanoparticle and zoom-in view of the interface region and the Rh domain surface.

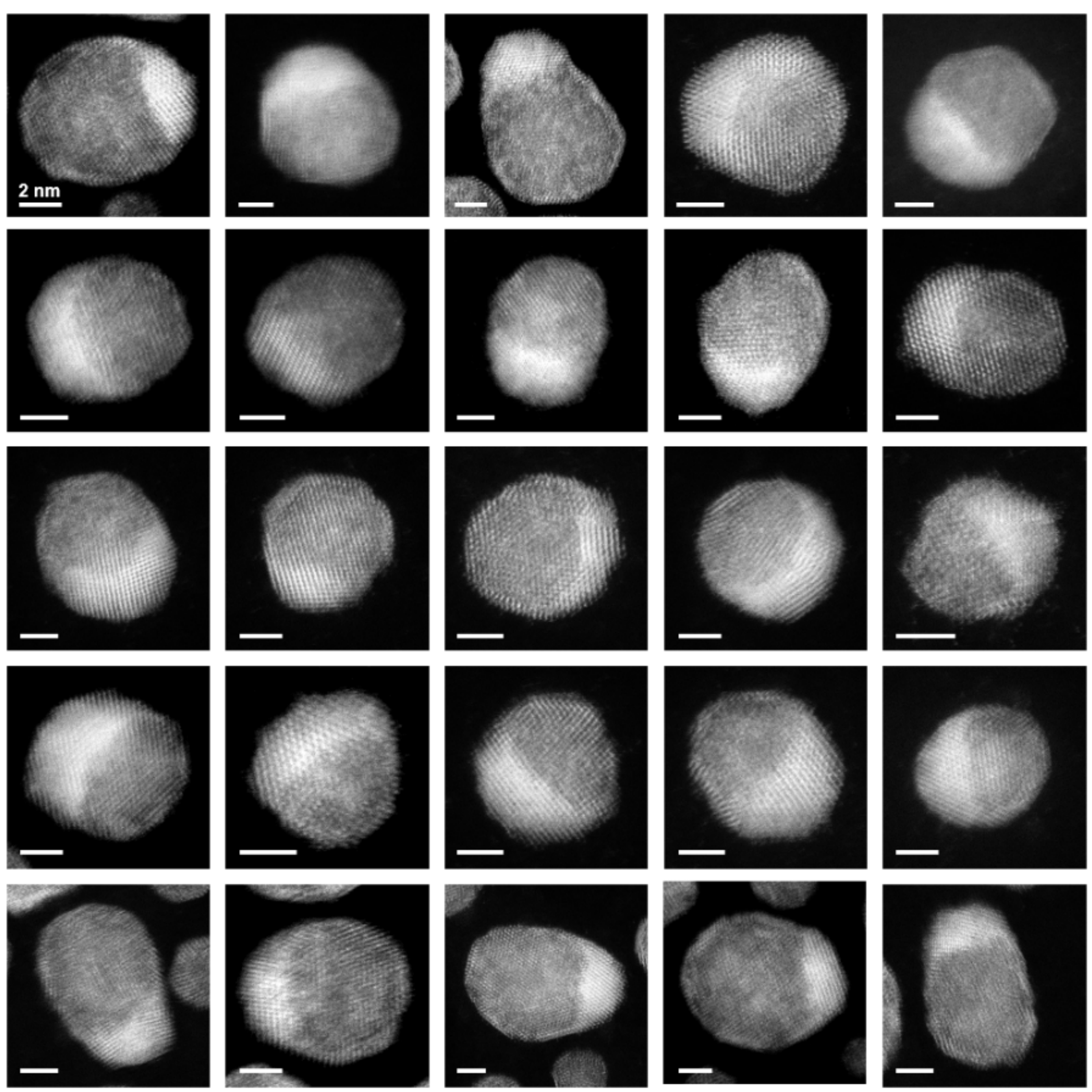

**Fig. S6.**
HAADF-STEM images of AuRh nanoparticles.

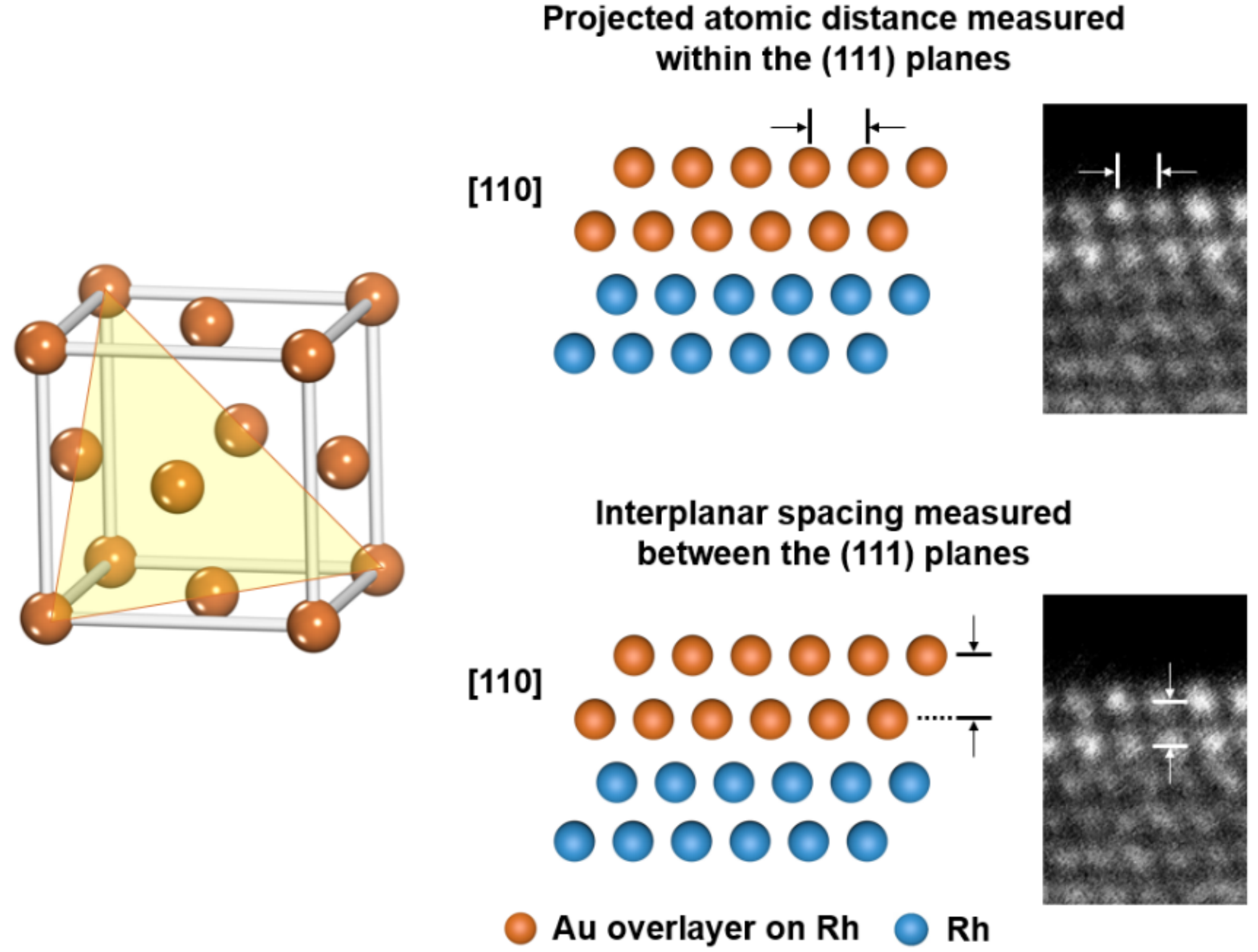

**Fig. S7.**
Measurement of the projected atomic distance within the (111) planes and the interplanar spacing between the (111) planes on the nanoparticle surface.

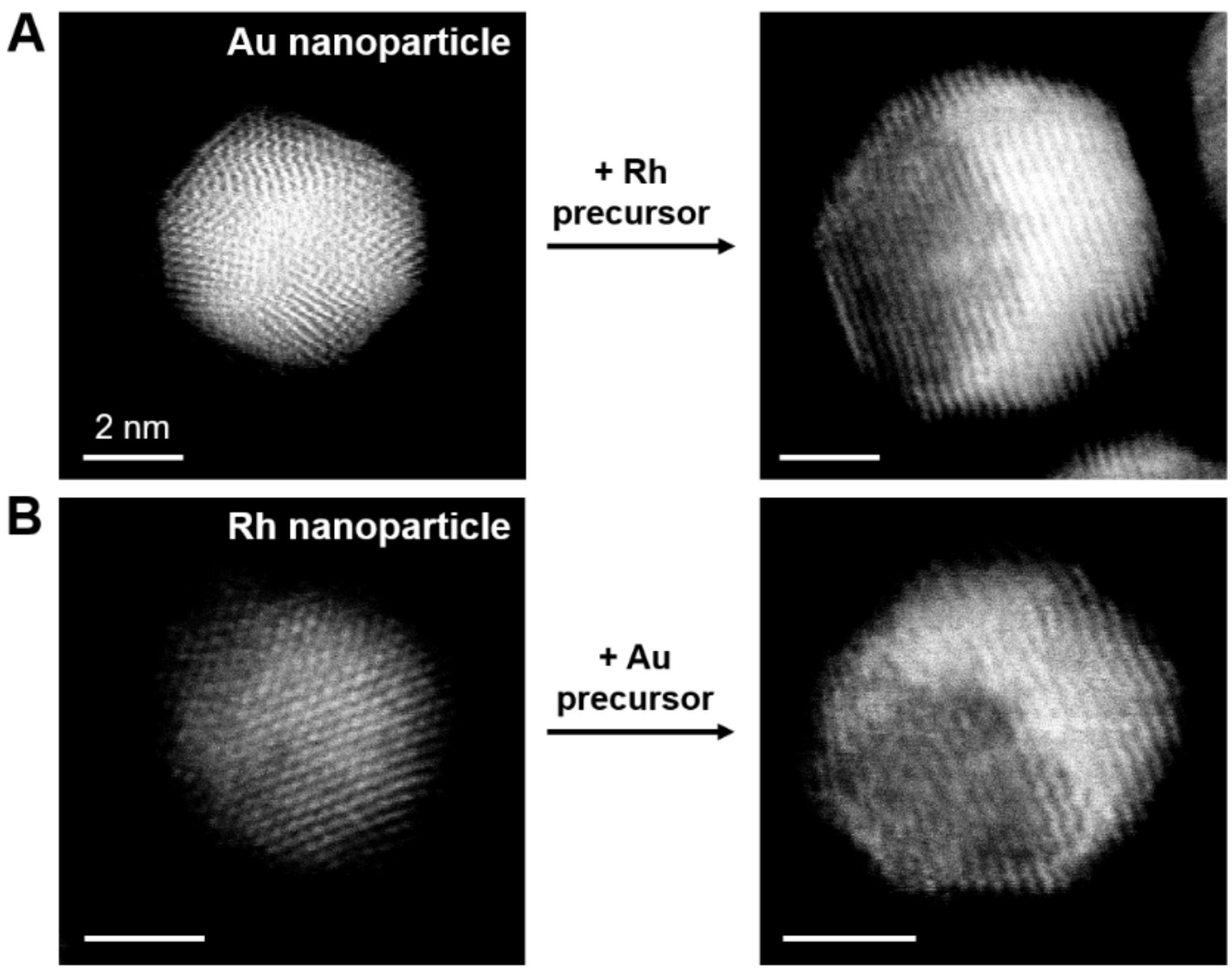

**Fig. S8.**
HAADF-STEM images of AuRh nanoparticles synthesized in a stepwise manner. (A) Au nanoparticles were synthesized first followed by adding Rh precursor to generate the AuRh nanoparticles. (B) Rh nanoparticles were synthesized first followed by adding Au precursor to generate the AuRh nanoparticles. The AuRh structures obtained via both sequences have shell dimer configurations.

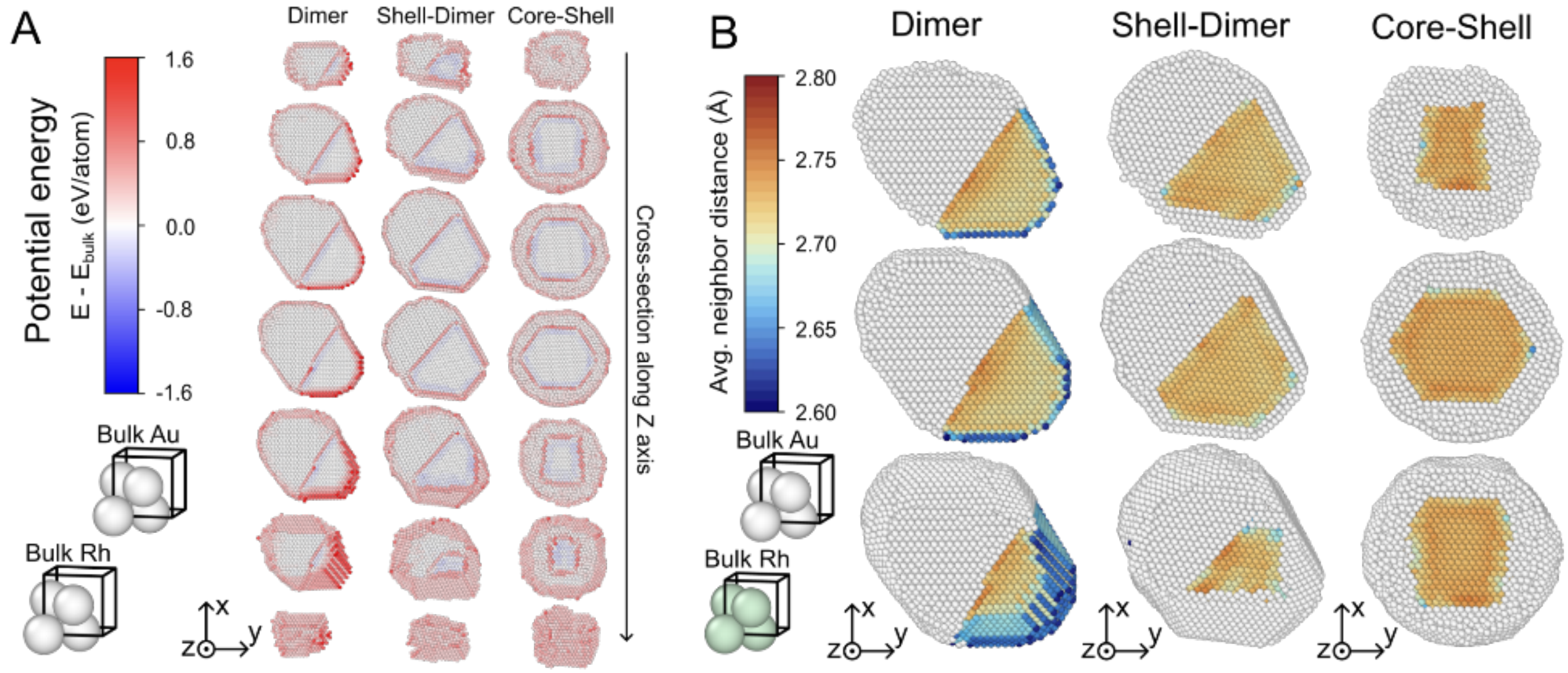

**Fig. S9.**
(A) Sequential cross-sections along the z axis show the three-dimensional organization of each morphology, colored by per-atom potential energies relative to the corresponding bulk phases. (B) Strain effects in the Rh domains of Au-Rh nanoparticles with different architectures. The average Rh-Rh nearest neighbor distances are plotted for the final dimer, shell-dimer, and core-shell structures extracted from MD simulations. Au atoms are rendered in white.

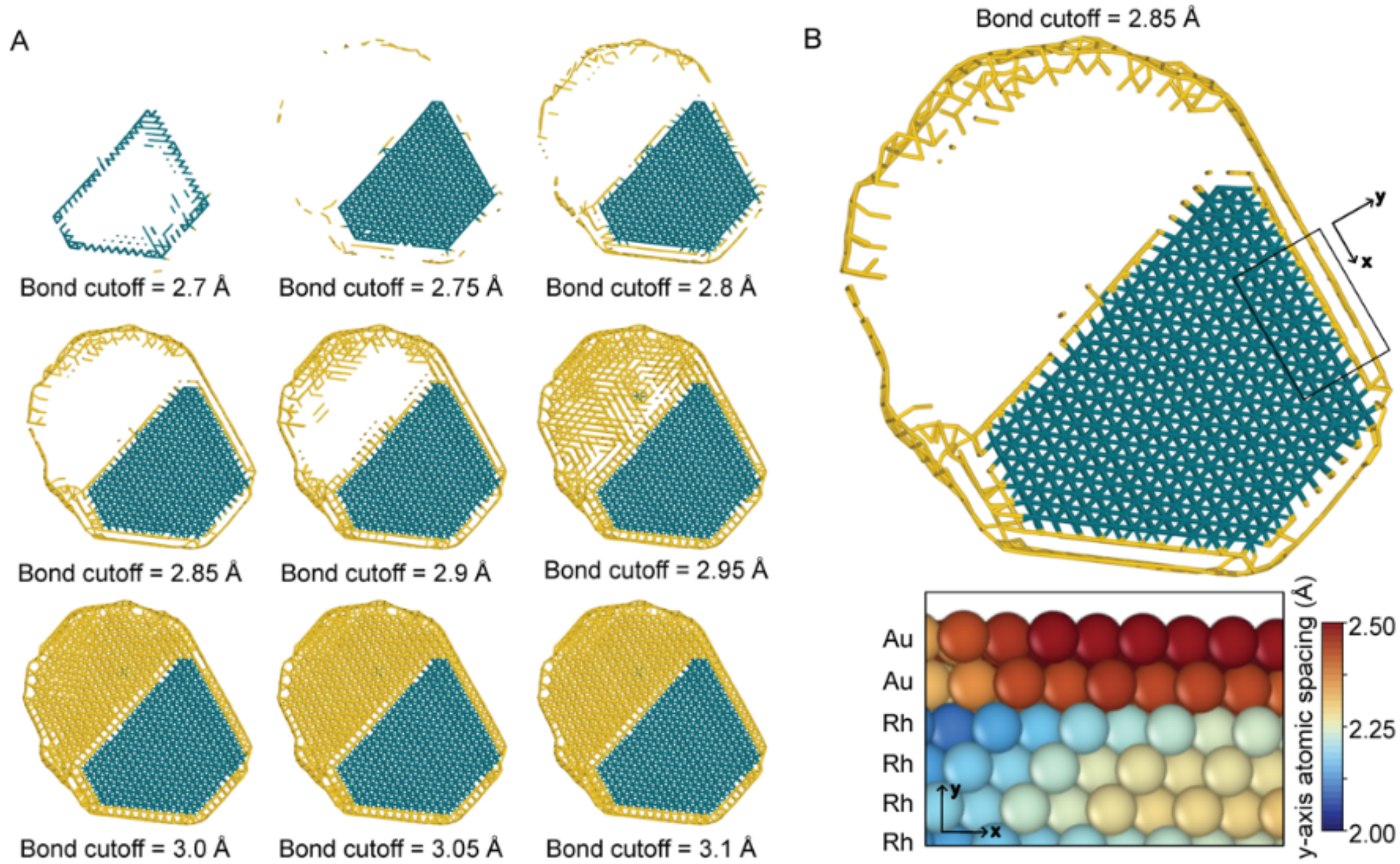

**Fig. S10.**

Bond-length anisotropy in an MD-simulated Au–Rh shell–dimer nanoparticle. (A) Bond networks identified using Au–Au distance cutoffs from 2.70 to 3.10 Å, showing the connectivity of the Au shell and Au-rich domain. (B) Bond network at a cutoff of 2.85 Å, with the enlarged region showing the local atomic arrangement and the variation in atomic spacing along the y direction. Cross-sectional views reveal directional strain in the Au overlayer covering the Rh domain. Au-Au bonds are compressed parallel to the Au-Rh interface and elongated perpendicular to the interface, consistent with the experimental observations. Additional bond contraction occurs in the outermost Au layers of the Au-rich domain, reflecting the tendency of nanoparticle surfaces to contract to reduce surface energy.

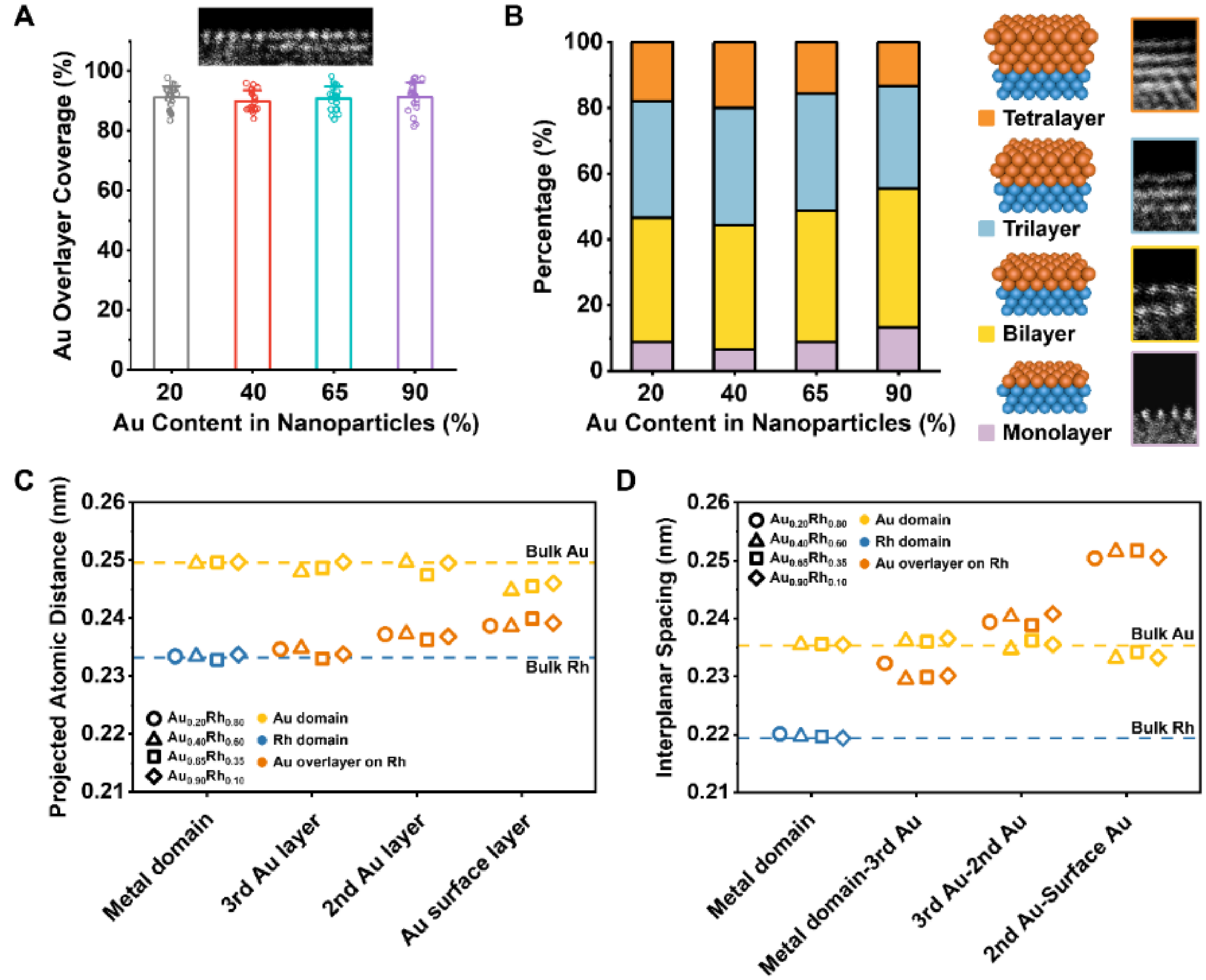

**Fig. S11.**
Effect of nanoparticle composition on the structure of Au overlayers. (A) Coverage of Au overlayer on the Rh domains of AuRh nanoparticles with different compositions. (B) Statistical distribution of overlayers with different numbers of Au atomic layers. (C) Strain analysis of the Au and Rh domains in the direction parallel to the nanoparticle surface. The projected atomic distance is measured within the (111) planes viewed along the [110] zone axis. The (111) planes are parallel to the nanoparticle surface. (D) Strain analysis of the Au and Rh domains in the direction perpendicular to the nanoparticle surface. The interplanar spacing is measured between the (111) planes viewed along the [110] zone axis.

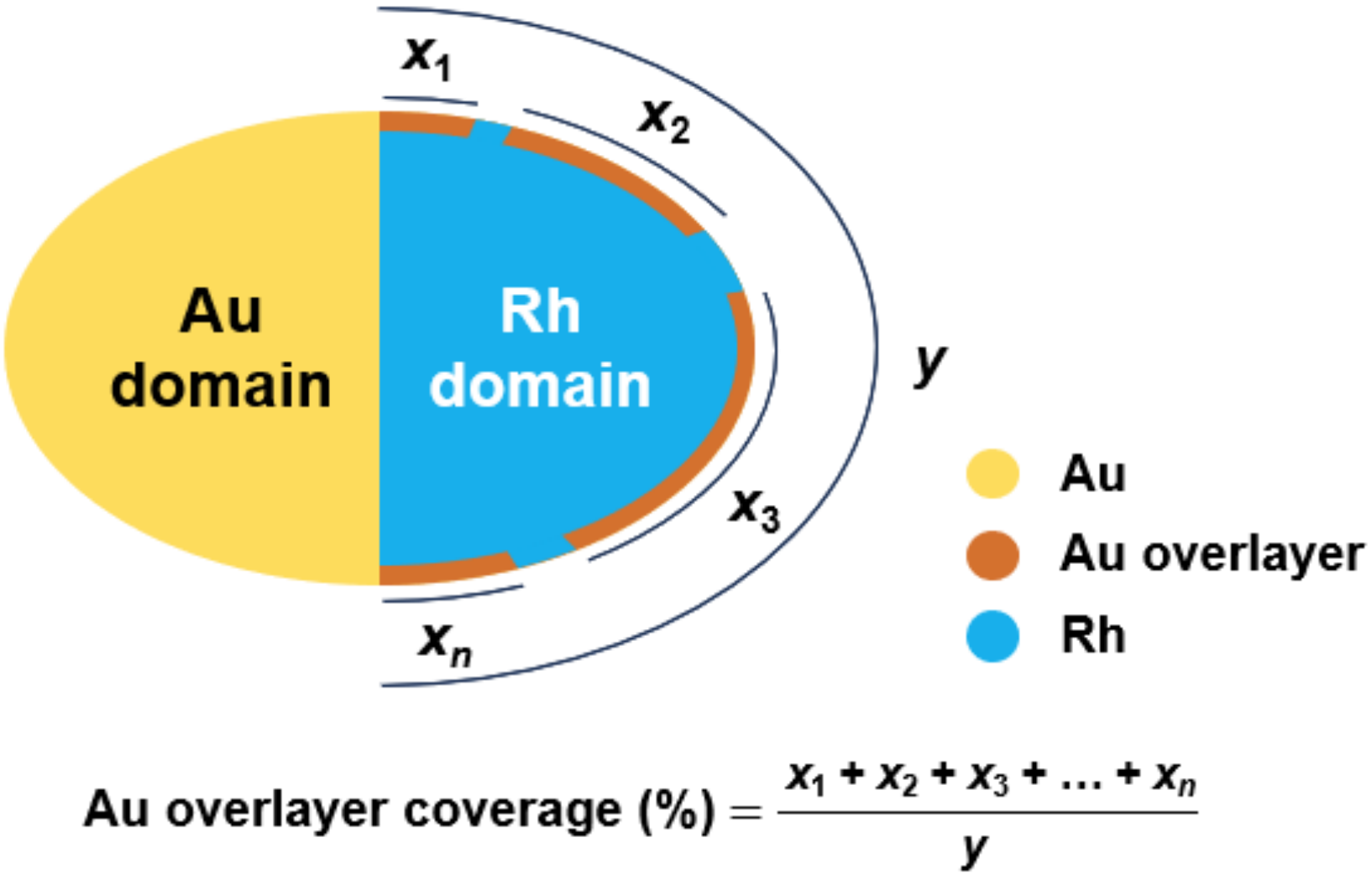

$$\text{Au overlayer coverage (\%)} = \frac{x_1 + x_2 + x_3 + \ldots + x_n}{y}$$

$x_i$ : Total length of Au overlayer

$y$ : Length of Rh domain contour

**Fig. S12.**
Schematic illustration of the calculation of Au overlayer coverage on the Rh domain.

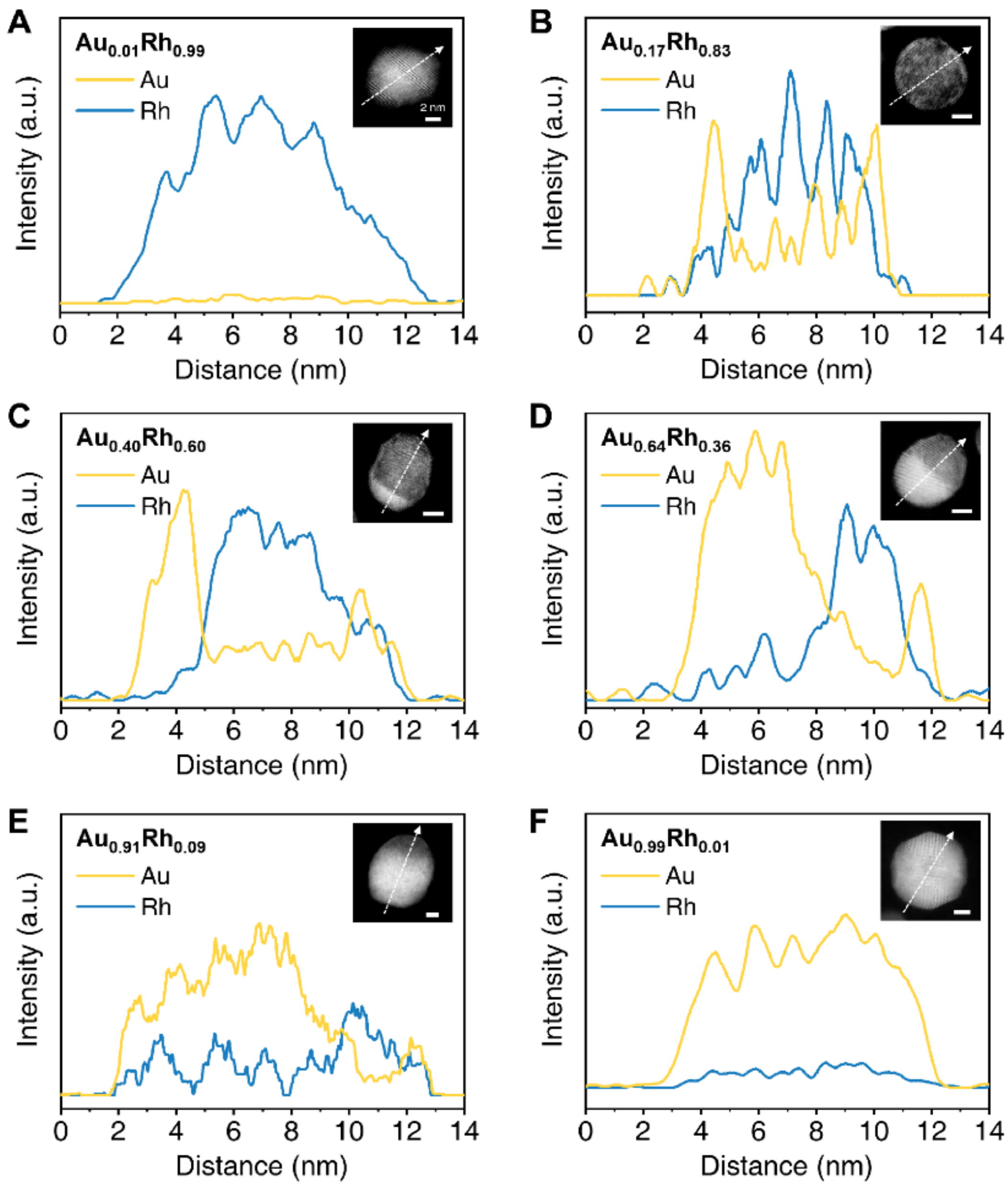

**Fig. S13.**
EDS line-scans of AuRh nanoparticles with different compositions. (A) Au/Rh=1/99, (B) Au/Rh=17/83, (C) Au/Rh= 40/60, (D) Au/Rh=64/36, (E) Au/Rh=91/9, and (F) Au/Rh=99/1. The white arrows in the inset HAADF-STEM images show the traces of the EDS line scans.

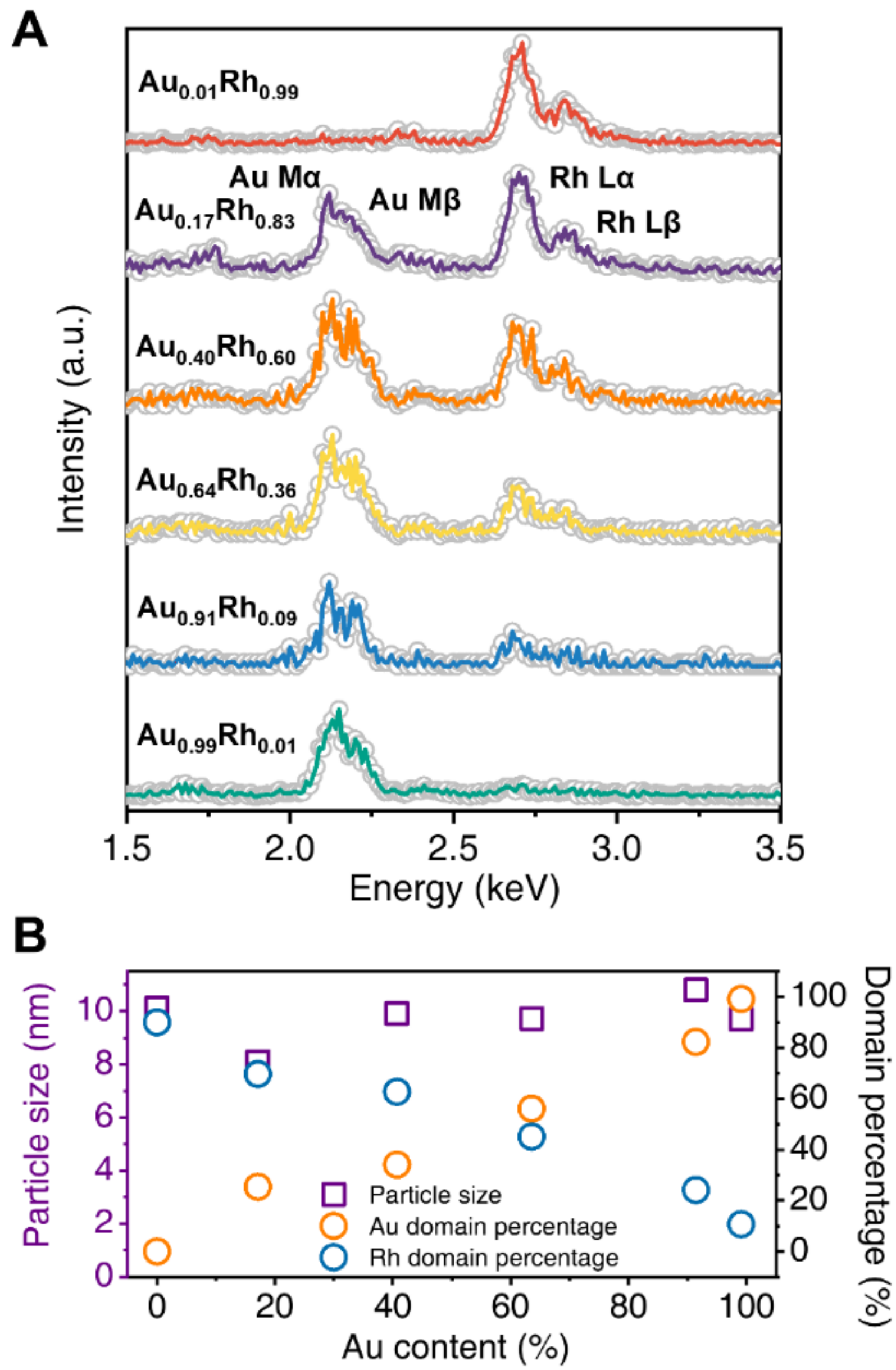

**Fig. S14.**
(A) EDS spectra of AuRh nanoparticles with different compositions. (B) Size and domain area percentage of the AuRh nanoparticles with different compositions.

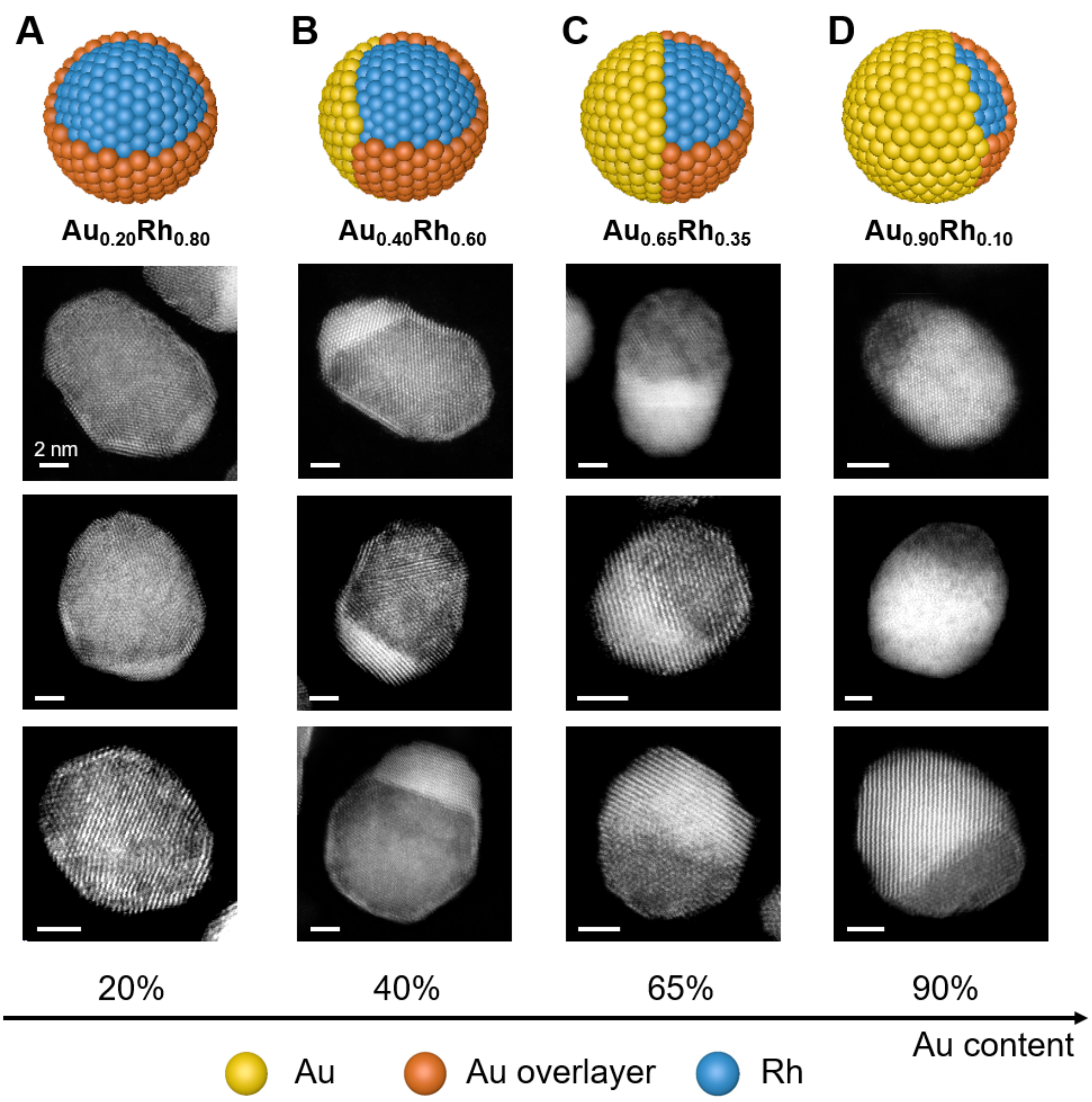

**Fig. S15.**
HAADF-STEM images of AuRh nanoparticles with different compositions. The Au contents in the nanoparticles are (A) ~20 at.%, (B) ~40 at.%, (C) ~65 at.%, and (D) ~90 at.%.

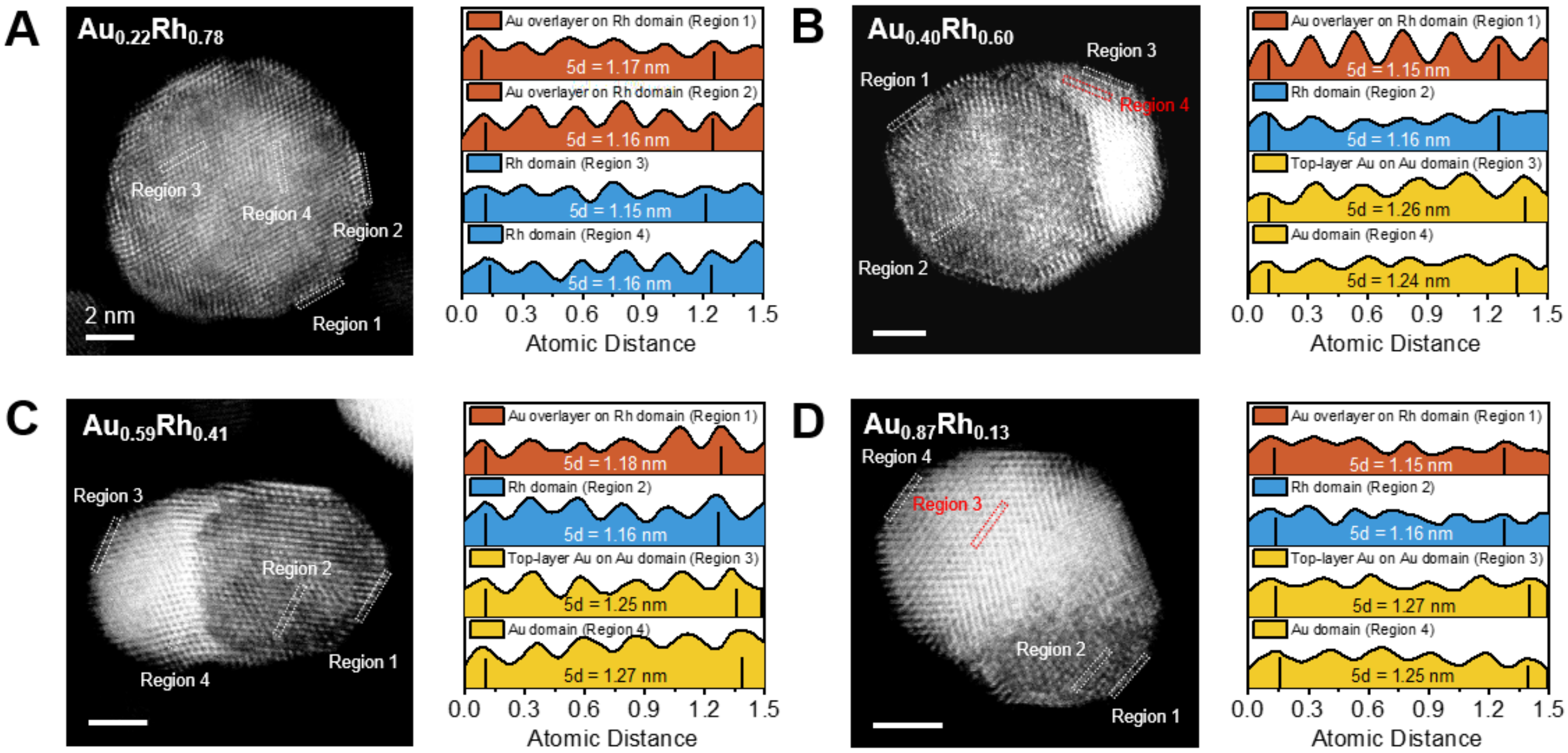

**Fig. S16.**

Atomic resolution HAADF-STEM images of AuRh nanoparticles and line-scan profiles extracted from the STEM images. (A) Au/Rh=22/78, (B) Au/Rh=40/60, (C) Au/Rh=59/41, and (D) Au/Rh=87/13. For (A), regions 1 and 2 are the Au overlayer, and regions 3 and 4 are the Rh domain with Au overlayer. For (B-D), regions 1 are the Au overlayers. Regions 2 are the Rh domains with Au overlayers. Regions 3 are the edge of Au domains. Regions 4 are the Au domains.

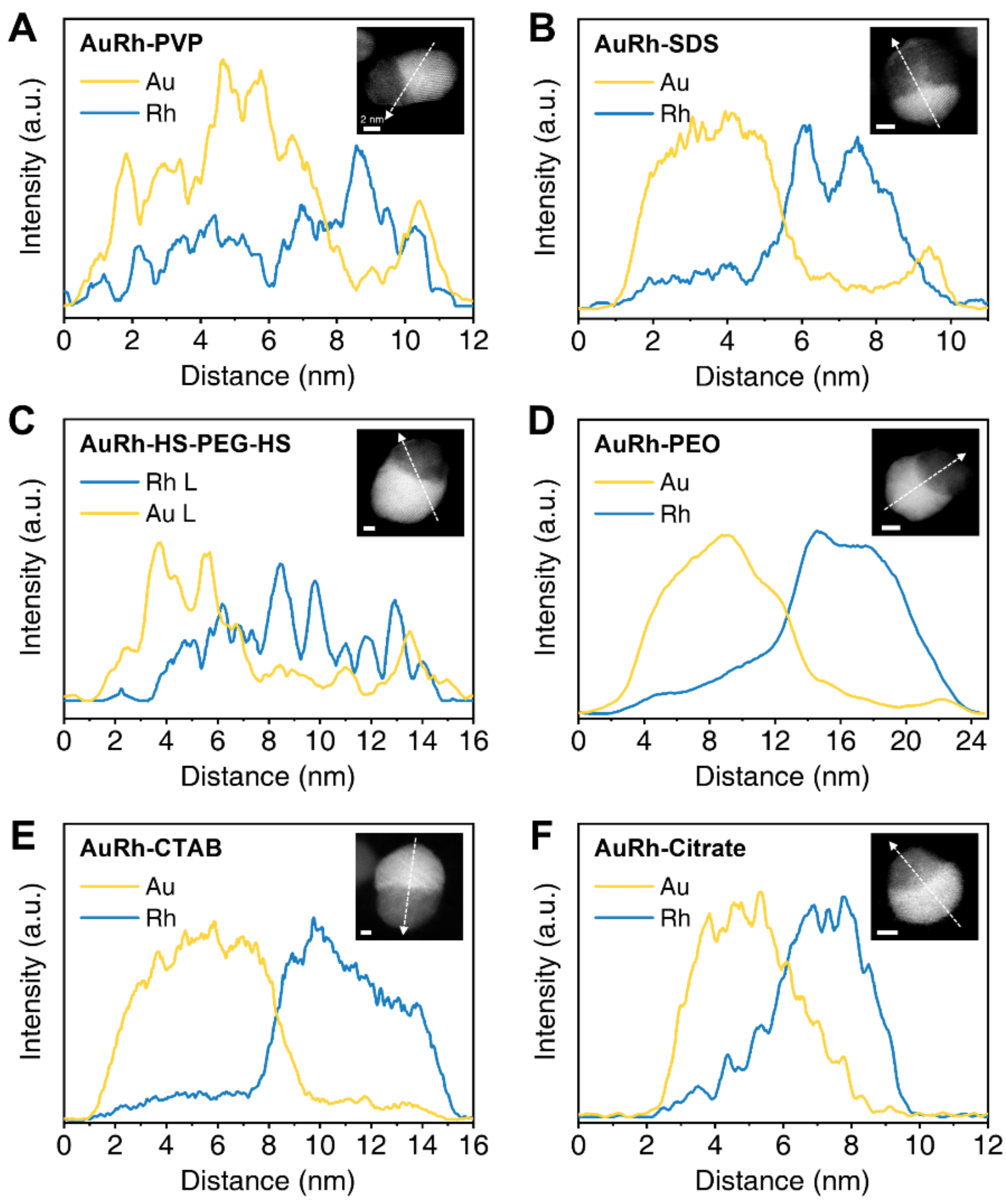

**Fig. S17.**
EDS line scans of AuRh nanoparticles modified by adsorbates originating from (A) PVP, (B) SDS, (C) HS-PEG-HS, (D) PEG, (E) CTAB, and (F) Citrate. The white arrows in the inset STEM images show the traces of the EDS line scans.

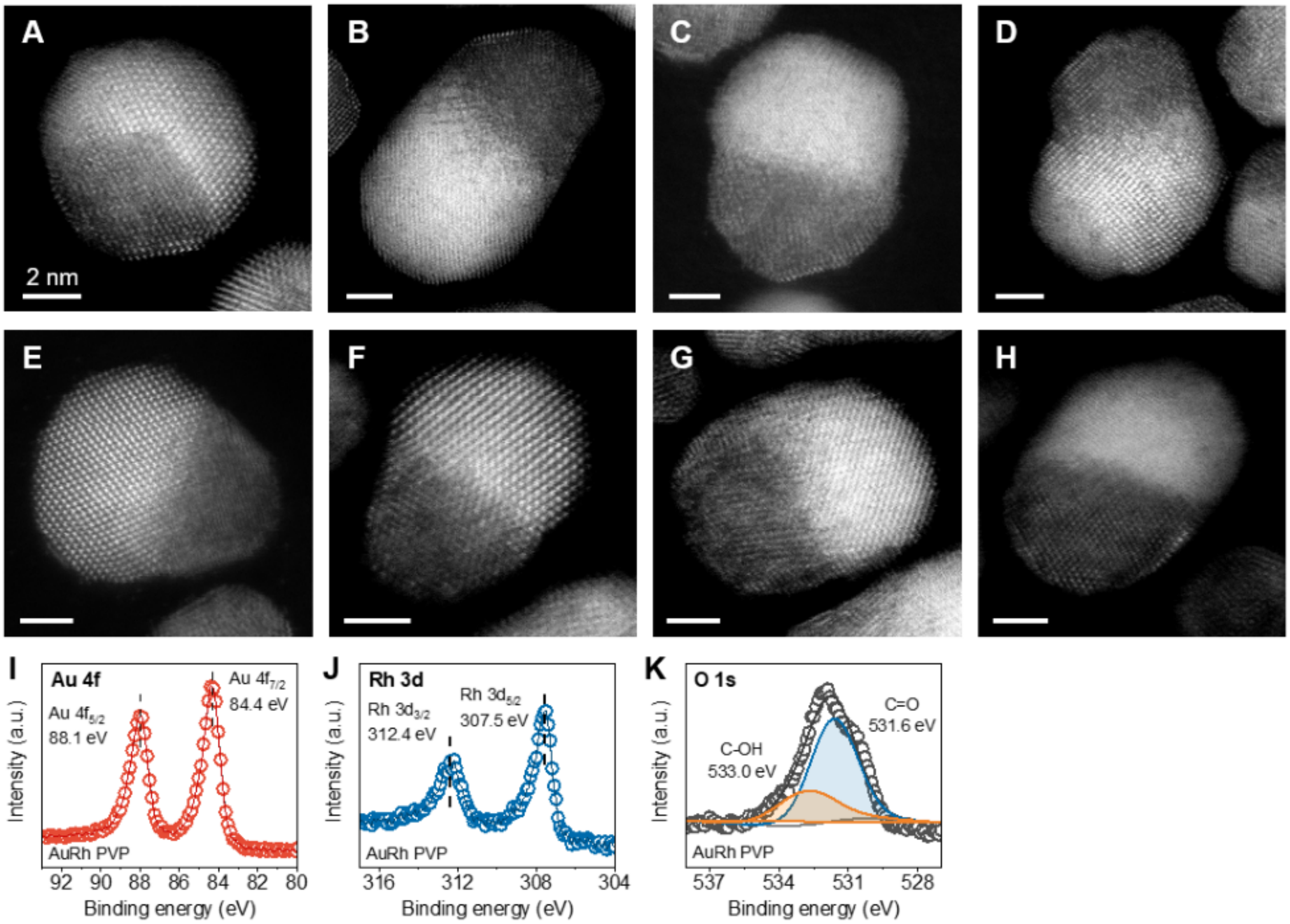

**Fig. S18.**
(A-H) HAADF-STEM images of AuRh nanoparticles decorated by PVP and annealed at 500 °C. (I-K) XPS spectra of the AuRh-PVP nanoparticles in the (I) Au 4f, (J) Rh 3d, and (K) O 1s regions.

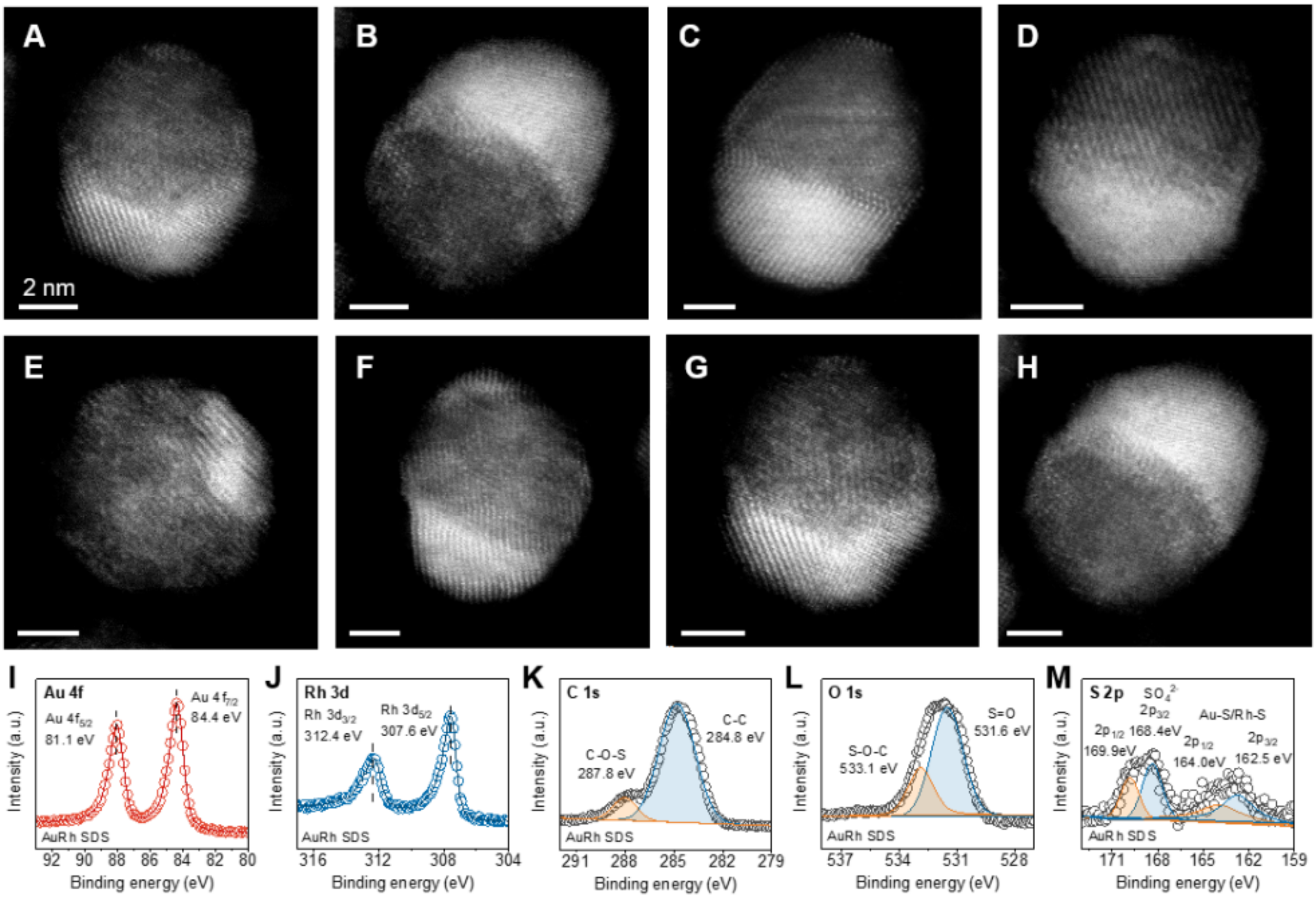

**Fig. S19.**
(A-H) HAADF-STEM images of AuRh nanoparticles decorated by SDS and annealed at 500 °C. (I-M) XPS spectra of the AuRh-SDS nanoparticles in the (I) Au 4f, (J) Rh 3d, (K) C 1s, (L) O 1s, and (M) S 2p regions.

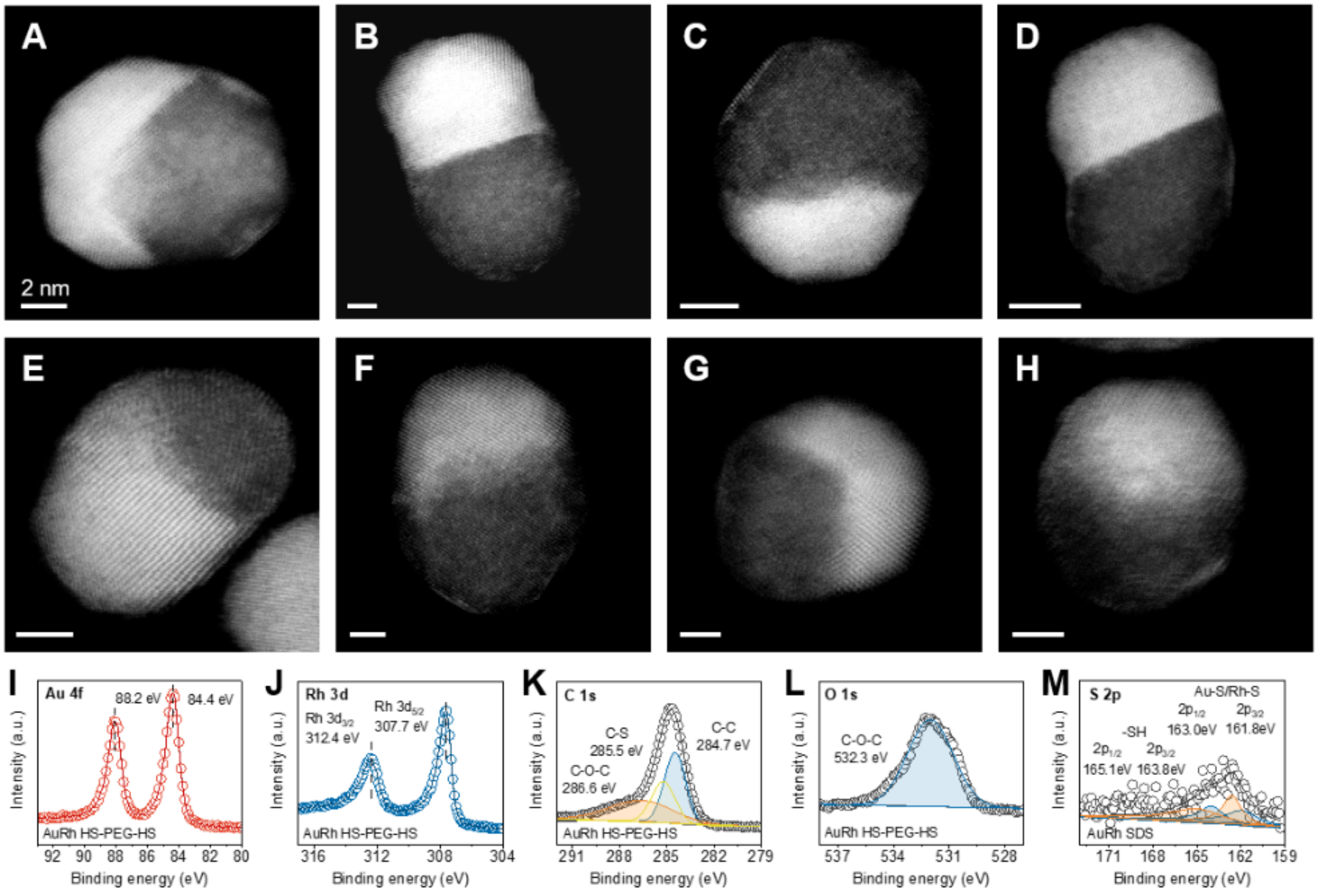

**Fig. S20.**

(A-H) HAADF-STEM images of AuRh nanoparticles decorated by HS-PEG-HS and annealed at 500 °C. (I-M) XPS spectra of the AuRh-HS-PEG-HS nanoparticles in the (I) Au 4f, (J) Rh 3d, (K) C 1s, (L) O 1s, and (M) S 2p regions.

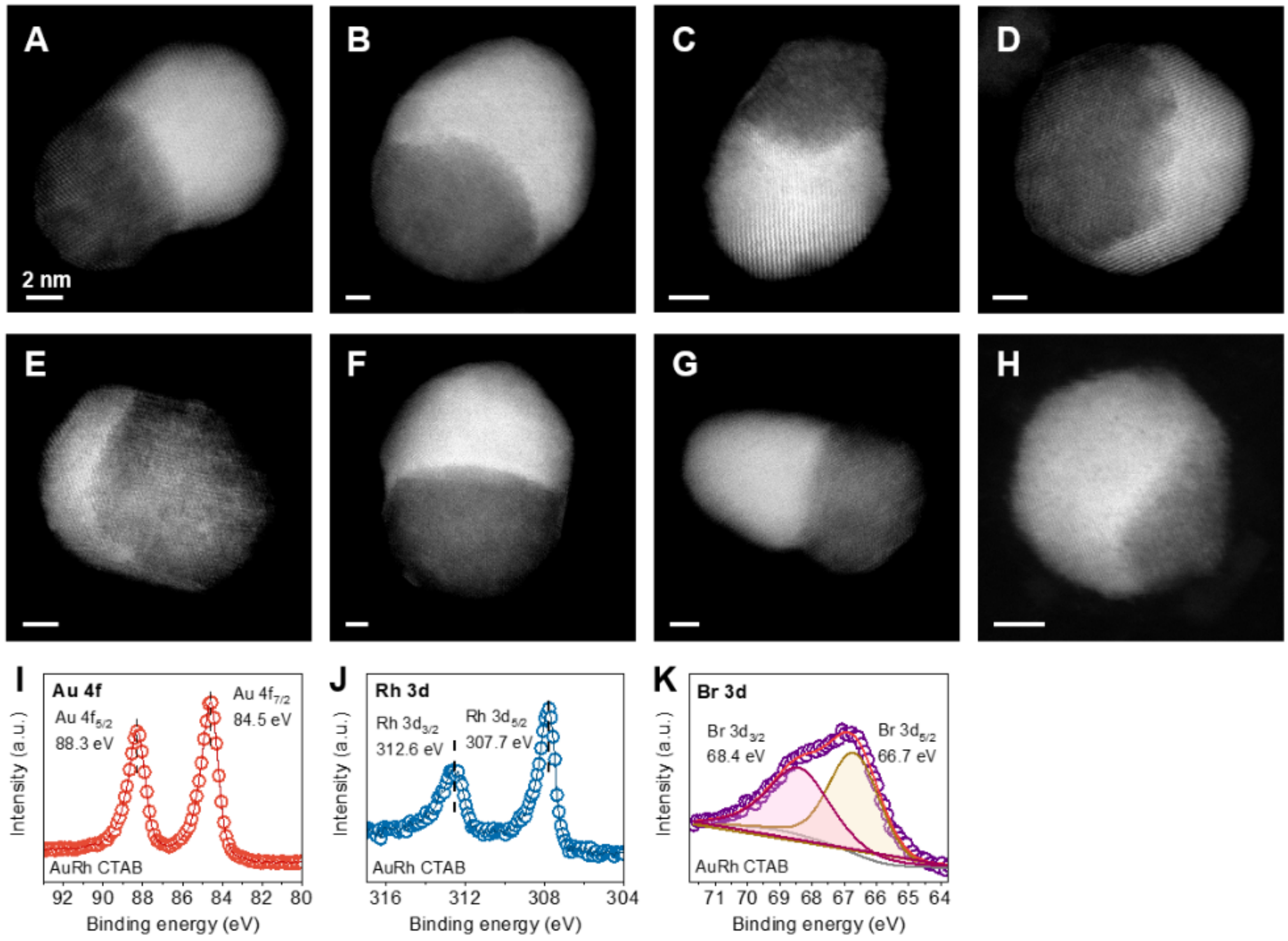

**Fig. S21.**
(A-H) HAADF-STEM images of AuRh nanoparticles decorated by CTAB and annealed at 500 °C. (I-K) XPS spectra of the AuRh-CTAB nanoparticles in the (I) Au 4f, (J) Rh 3d, and (K) Br 3d regions.

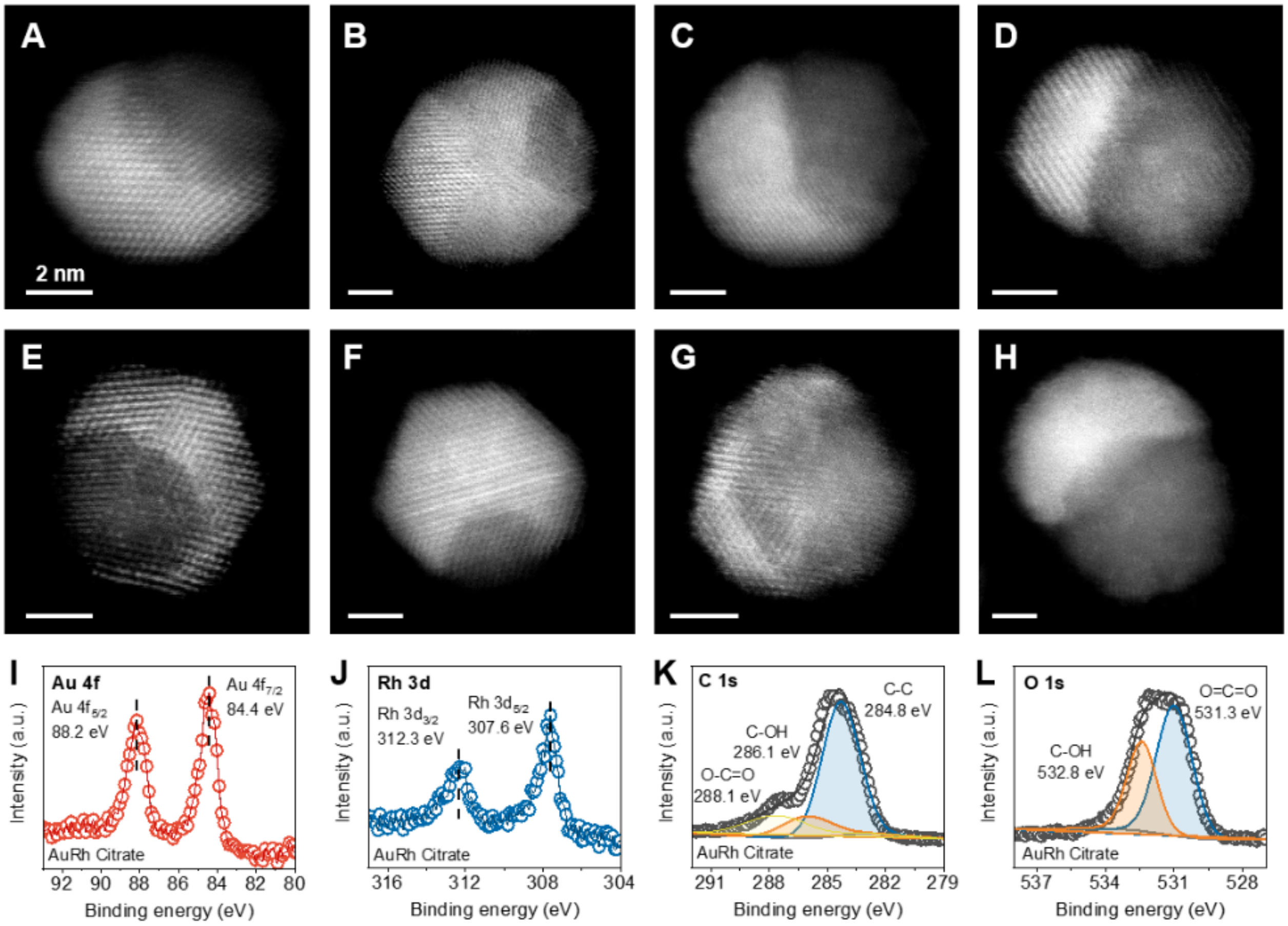

**Fig. S22.**
(A-H) HAADF-STEM images of AuRh nanoparticles decorated by trisodium citrate and annealed at 500 °C. (I-L) XPS spectra of the AuRh-Citrate nanoparticles in the (I) Au 4f, (J) Rh 3d, (K) C 1s, and (L) O 1s regions.

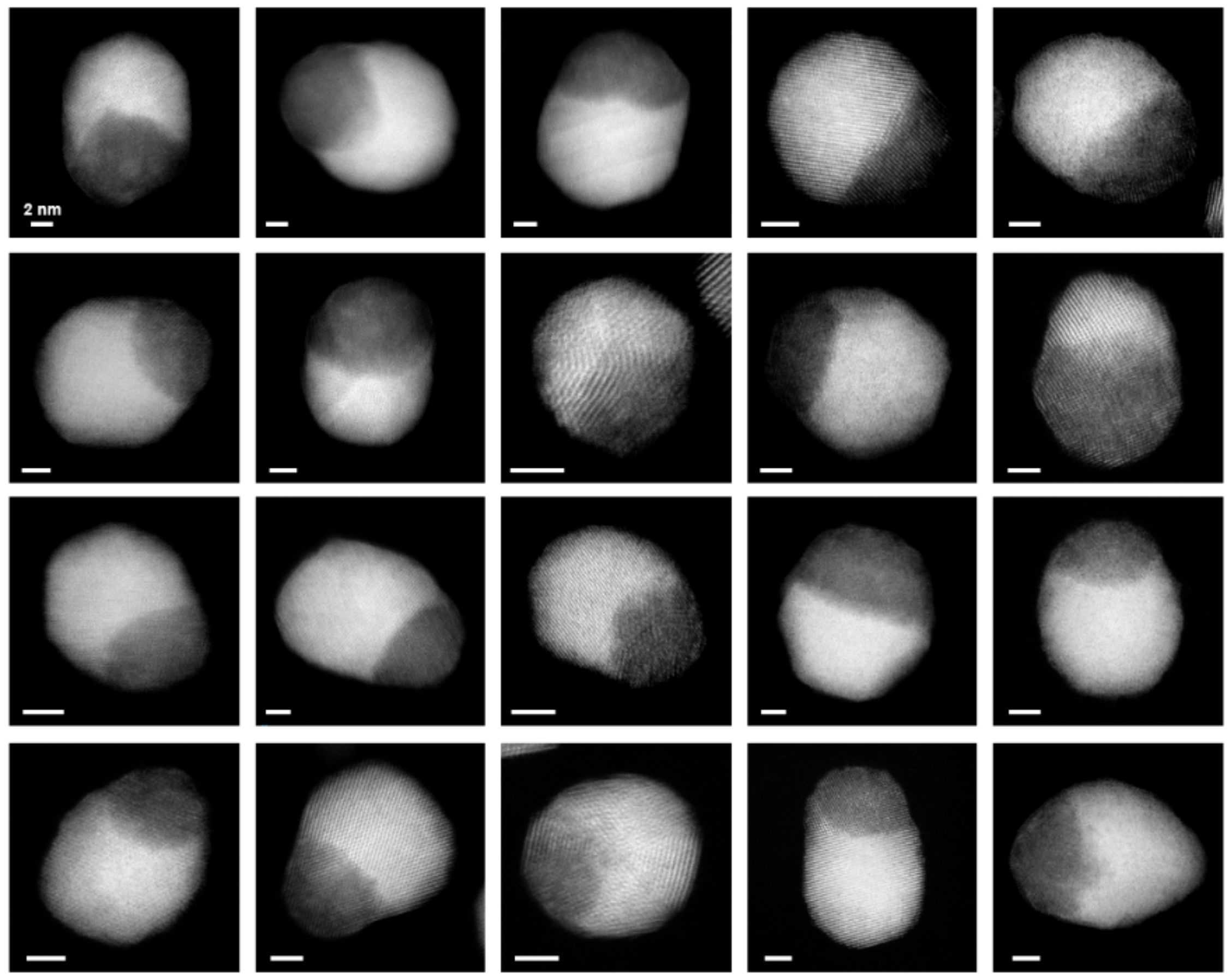

**Fig. S23.**
HAADF-STEM images of AuRh nanoparticles decorated by PEG and annealed at 500 °C.

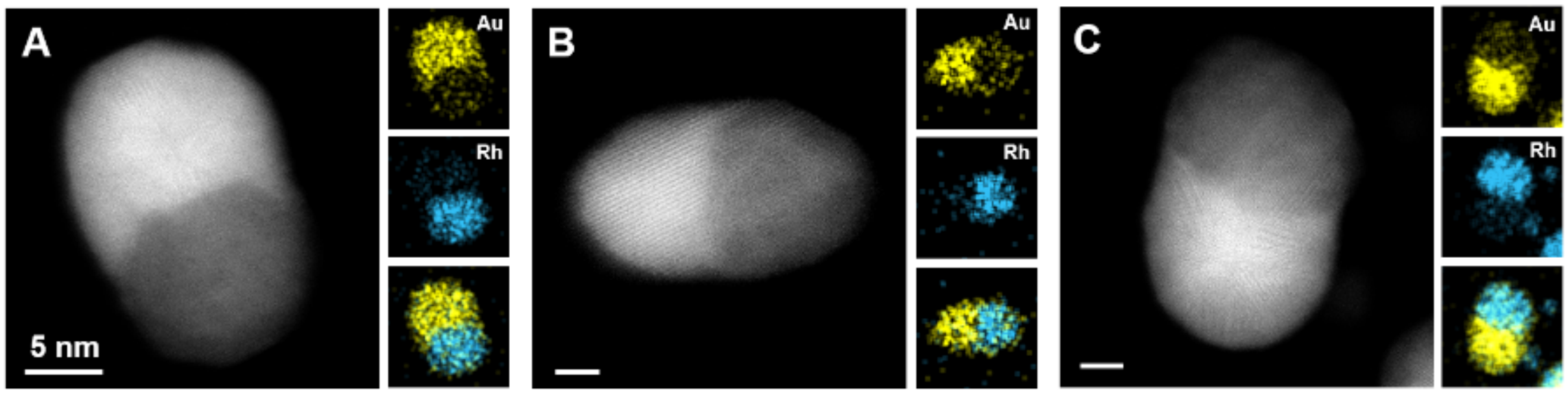

**Fig. S24.**
(A-C) HAADF-STEM images and EDS elemental mapping of AuRh nanoparticles decorated by PEG and annealed at 500 °C. Unlike surface-clean AuRh nanoparticles with Au overlayers observed on the Rh domain surface, Au overlayers are hardly observed in AuRh nanoparticles that have been modified by PEG and annealed at 500 °C.

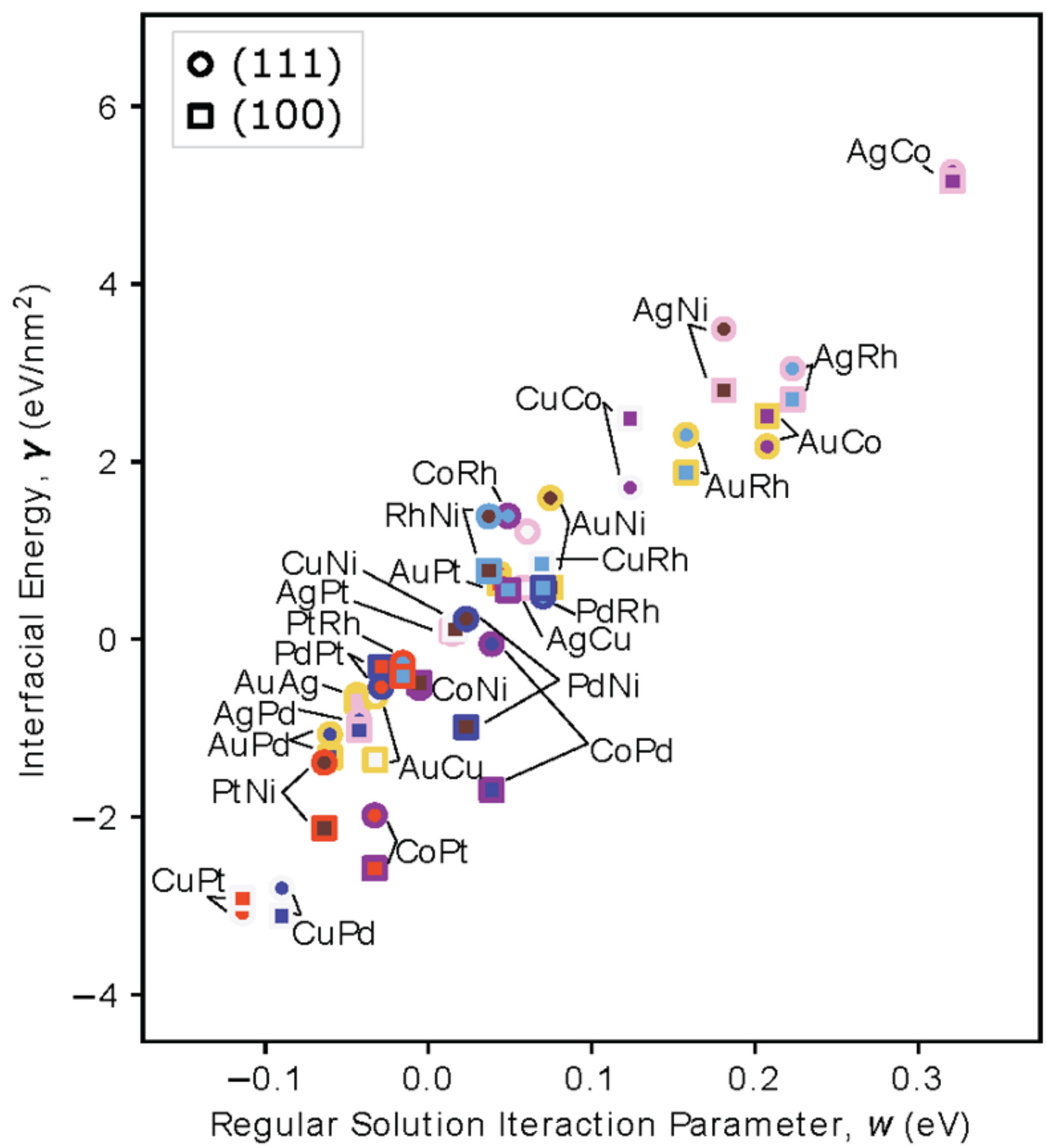

**Fig. S25.**
Interfacial tension, γ, and regular solution interaction parameter, w, calculated for binary combinations of Au, Ag, Cu, Co, Ni, Pd, Pt, and Rh. Interfacial tensions were calculated for coherent interfaces oriented normal to either the (111) or (100) direction. Circles and squares denote the (111) and (100) interfaces, respectively.

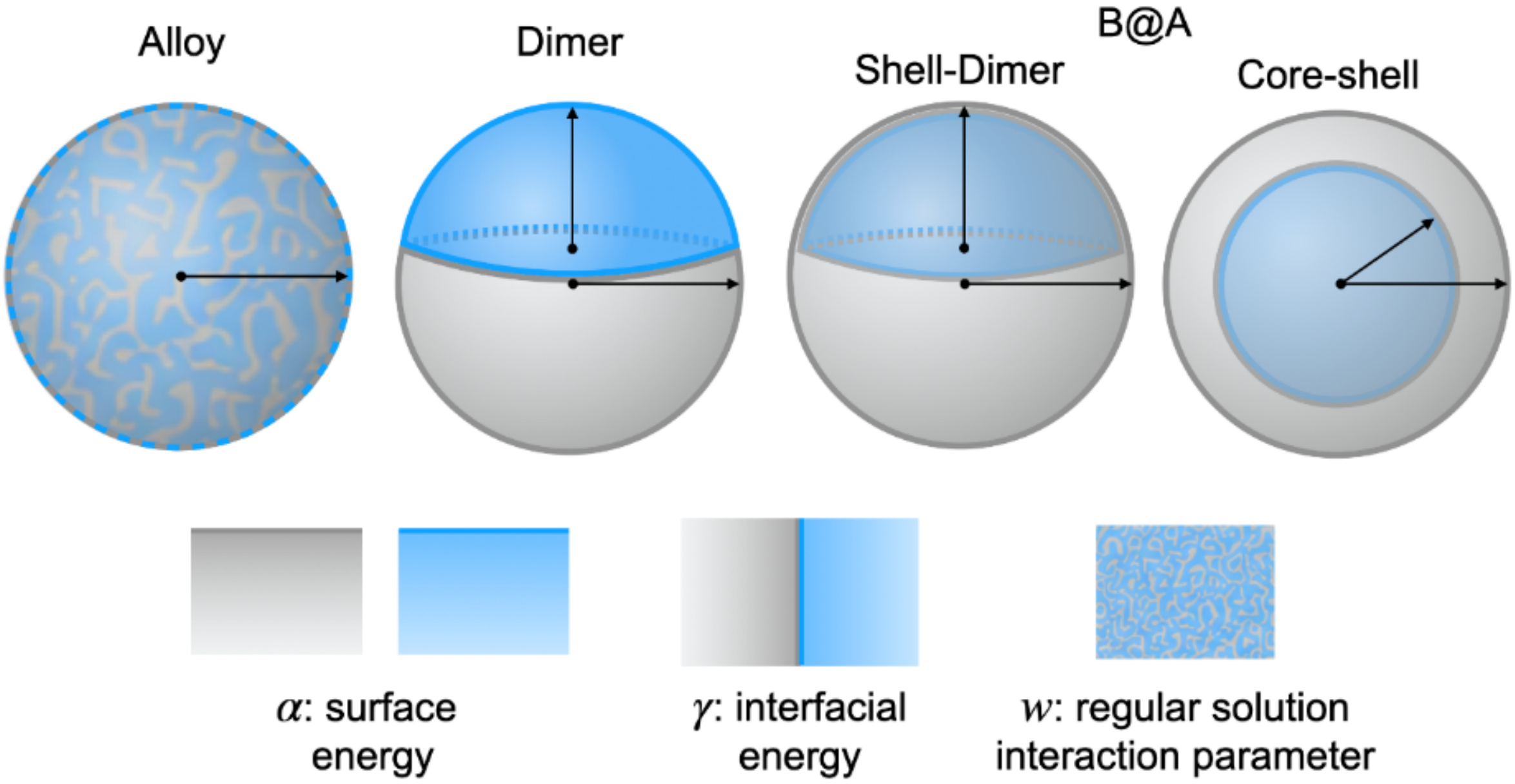

**Fig. S26.**
Scheme of alloy, dimer, shell-dimer, and core-shell nanoparticle architectures, showcasing the differences in the interfaces present in each geometry. Each surface, interface, and mixing of elements has an associated energy penalty or reward.

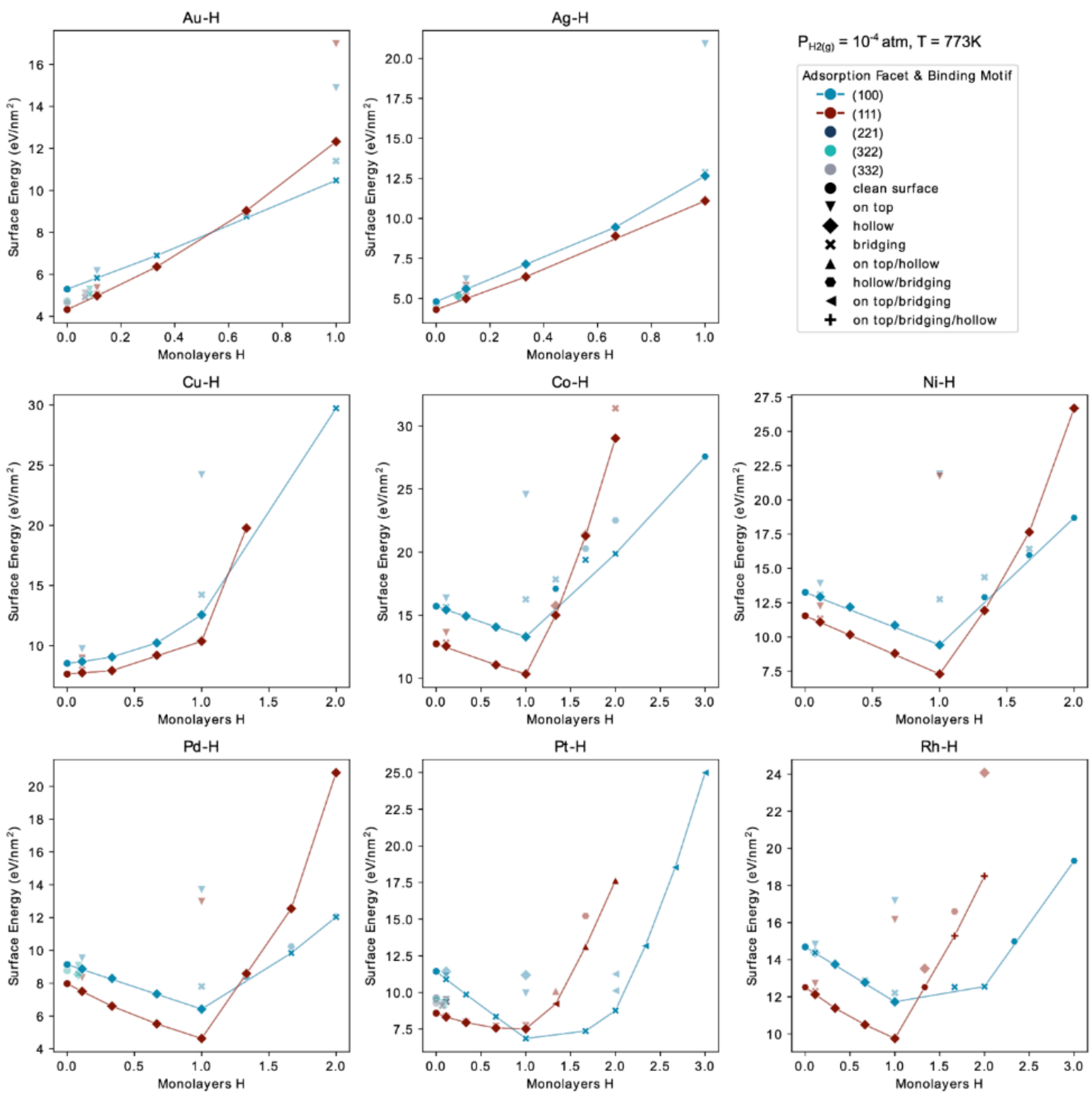

**Fig. S27.**
Calculated surface energies of H-passivated metal surfaces at partial pressures of $PH_2 = 10^{-4}$ atm and temperatures of 773K (500°C). Results are shown for the (100) and (111) facets of each metal over the full range of considered H coverages. A coverage of 1 monolayer corresponds to one adsorbed H atom per surface metal atom. Solid lines trace the lower-energy convex hull for each facet, with the minimum identifying the most stable surface coverage under the specified conditions. Marker shapes indicate the H adsorption motifs used in each calculation, including on-top, bridging, hollow, and mixed configurations.

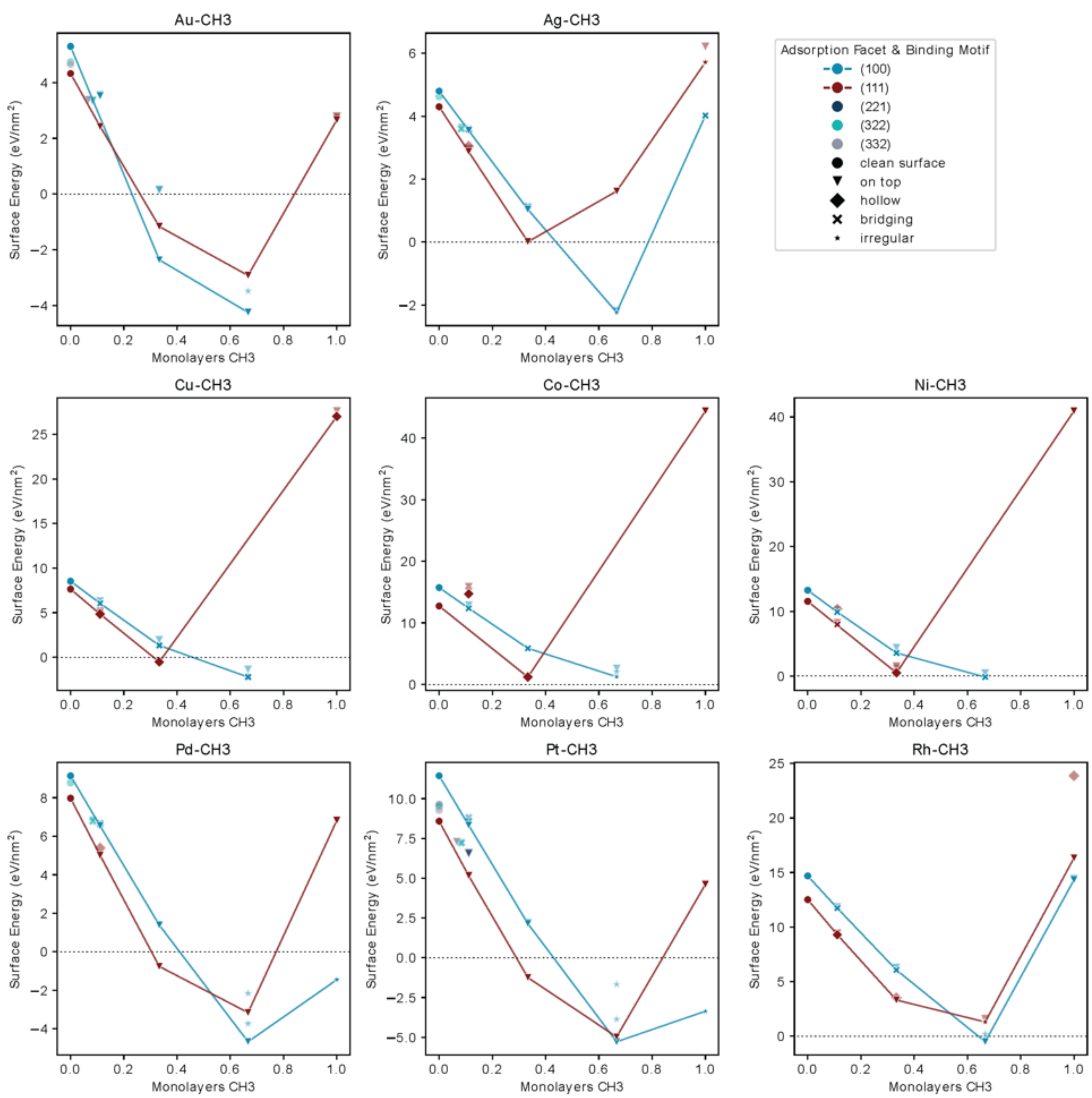

**Fig. S28.**
Calculated surface energies of $CH_3$-passivated metal surfaces. Results are shown for the (100) and (111) facets of each metal over the full range of considered $CH_3$ coverages. A coverage of 1 monolayer corresponds to one adsorbed $CH_3$ molecule per surface metal atom. Solid lines trace the lower-energy convex hull for each facet, with the minimum identifying the most stable surface coverage under the specified conditions. Marker shapes indicate the $CH_3$ adsorption motifs used in each calculation, including on-top, bridging, hollow, and mixed configurations.

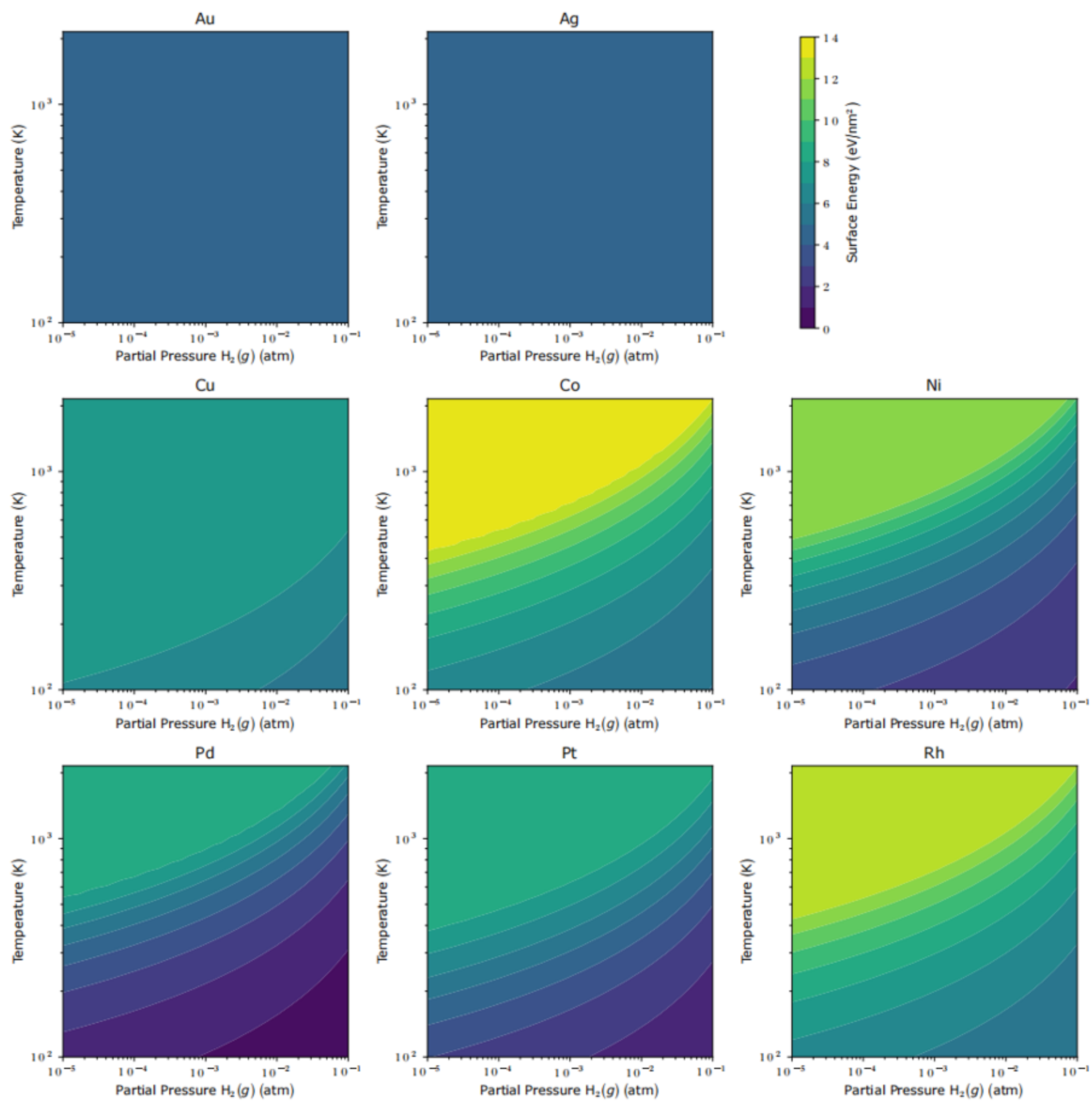

**Fig. S29.**
Calculated surface energies of Wulff constructions of metal surfaces under different temperature and $H_2$ partial pressure conditions.

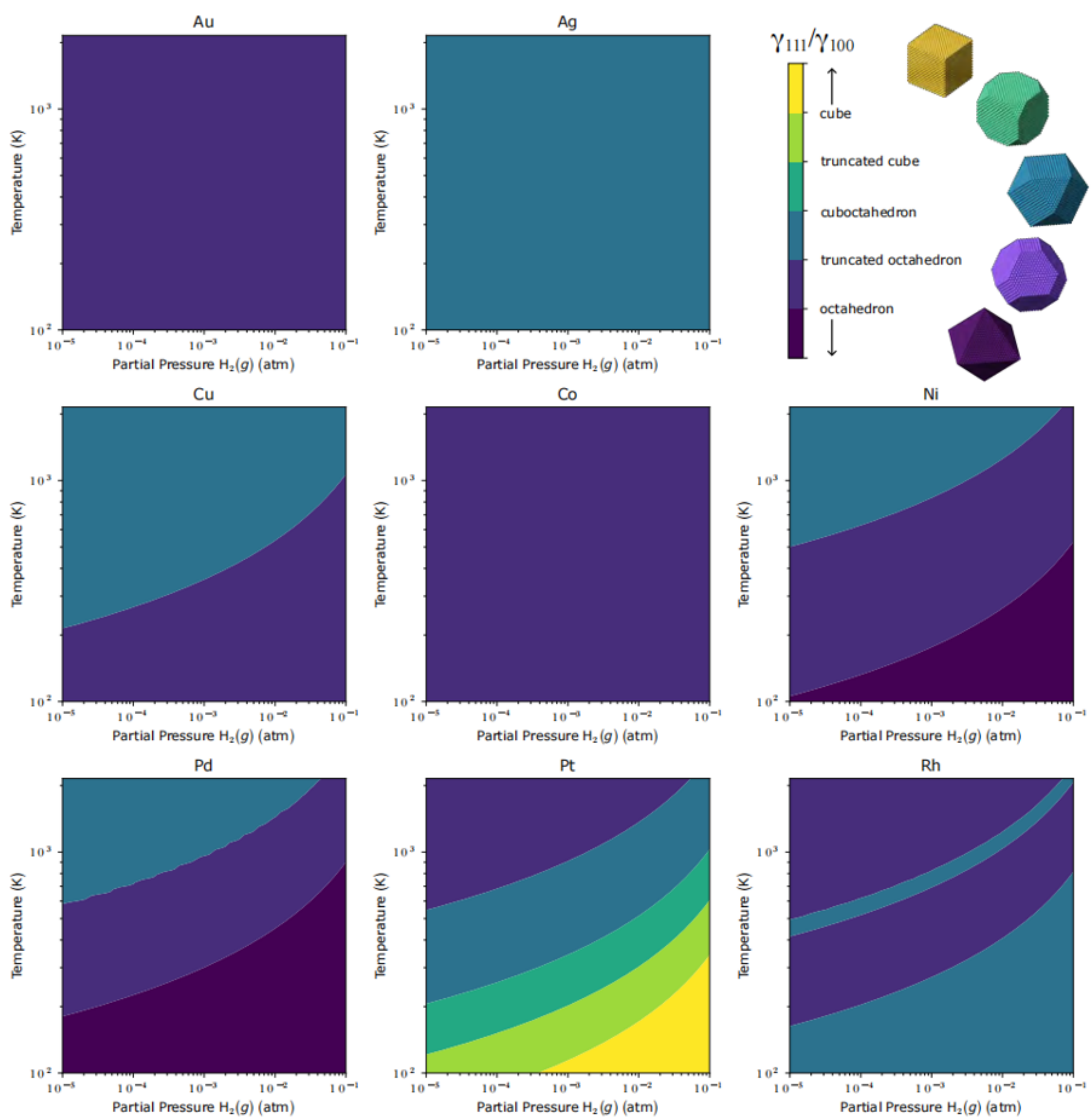

**Fig. S30.**
Calculated Wulff shapes of metal nanoparticles under different temperature and $H_2$ partial pressure conditions.

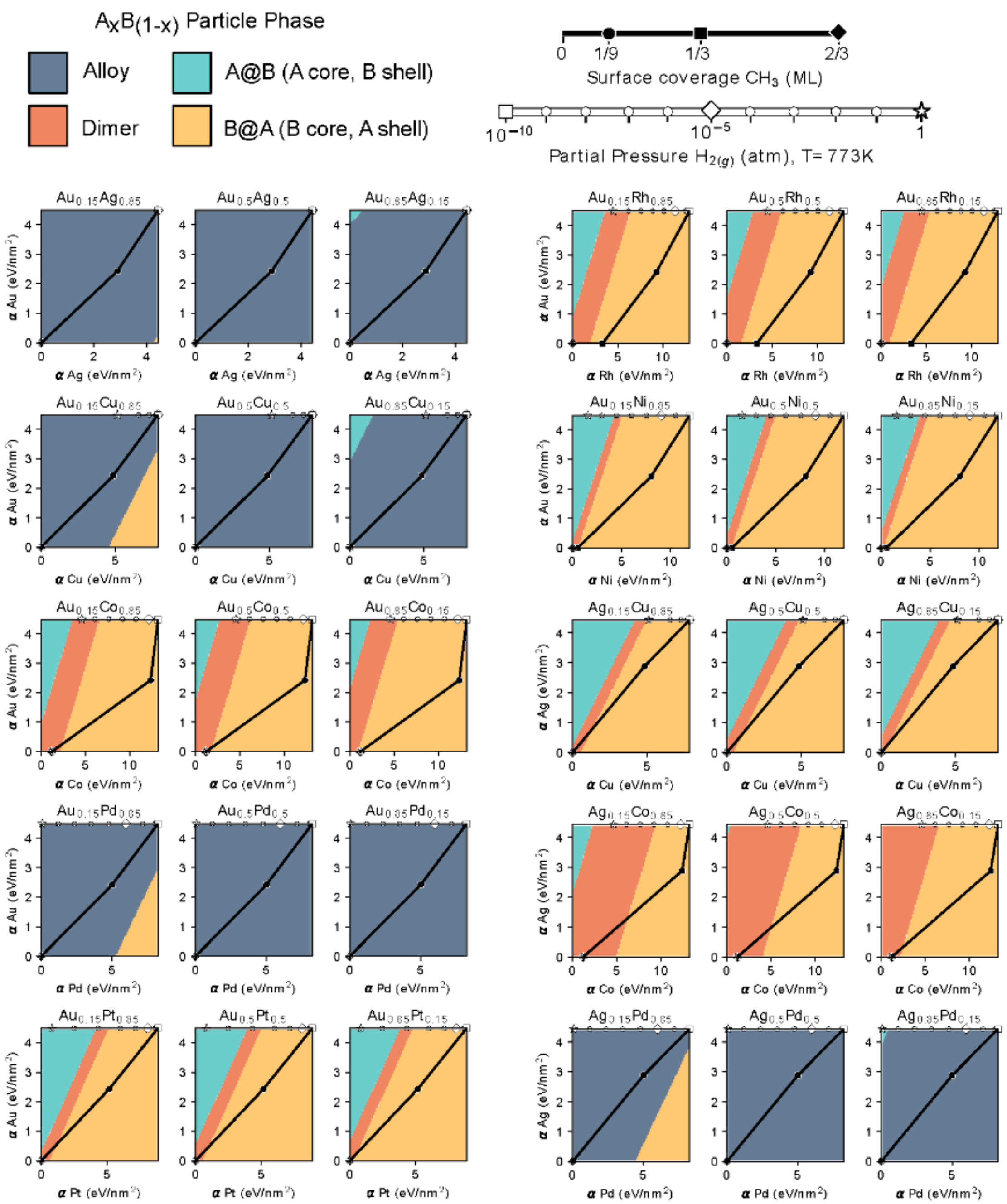

**Fig. S31.**
Thermodynamic phase diagrams for 10 nm binary metal nanoparticles at 500 °C. Predicted phase stability is shown for A-B nanoparticles (A-B = Au-Ag, Au-Cu, Au-Co, Au-Pd, Au-Pt, Au-Rh, Au-Ni, Ag-Cu, Ag-Co, and Ag-Pd) as a function of the surface energies of metals A and B. Colored regions indicate the predicted equilibrium particle morphology: alloy, dimer, or shell (core-shell and shell-dimer). For each metal pair, diagrams are shown for compositions $A_{0.15}B_{0.85}$ (left), $A_{0.5}B_{0.5}$ (middle), and $A_{0.85}B_{0.15}$ (right). Black lines show changes in surface energy with increasing $CH_3$ coverage, and white lines show changes with increasing $H_2$ partial pressure; symbols correspond to the conditions indicated in the legends. The upper-right corner of each diagram represents the unpassivated, clean-surface limit.

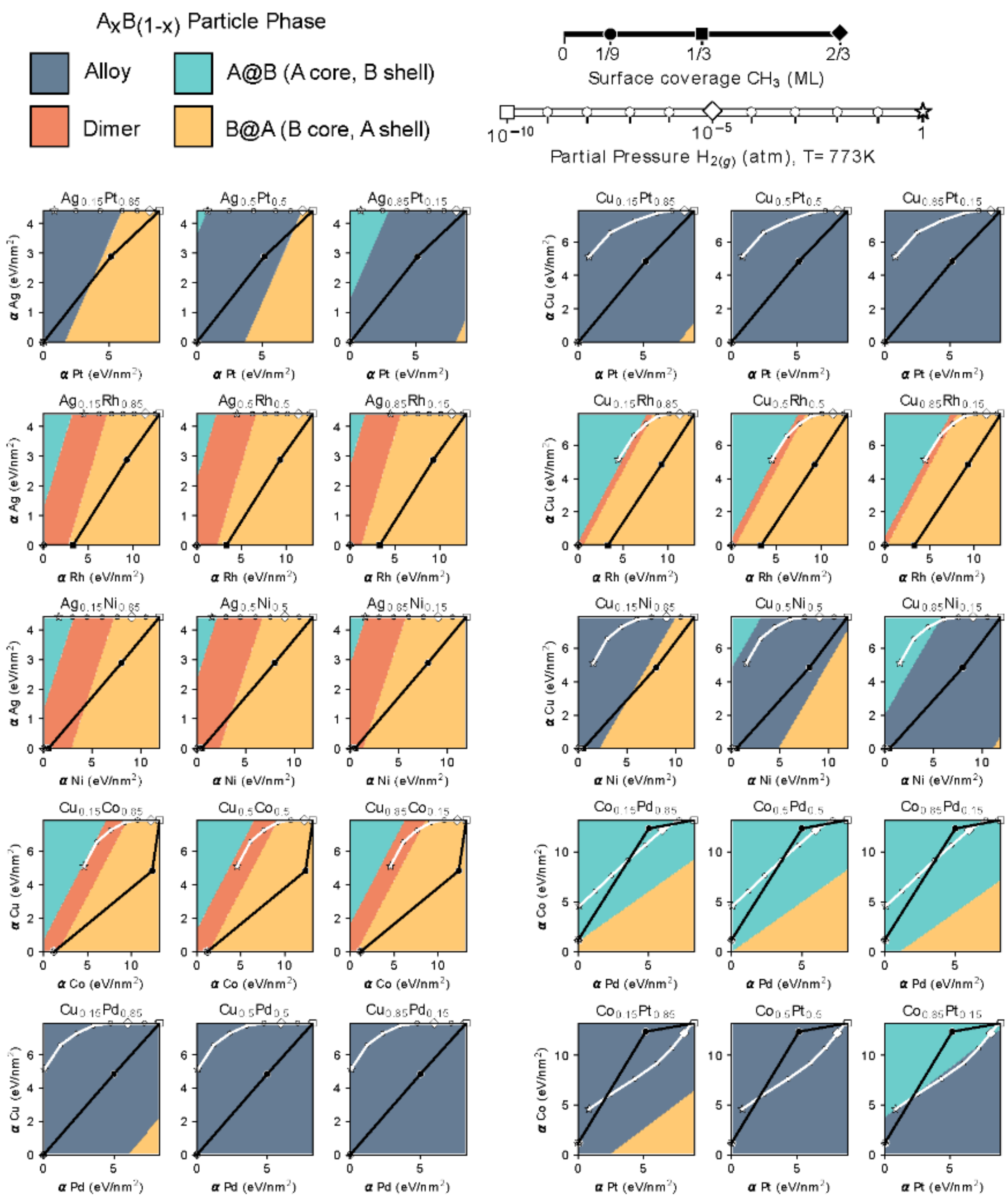

**Fig. S32.**

Thermodynamic phase diagrams for 10 nm binary metal nanoparticles at 500 °C. Predicted phase stability is shown for A-B nanoparticles (A-B = Ag-Pt, Ag-Rh, Ag-Ni, Cu-Co, Cu-Pd, Cu-Pt, Cu-Rh, Cu-Ni, Co-Pd, and Co-Pt) as a function of the surface energies of metals A and B. Colored regions indicate the predicted equilibrium particle morphology: alloy, dimer, or shell (core-shell and shell-dimer). For each metal pair, diagrams are shown for compositions $A_{0.15}B_{0.85}$ (left), $A_{0.5}B_{0.5}$ (middle), and $A_{0.85}B_{0.15}$ (right). Black lines show changes in surface energy with increasing $CH_3$ coverage, and white lines show changes with increasing $H_2$ partial pressure; symbols correspond to the conditions indicated in the legends. The upper-right corner of each diagram represents the unpassivated, clean-surface limit.

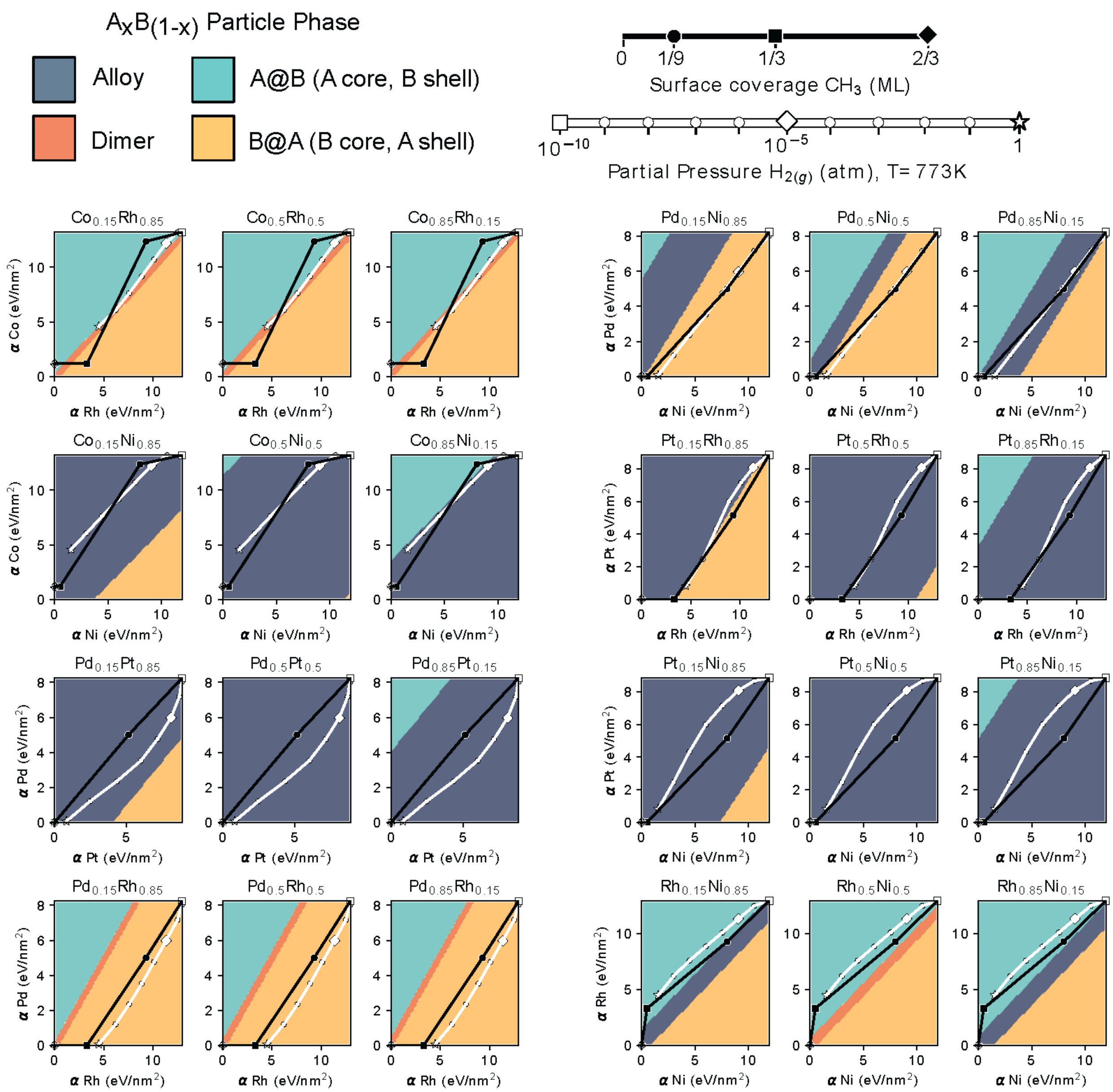

**Fig. S33.**
Thermodynamic phase diagrams for 10 nm binary metal nanoparticles at 500 °C. Predicted phase stability is shown for A-B nanoparticles (A-B = Co-Rh, Co-Ni, Pd-Pt, Pd-Rh, Pd-Ni, Pt-Rh, Pt-Ni, and Rh-Ni) as a function of the surface energies of metals A and B. Colored regions indicate the predicted equilibrium particle morphology: alloy, dimer, or shell (core-shell and shell-dimer). For each metal pair, diagrams are shown for compositions $A_{0.15}B_{0.85}$ (left), $A_{0.5}B_{0.5}$ (middle), and $A_{0.85}B_{0.15}$ (right). Black lines show changes in surface energy with increasing $CH_3$ coverage, and white lines show changes with increasing $H_2$ partial pressure; symbols correspond to the conditions indicated in the legends. The upper-right corner of each diagram represents the unpassivated, clean-surface limit.

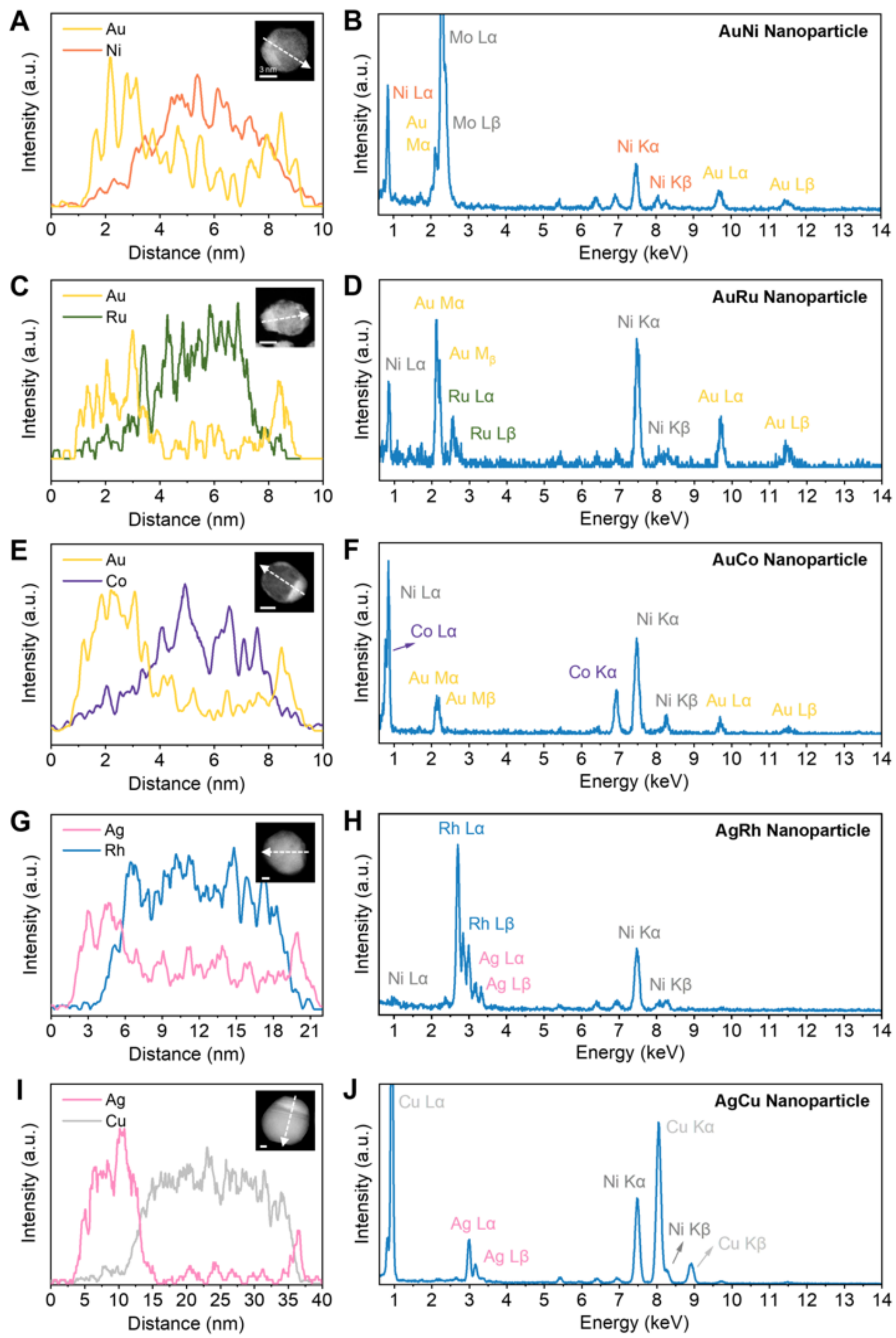

**Fig. S34.**
EDS line scans and EDS spectra of (A,B) AuNi, (C,D) AuRu, (E,F) AuCo, (G,H) AgRh, and (I,J) AgCu nanoparticles. The white arrows in the inset HAADF-STEM images show the traces of the EDS line scans.

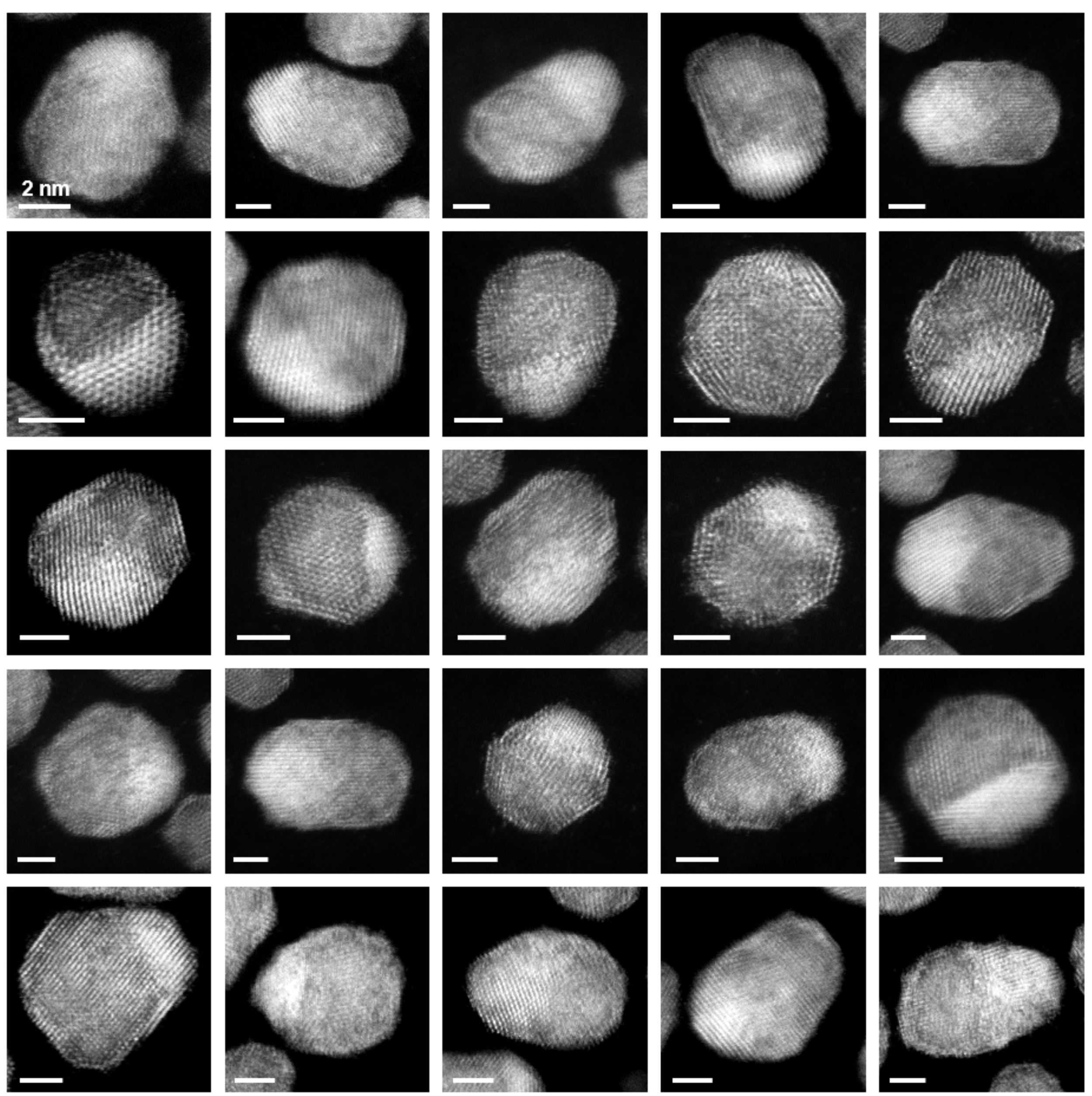

**Fig. S35.**
HAADF-STEM images of AuRu nanoparticles.

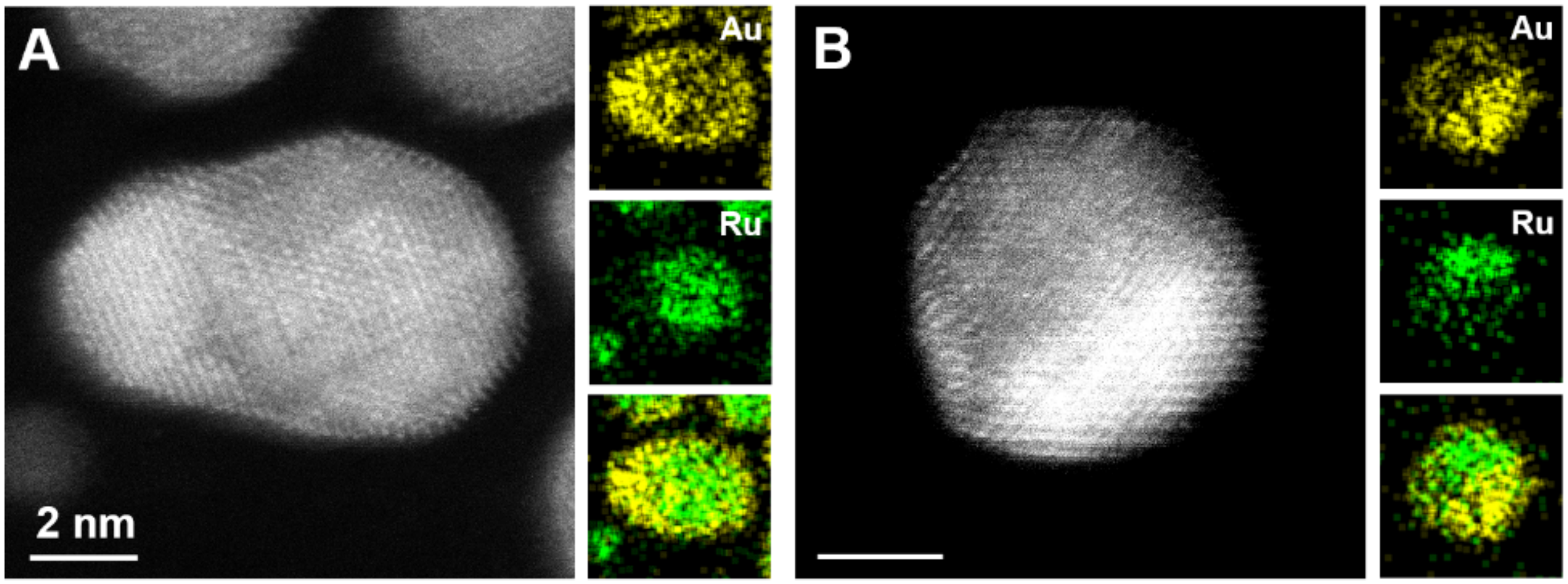

**Fig. S36.**
HAADF-STEM images and EDS elemental mapping of AuRu nanoparticles. Au and Ru are immiscible with each other, forming a phase-separated structure with an ultrathin Au overlayer on the surface of the Ru domain.

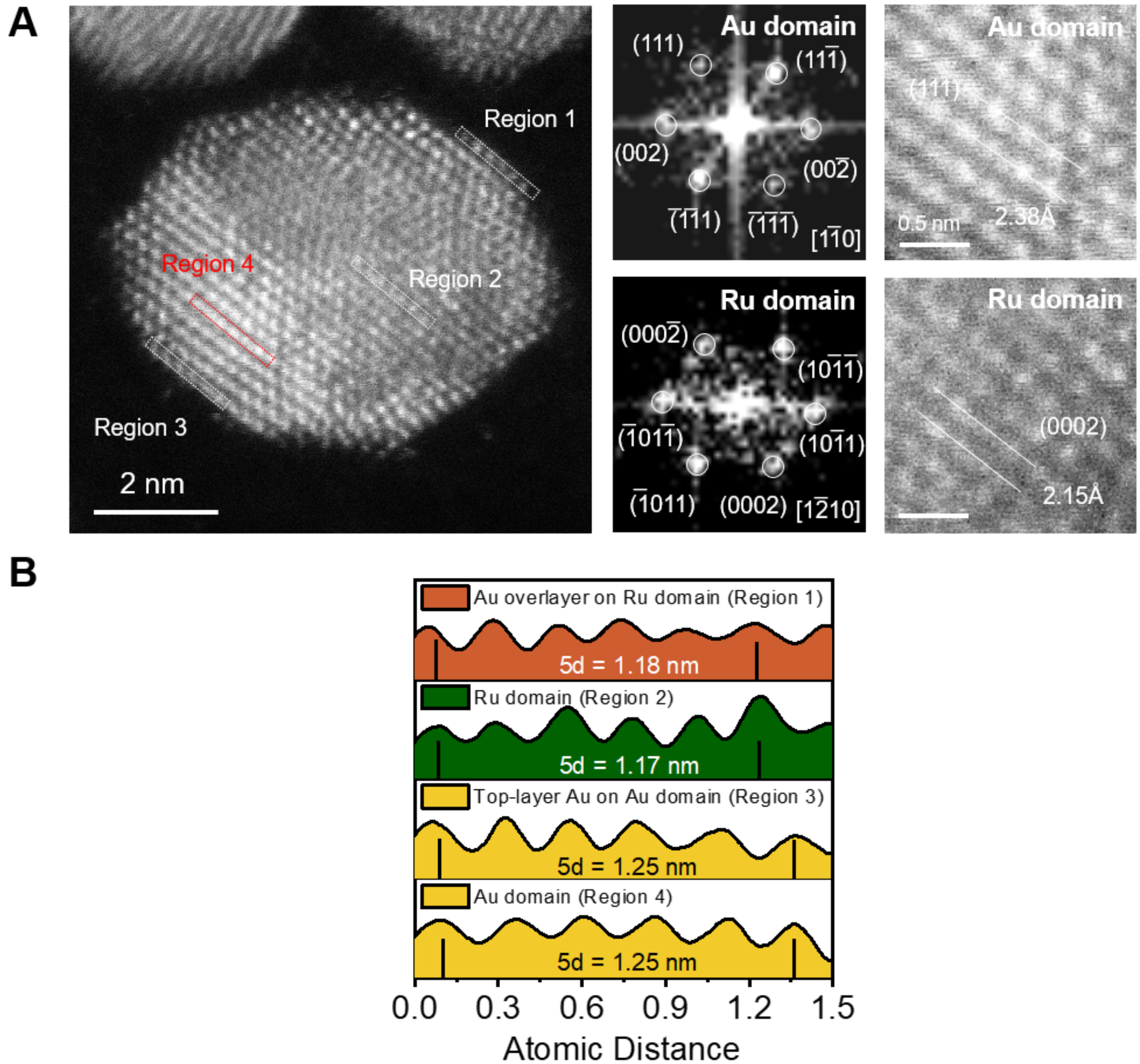

**Fig. S37.**
(A) Atomic resolution HAADF-STEM images of AuRu nanoparticles and fast Fourier transform (FFT) analysis of the Au and Ru domains. (B) Line-scan profiles extracted from the STEM images of AuRu nanoparticles. Region 1 is the Au overlayer. Region 2 is the Ru domain with Au overlayer. Region 3 is the edge of Au domain. Region 4 is the Au domain. Au and Ru adopt face-centered cubic and hexagonal close-packed (hcp) structures, respectively. The (111) interplanar spacing in the Au domain is approximately 2.38 Å and the (0002) interplanar spacing in the Ru domain is approximately 2.15 Å. FFT analysis of the Au and Ru domains show characteristic diffraction spots consistent with the [110] zone axis of an Au fcc structure and the [1210] zone axis of a Ru hcp structure, confirming that the Au and Ru domains are crystalline and possess fcc and hcp structures, respectively.

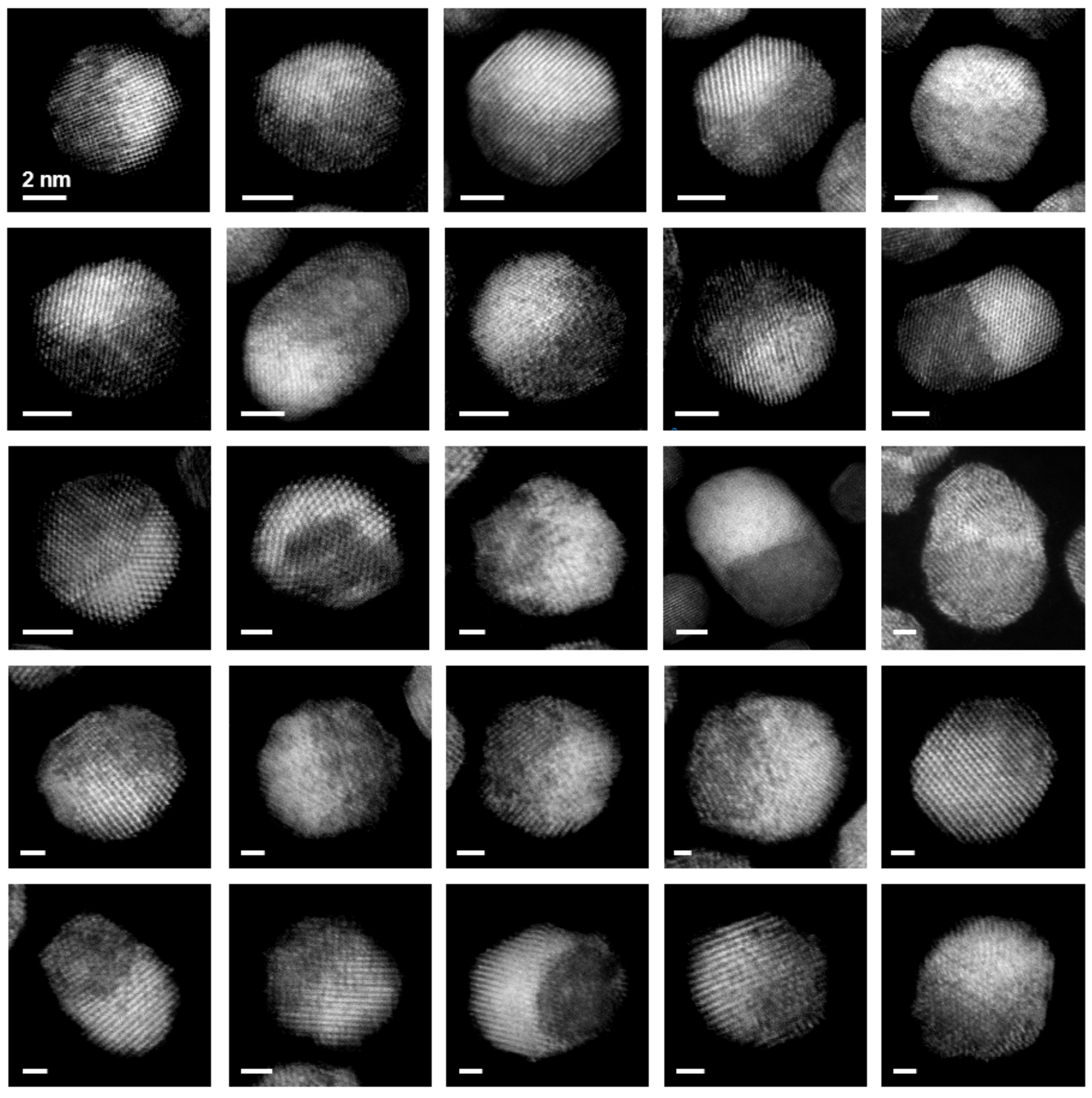

**Fig. S38.**
HAADF-STEM images of AuRu nanoparticles decorated by PEG and annealed at 500 °C.

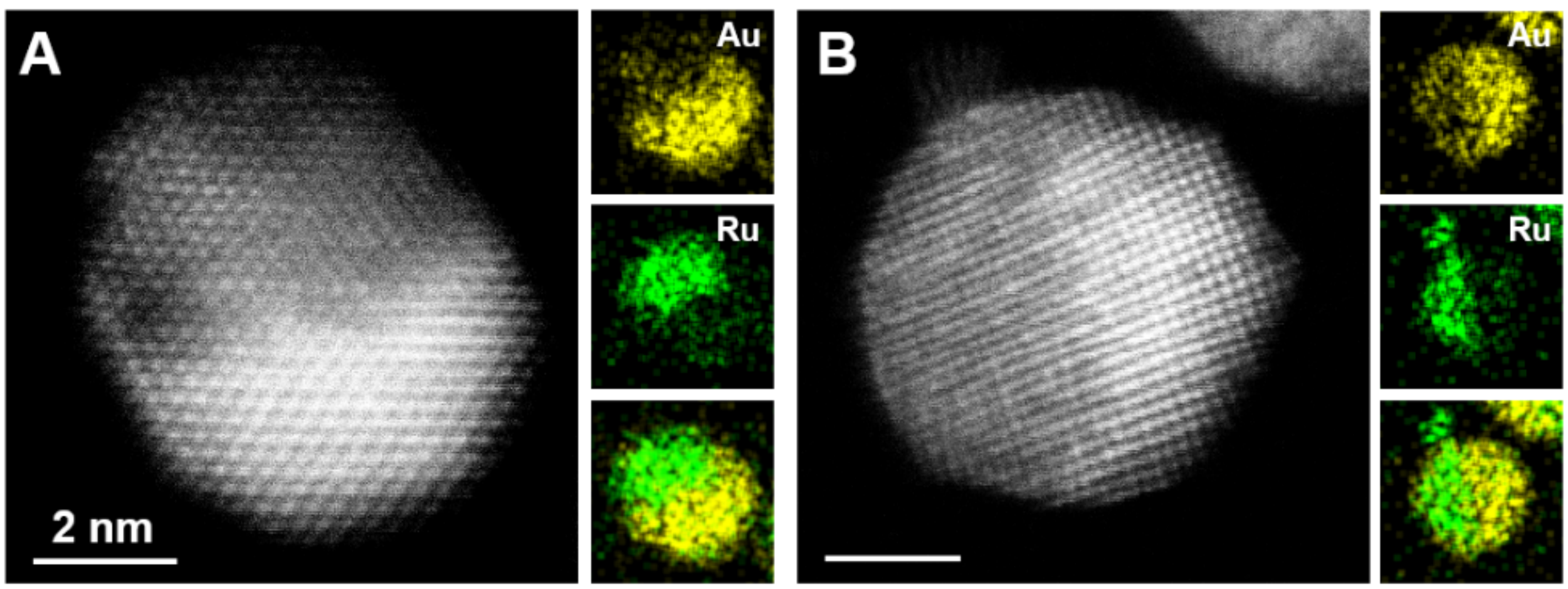

**Fig. S39.**
HAADF-STEM images and EDS elemental mapping of AuRu nanoparticles decorated by PEG and annealed at 500 °C. Unlike surface-clean AuRu nanoparticles with Au overlayers formed on the Ru domains, Au overlayers are hardly observed in AuRu nanoparticles that have been modified by PEG and annealed at 500 °C.

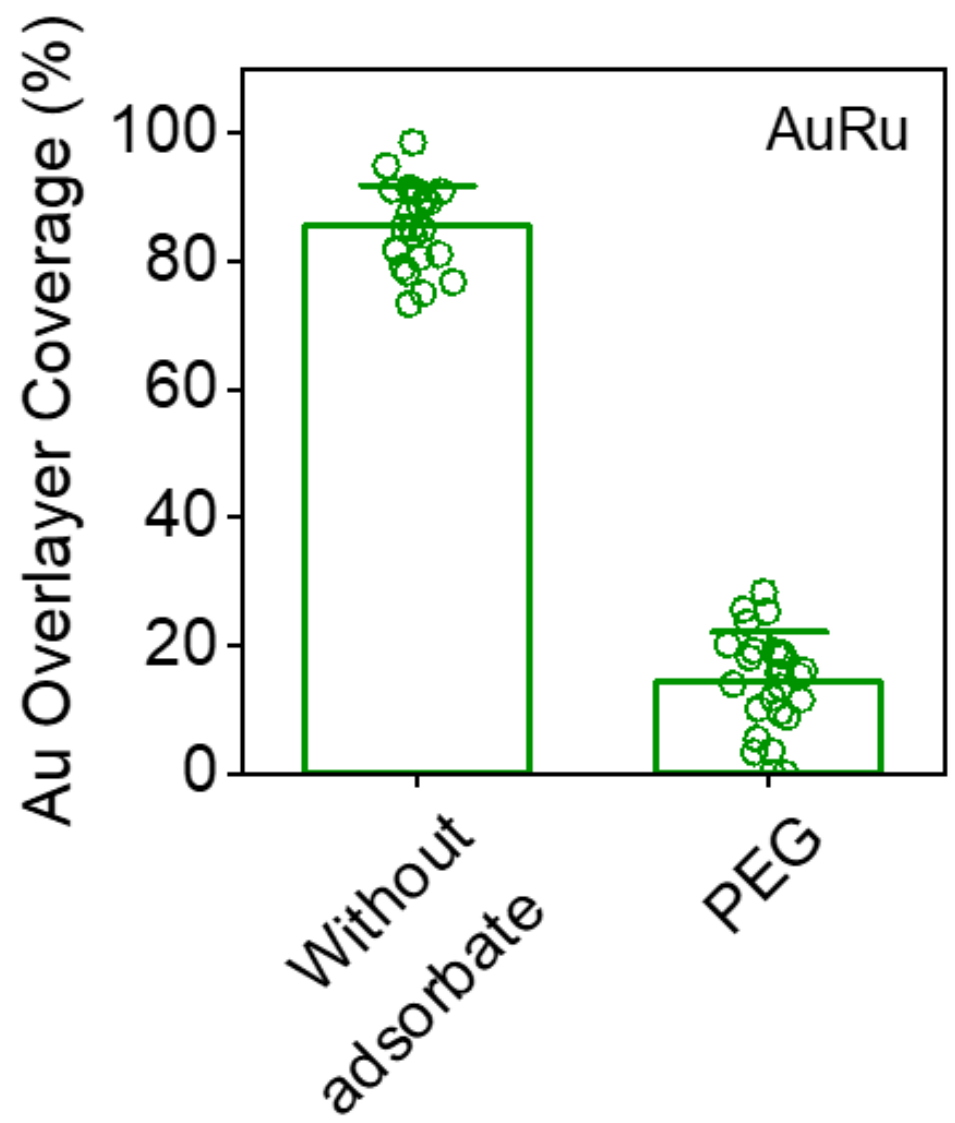

**Fig. S40.**
Au overlayer coverage on the Ru domains of adsorbate-free and passivated AuRu nanoparticles.

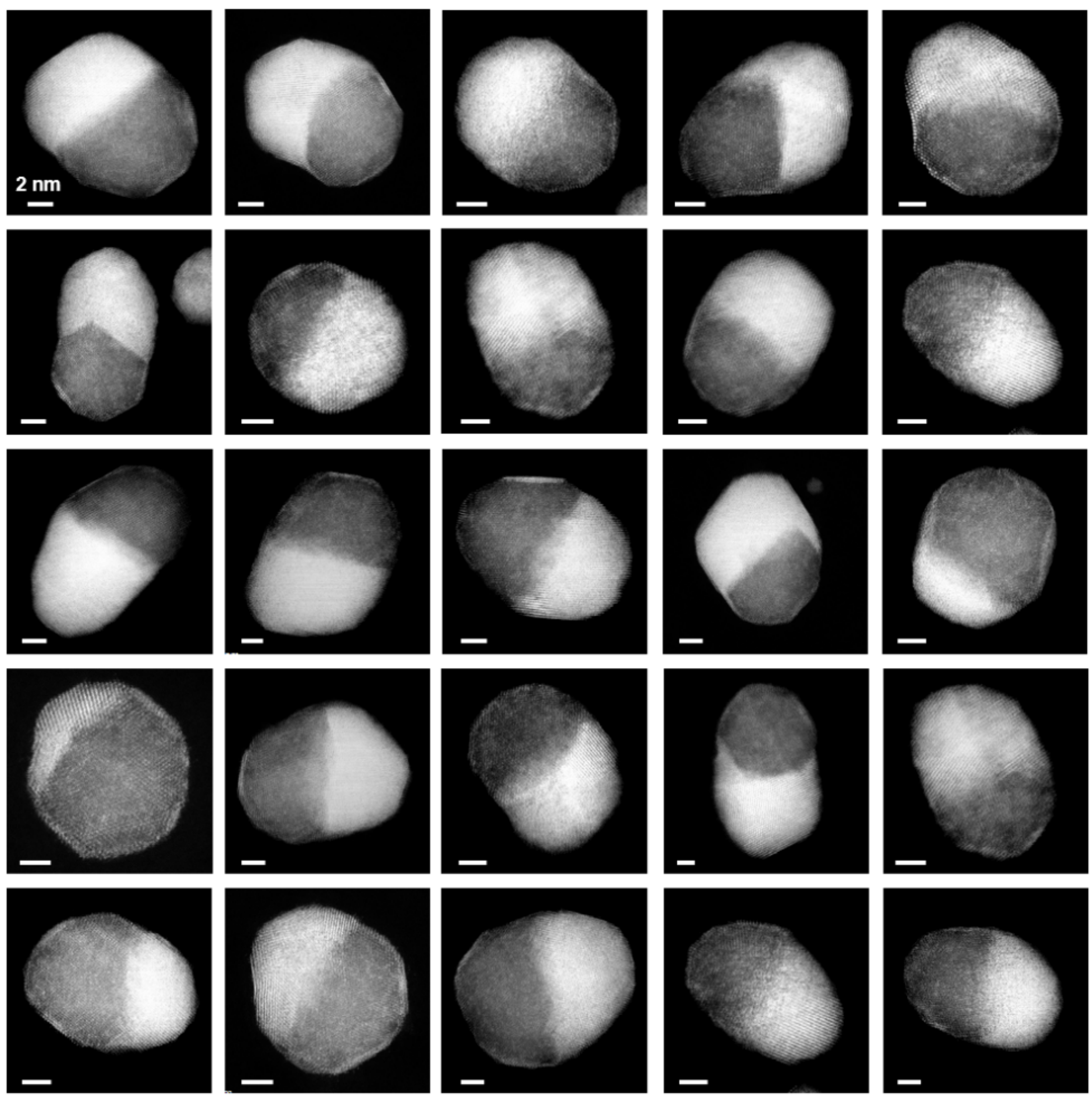

**Fig. S41.**
HAADF-STEM images of AuNi nanoparticles.

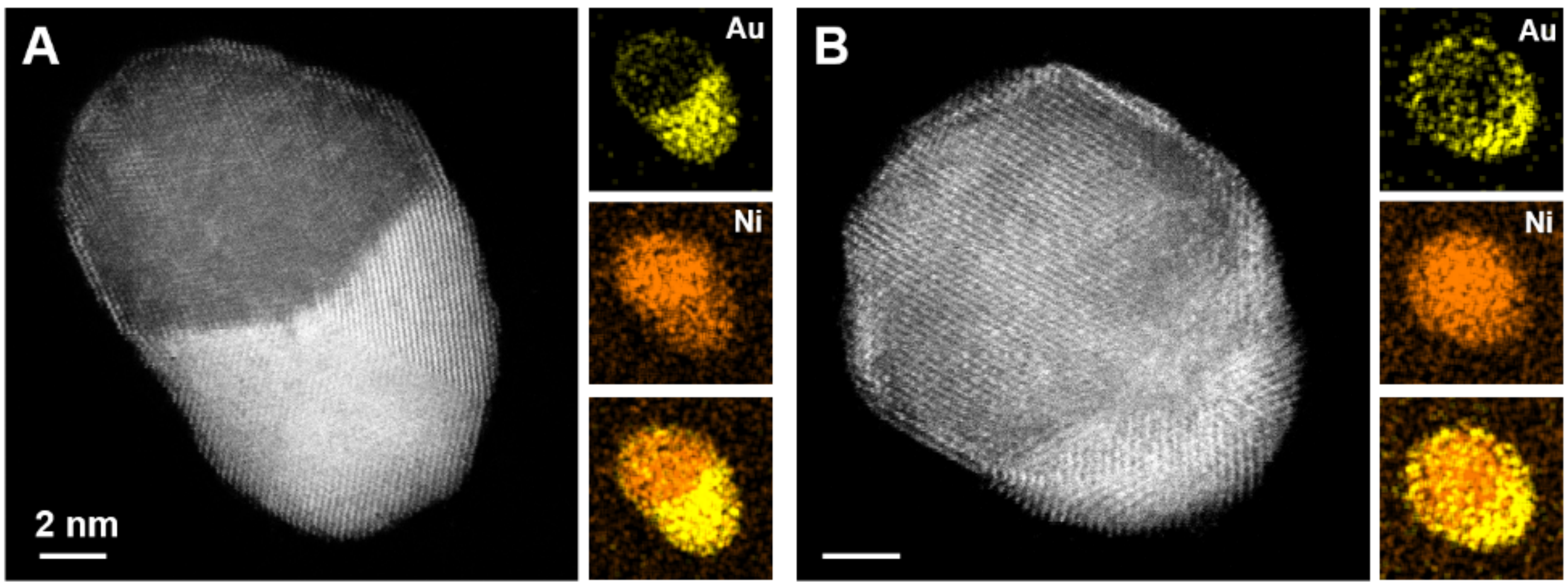

**Fig. S42.**
HAADF-STEM images and EDS elemental mapping of AuNi nanoparticles. Au and Ni are immiscible with each other, forming a phase-separated structure with an ultrathin Au overlayer on the surface of the Ni domain.

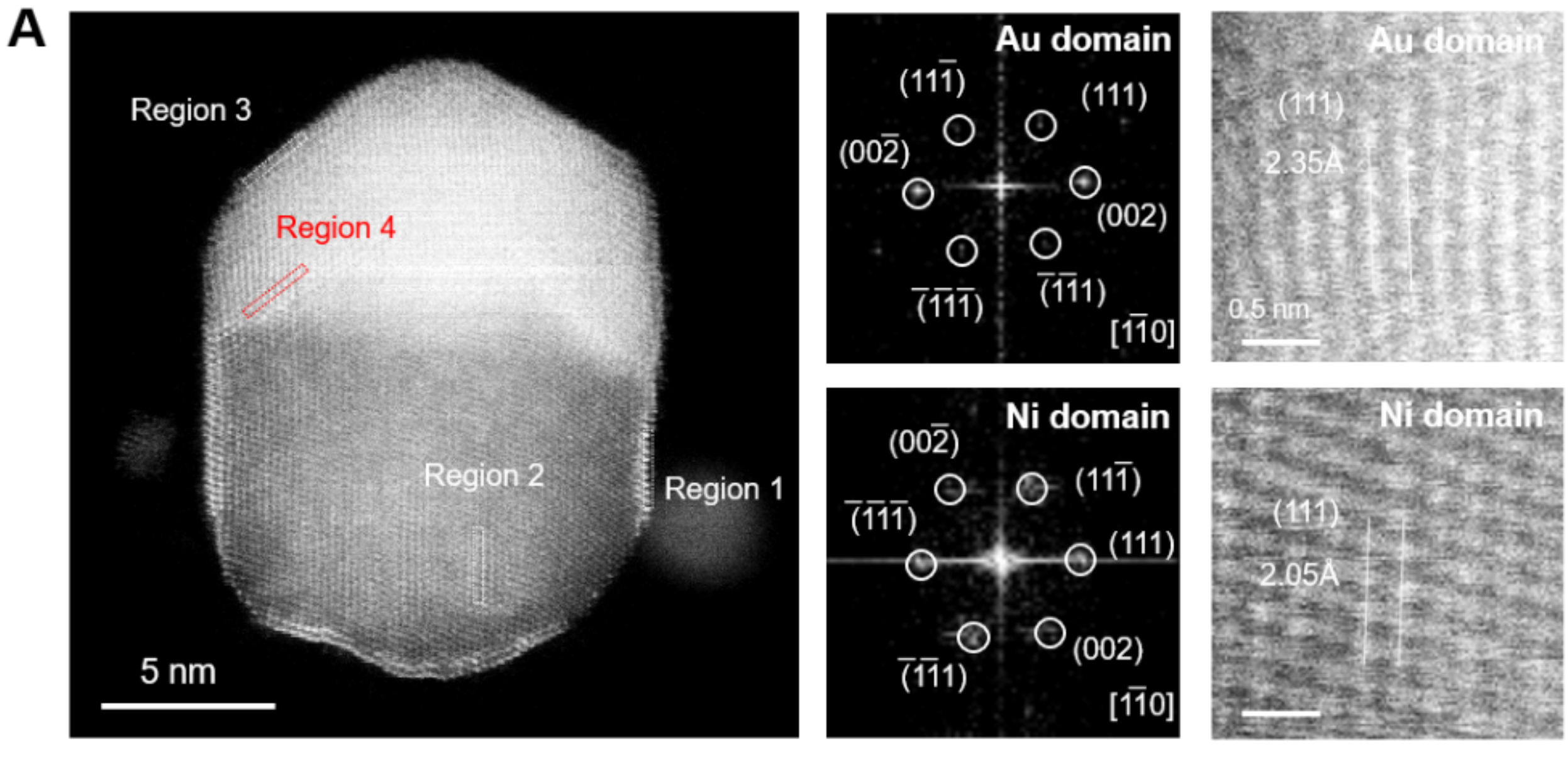

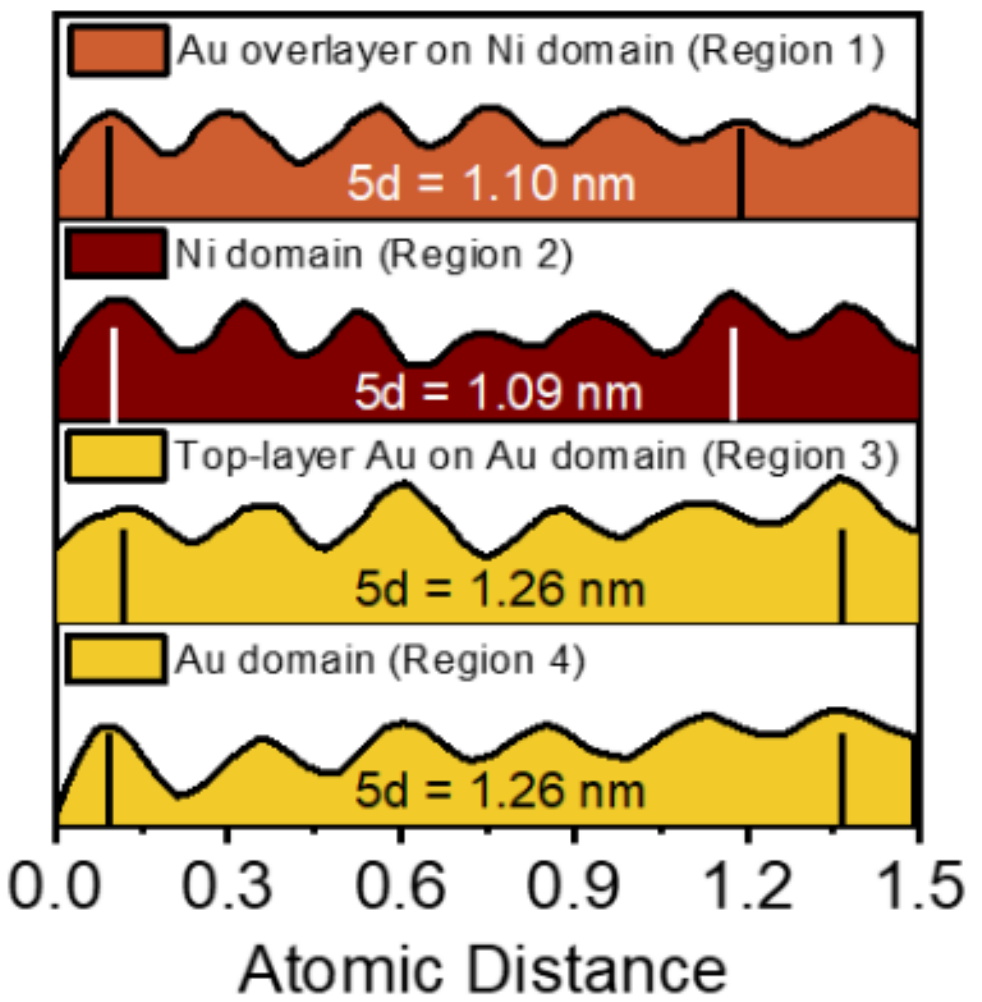

**Fig. S43.**
(A) Atomic resolution HAADF-STEM images of AuNi nanoparticles and fast Fourier transform (FFT) analysis of the Au domain and Ni domain. (B) Line-scan profiles extracted from the STEM images of AuNi nanoparticles. Region 1 is the Au overlayer. Region 2 is the Ni domain with Au overlayer. Region 3 is the edge of Au domain. Region 4 is the Au domain. Both Au and Ni adopt a face-centered cubic (fcc) structure but possess different unit cell parameters (Au: 4.07 Å; Ni: 3.52 Å). The measured (111) interplanar spacings are approximately 2.35 Å for the Au domain and 2.05 Å for the Ni domain. FFT analysis of the Au and Ni domains exhibit characteristic diffraction spots consistent with the [110] zone axis of an Au fcc and a Ni fcc structures, confirming that the Au and Ni domain are crystalline with an fcc structure.

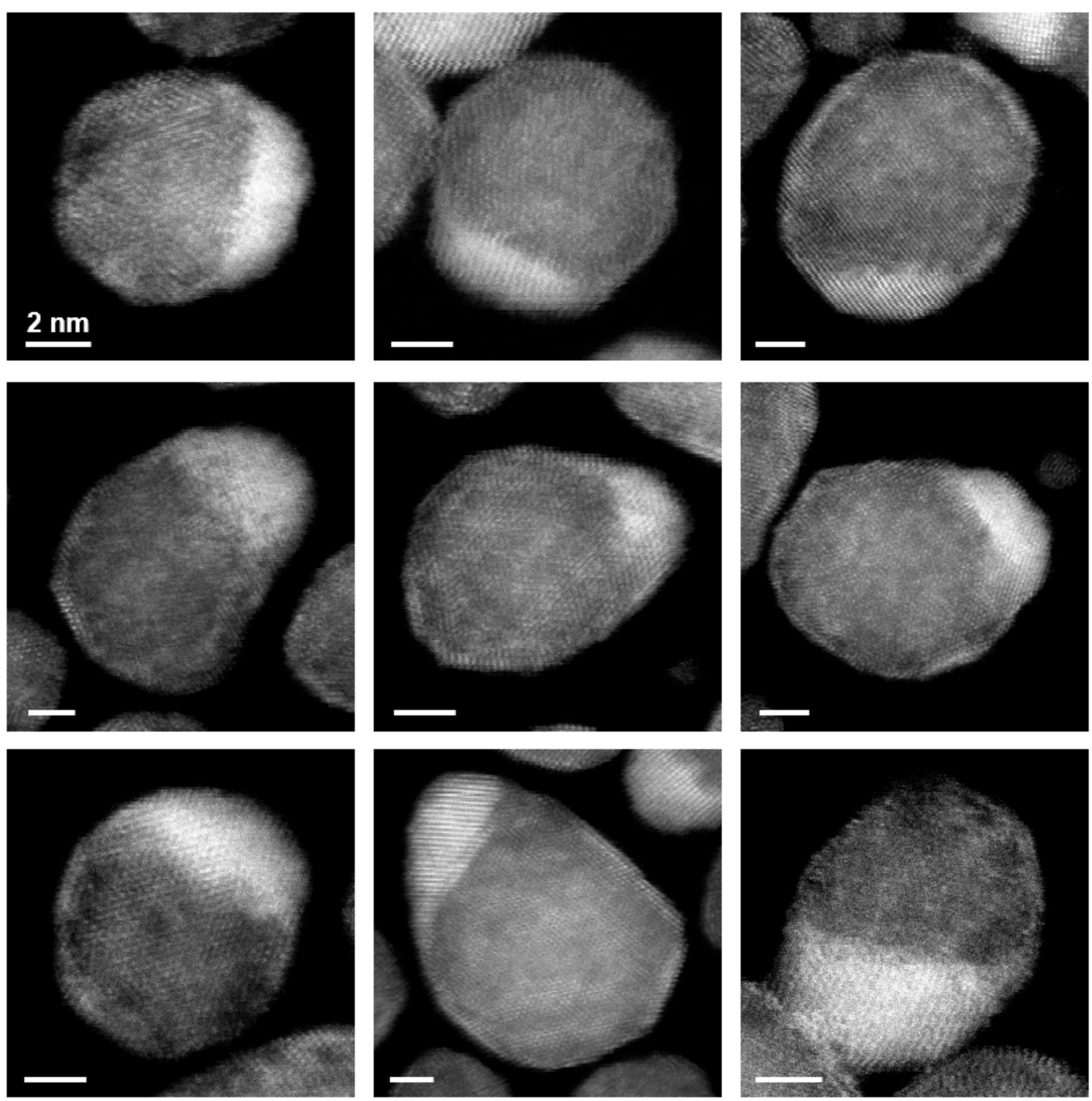

**Fig. S44.**
HAADF-STEM images of AuCo nanoparticles.

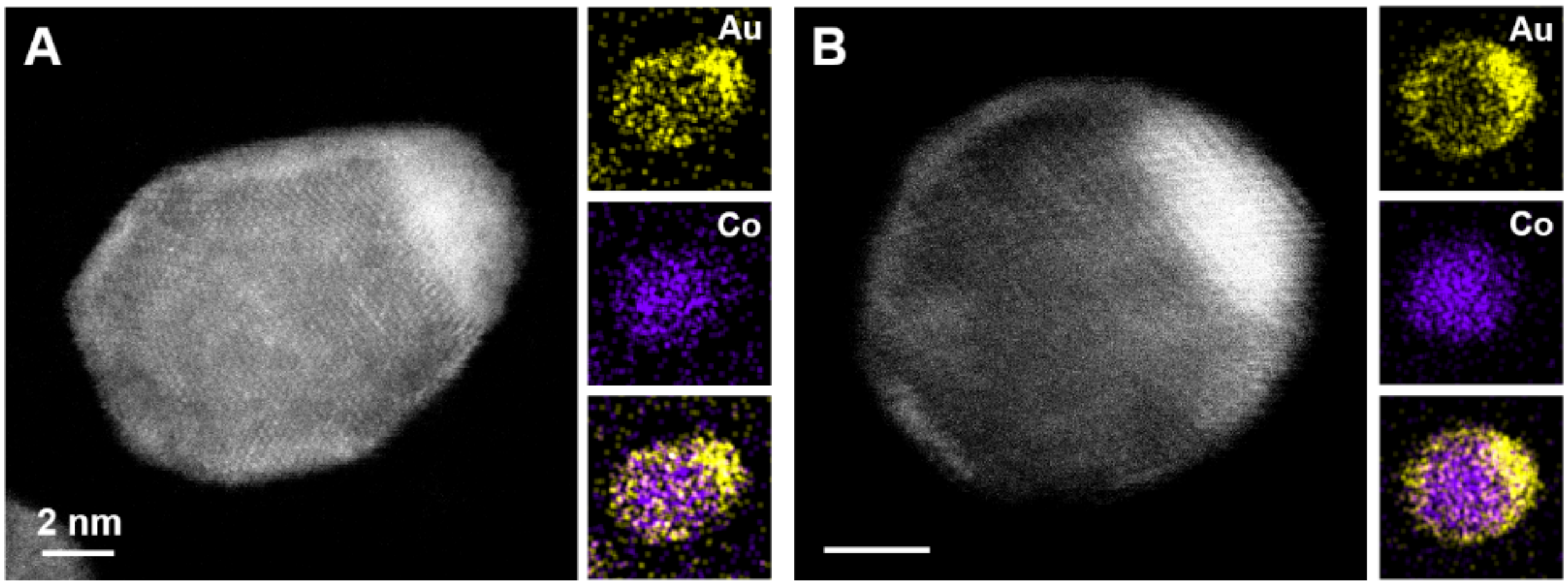

**Fig. S45.**
HAADF-STEM images and EDS elemental mapping of AuCo nanoparticles. Au and Co are immiscible with each other, forming a phase-separated structure with an ultrathin Au overlayer on the surface of the Co domain.

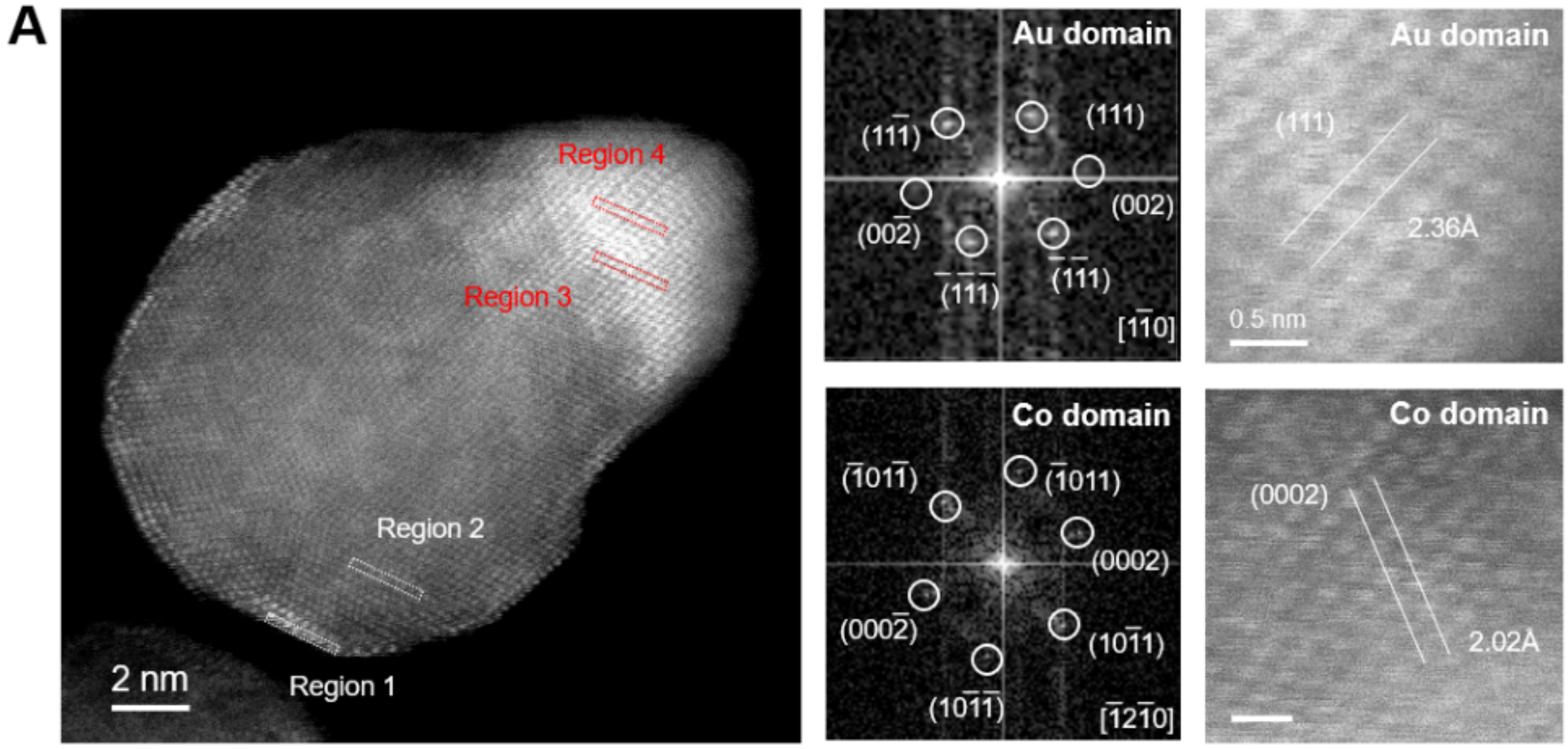

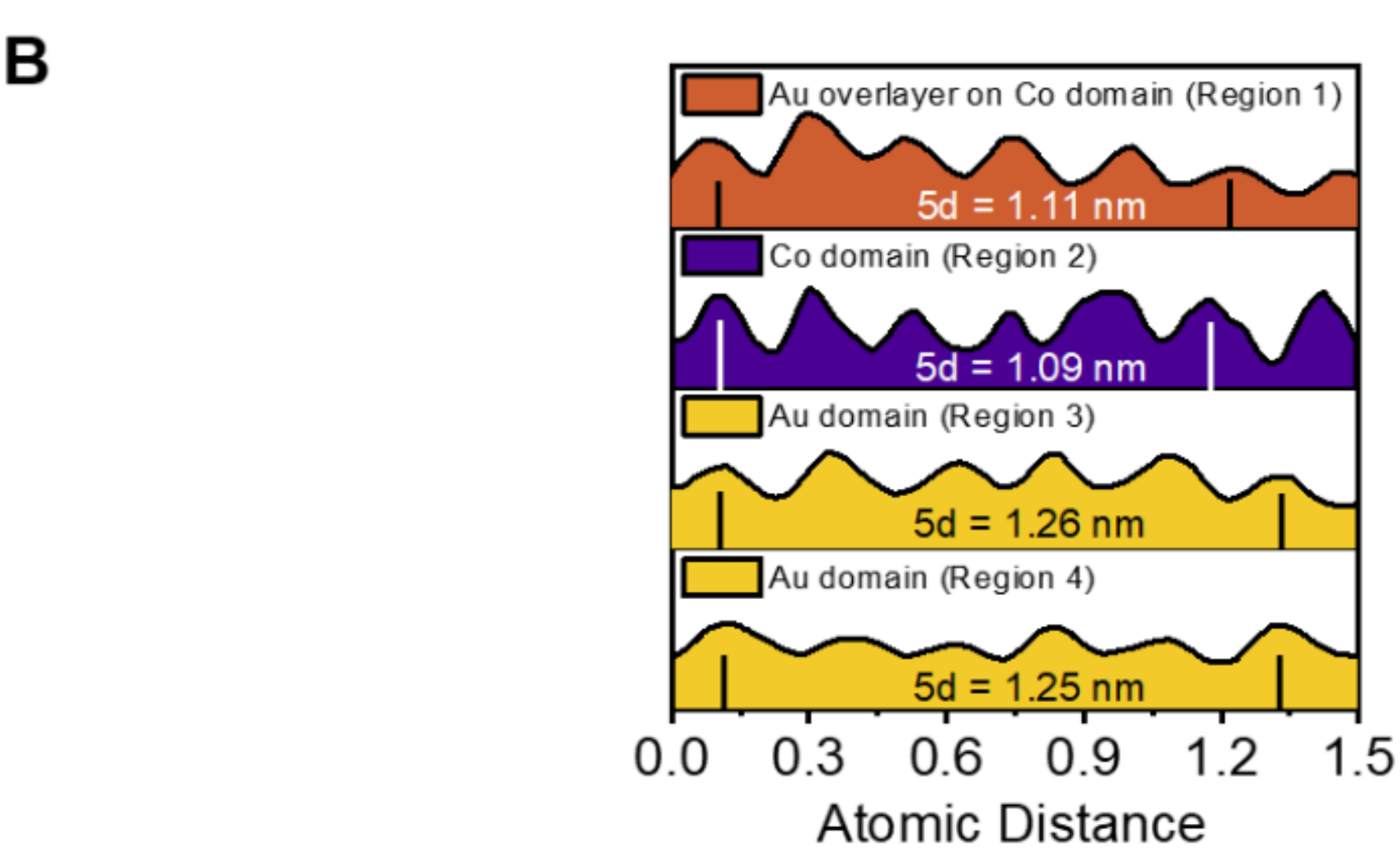

**Fig. S46.**
(A) Atomic-resolution HAADF-STEM images of AuCo nanoparticles and fast Fourier transform (FFT) analysis of the Au domain and Co domain. (B) Line-scan profiles extracted from the STEM images. Region 1 is the Au overlayer. Region 2 is the Co domain with Au overlayer. Regions 3 and 4 are the Au domain. Au and Co exhibit fcc and hcp structures, respectively. In the STEM images, the (111) interplanar spacing in the Au domain is approximately 2.36 Å and the (0002) interplanar spacing in the Co domain is approximately 2.02 Å. FFT analysis of the Au and Co domains display characteristic diffraction spots consistent with the [110] zone axis of an Au fcc structure and the [1210] zone axis of a Co hcp structure. This confirms that the Au and Co domains are crystalline and possess fcc and hcp structures, respectively.

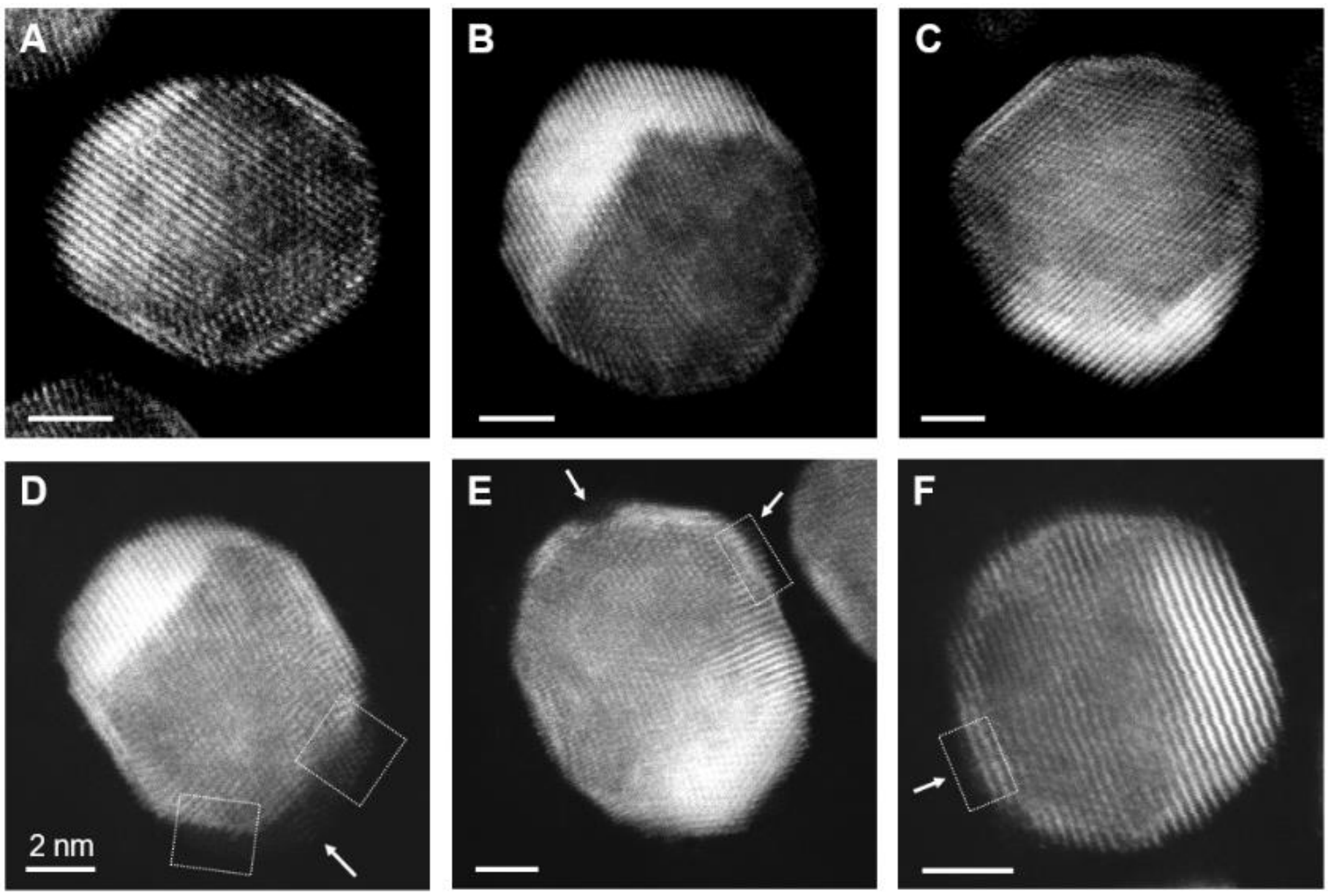

**Fig. S47.**
HAADF-STEM image of (A-C) as-synthesized AuCo nanoparticles and (D-F) AuCo nanoparticles after being exposed to air at room temperature for 30 mins. Dashed rectangles and white arrows highlight the cobalt oxide segregated to the nanoparticle surface.

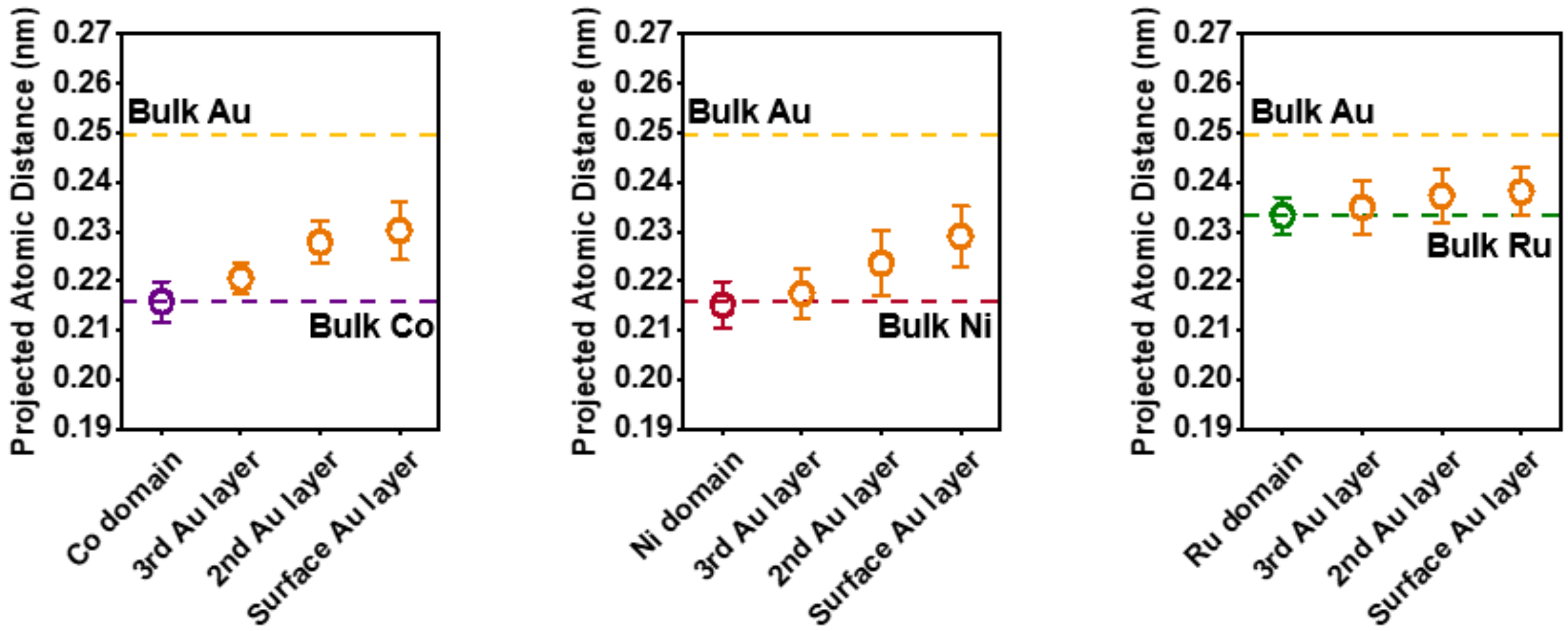

**Fig. S48.**

Strain analysis of the Au overlayers in AuCo, AuNi, and AuRu nanoparticles in the direction parallel to the nanoparticle surface. The projected atomic distance in the Ni domain is measured within the (111) planes viewed along the [110] zone axis. The projected atomic distance in the Co or Ru domain is measured within the (0002) planes viewed along the [1210] zone axis. For bulk Au fcc structure, the projected atomic distance within the (111) planes is 0.2497 nm. The corresponding projected atomic distance of hcp Co and Ru are 0.2158 nm and 0.2332 nm, respectively, and the projected atomic distance of fcc Ni is 0.2157 nm. To accommodate the atoms in the underlying Co, Ni, or Ru domains, the Au overlayers on Co, Ni, and Ru domains are subjected to varying degrees of compressive strains. Consequently, the projected atomic distance in the Au overlayers increases from the inner layer to the outer layer, but the atomic distance in the outmost layer remains smaller than that of bulk Au.

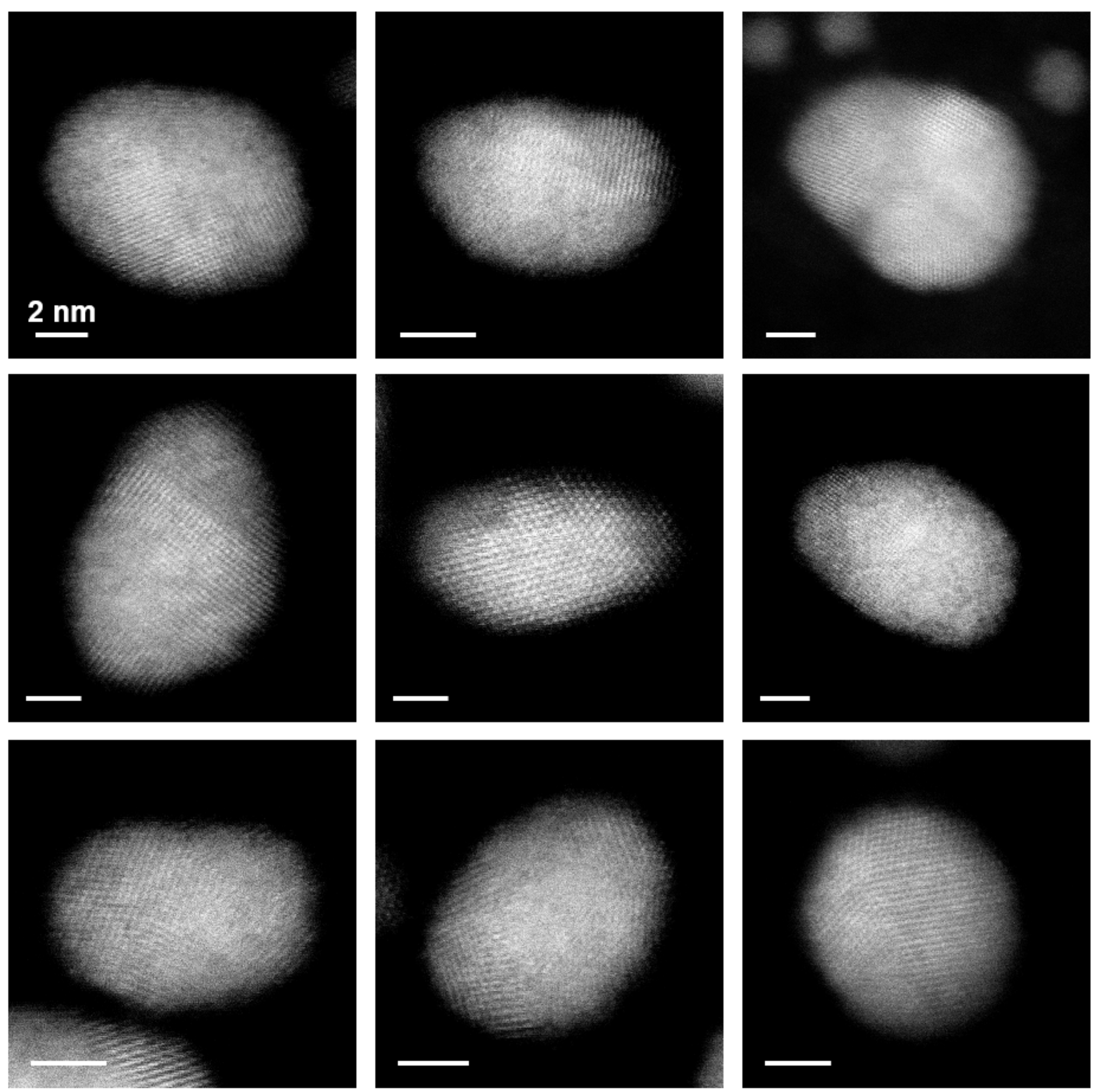

**Fig. S49.**
HAADF-STEM images of CuRh nanoparticles.

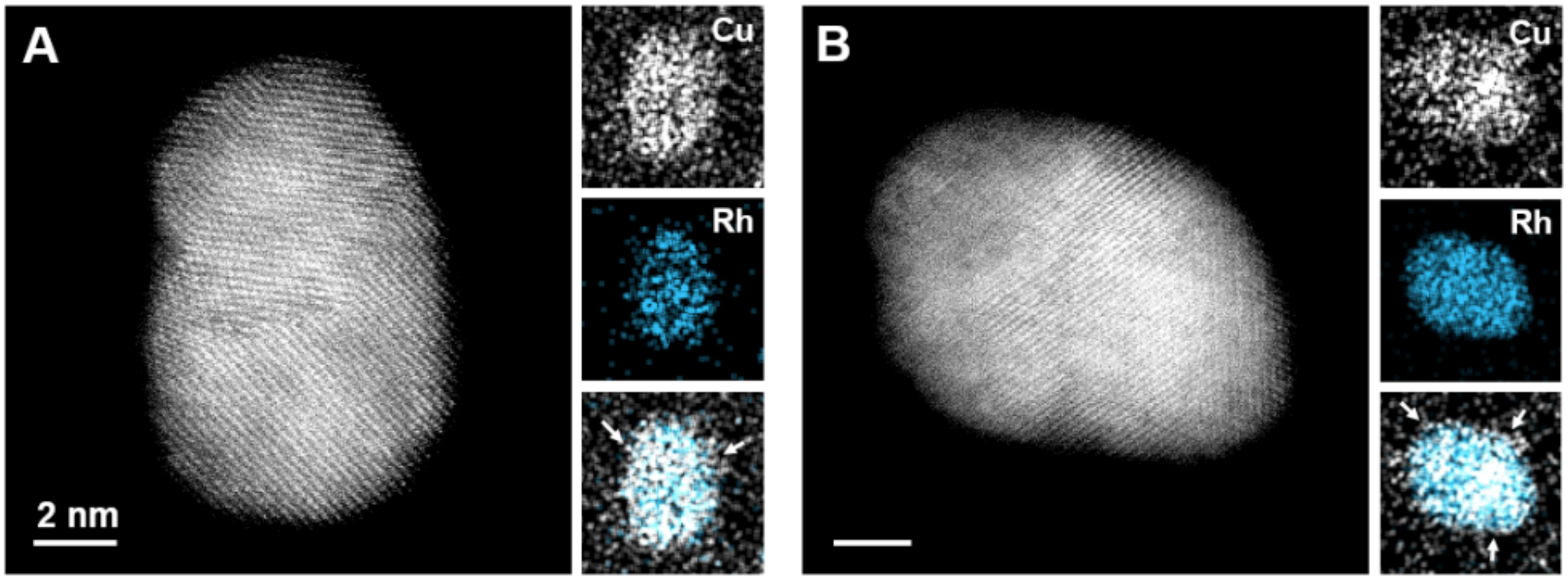

**Fig. S50.**
HAADF-STEM images and EDS elemental mapping of CuRh nanoparticles. White arrows indicate the surface segregation of Cu in the nanoparticles. EDS elemental mapping indicates partial alloying between Cu and Rh. Due to the lower surface energy of Cu compared to Rh, surface segregation of Cu is observed in CuRh nanoparticles.

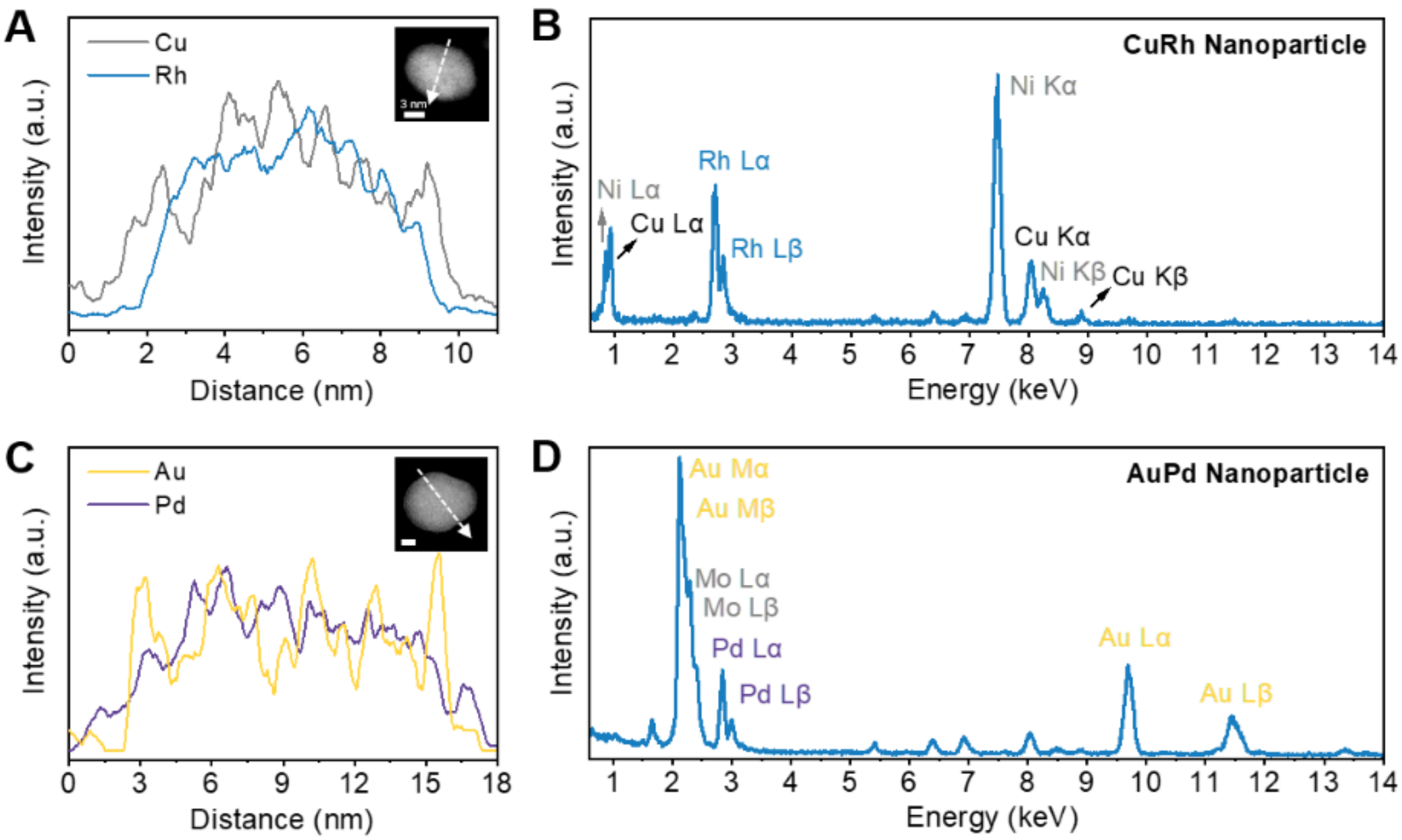

**Fig. S51.**
EDS line scans and EDS spectra of (A,B) CuRh and (C,D) AuPd nanoparticles. The white arrows in the inset HAADF-STEM images show the traces of EDS line scans.

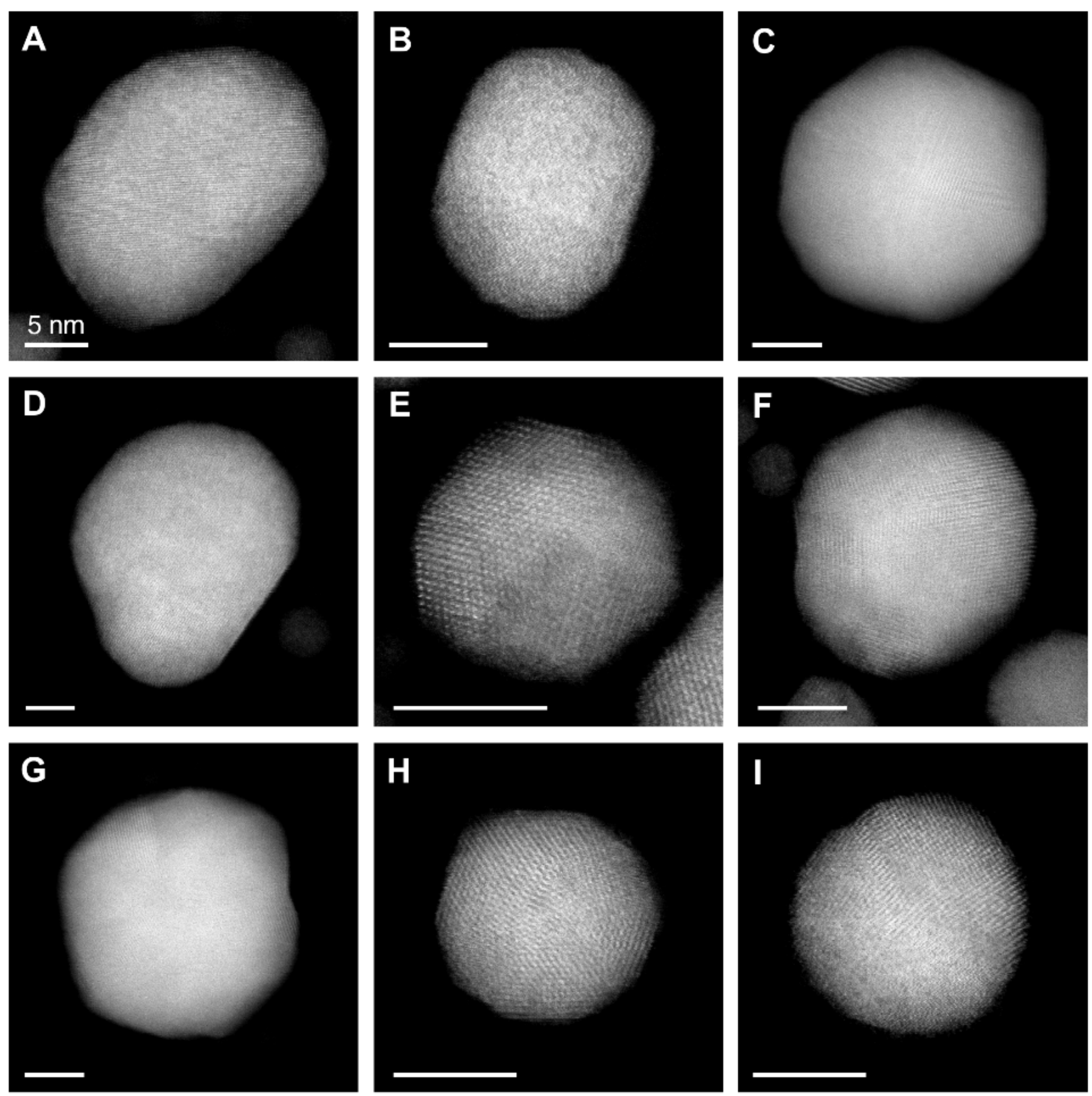

**Fig. S52.**
HAADF-STEM images of AuPd nanoparticles.

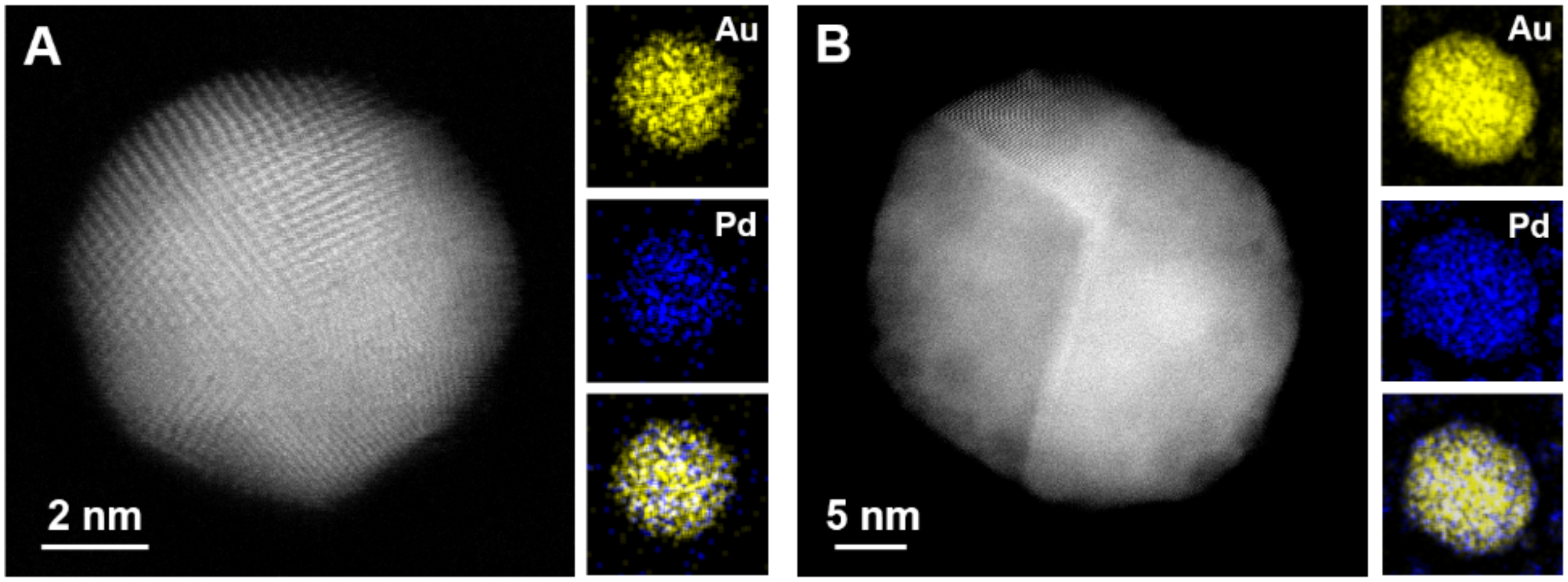

**Fig. S53.**
HAADF-STEM images and EDS elemental mapping of AuPd nanoparticles. EDS elemental mapping confirms the high miscibility between Au and Pd. Surface segregation has not been found in AuPd nanoparticles.